\documentclass[12pt]{article}

\usepackage[margin=1in]{geometry}
\usepackage{setspace}
\usepackage{graphicx}
\usepackage{natbib}
\usepackage{xcolor}
\usepackage{amsmath}
\usepackage{mathrsfs}
\usepackage{amsfonts}
\usepackage{dsfont}
\usepackage{diagbox}
\usepackage{makecell}
\usepackage{color}
\usepackage{hyperref}
\usepackage{multirow}
\usepackage{mathtools}
\usepackage{comment}
\usepackage{mathtools}
\usepackage{booktabs}
\usepackage{float}

\newcommand{\eqaldot}{\mathrel{=\mkern-0.25mu:}}
\usepackage{mathabx}
\usepackage{calligra}

\usepackage[perpage]{footmisc}

\DeclareMathOperator{\tf}{\mathfrak{t}}
\newcommand{\de}{\mathrm{d}}
\newcommand{\M}{\mathds{M}}
\newcommand{\ee}{\mathds{E}}
\newcommand{\x}{{\mathbf x}}
\newcommand{\xl}{{\mathbf x}^{\mathds{L}}}
\newcommand{\X}{\breve{X}}

\newcommand{\R}{\mathds R}

\newcommand{\dL}{d_{\mathds{L}}}
\newcommand{\LL}{\mathds{L}}
\newcommand{\rL}{r_{\LL}}
\newcommand{\1}{{\mathbf 1}}

\DeclareMathOperator{\cov}{\mathds{C}ov}

\newcommand{\mo}{\mathfrak{o}}

\graphicspath{{./Figures/}}

\begin{document}

\title{
Marked spatial point processes: current state and extensions to point processes on linear networks 
}

\maketitle
\begin{center}
{{\bf Matthias Eckardt$^{a}$} and {\bf Mehdi Moradi}$^{b}$}\\
\noindent $^{\text{a}}$ Chair of Statistics, Humboldt-Universit\"{a}t zu Berlin, Spandauer Strasse 1 , Berlin, Germany\\
\noindent $^{\text{b}}$ Department of Mathematics and Mathematical Statistics,  Ume\r{a} University, Ume\r{a}, Sweden\\
\end{center}

\begin{abstract}

Within the applications of spatial point processes, it is increasingly becoming common that events are labeled by marks, prompting an exploration beyond the spatial distribution of events by incorporating the marks in the undertaken analysis. In this paper, we first consider marked spatial point processes in $\R^2$, where marks are either integer-valued, real-valued, or object-valued, and review the state-of-the-art to analyze the spatial structure and type of interaction/correlation between marks. More specifically, we review cross/dot-type summary characteristics, mark-weighted summary characteristics, various mark correlation functions, and frequency domain approaches. Second, we propose novel cross/dot-type higher-order summary characteristics,  mark-weighted summary characteristics, and mark correlation functions for marked point processes on linear networks. Through a simulation study, we show that ignoring the underlying network gives rise to erroneous conclusions about the interaction/correlation between marks. Finally, we consider two applications: the locations of two types of butterflies in Melbourne, Australia, and the locations of public trees along the street network of Vancouver, Canada, where trees are labeled by their diameters at breast height.
\end{abstract} 


\section{Introduction}\label{sec:intro}
Significant advancements in data collection and storage capacities have led to vast point pattern data availability from diverse sources, which often give access to (precise) spatial locations and time occurrences of events together with further valuable point-specific information, i.e. marks. 
In some cases, spatial locations might be restricted by the entities of some spatially embedded relational systems, which consequently limit where
events could occur. Thus, the state space that accommodates the events plays a vital role in advanced statistical analysis.
Not surprisingly, the growing availability and accessibility of point pattern data, which find applications in distinct scientific fields, generally stimulated an increasing interest in developing suitable statistical/mathematical tools for the analysis of point patterns.
Within the literature, applications often include astronomy \citep{feigelson2012modern},
zoology \citep{russell2016dynamic},
ecology \citep{daniel2020efficient}, forestry \citep{Gavrikov1995, yazigi2021modeling}, 
geology \citep{flagg2020modeling}, and health \citep{bayisa2023regularised}.  Within these applications, typical examples of marks, which are point specific, include, e.g., the shape of galaxies, animal sightings of (non)invasive species, the diameter of trees at breast height (dbh), habitat characteristics, the magnitude of earthquakes, and exposure levels to pollutants. 
In any such application, the interesting aim is not only to make inferences about the spatial distribution of events and their potential interaction but also to understand the association among the corresponding marks. 
For instance, in the case of forestry and ecology,  
spatial variation of dbh measurements for pairs of distinct trees and 
spatial correlation between sighting locations for pairs of different animal species might be interesting. Note that understanding how the dbh values differ across spatial locations might provide valuable insights into tree growth patterns, resource distribution, and ecological interactions within the forest ecosystem. Also, studying sighting locations might uncover relationships and dependencies between animal distributions, shedding light on potential habitat preferences, interspecies dynamics, and ecological coexistence patterns.
These objectives are usually addressed by employing different so-called mark summary characteristics, which are expected to
account for the specificity of the marks and, potentially, the 
constrained spatial domain of the points. This paper discusses the existing methodologies and state-of-the-art for marked spatial point processes in $\R^2$ and proposes some novelties in the context of point processes on linear networks.

The literature for spatial point processes usually employs diverse exploratory tools, such as summary characteristics, to e.g. investigate both pairwise and higher-order interactions between points, as well as to validate fitted models.
As a general construction principle, any such summary characteristic quantitatively assesses the average interrelation between points and/or marks within a specific interpoint distance.
Initially, the main attention was paid to unmarked stationary and homogeneous point processes, where the spatial distribution of points is translation invariant, and the intensity function remains constant over the corresponding state space, leading to the development of the $K$-function \citep{Ripley1976}, used to study the pairwise relationships between points, and the $J$-function \citep{VanLieshoutBaddeley1999}, which goes beyond the second-order analysis. Both these summary characteristics are used to identify clustering and/or inhibition among points.  
These were later extended to marked stationary point processes, developing cross-type second/higher-order summary characteristics for point processes with integer-valued marks \citep{Lotwick1982, Lieshout2006}, and mark-weighted second-order summary characteristics for real-valued marks \citep{pettinen1992forest, Schlather2001}. 
A different line of research focused on the pairwise association/variation of real-valued marks, including the mark variogram \citep{cressie93,markvar, Stoyan2000}, which has similarities with the (semi-)variogram commonly used in geostatistical contexts, and Stoyan's mark correlation function \citep{StoyanStoyan1994}. Similar to the classic point process characteristics, any such tool is constructed based on the marks for pairs of points at arbitrary interpoint distances. While, in practice, points might exhibit inhomogeneity, all of these summary characteristics for real-valued marks are only defined for stationary point processes, and their inhomogeneous versions remain an open topic for future research. 
Other approaches to investigating the dependencies between marks and spatial locations are proposed by \citet{Schlather2004, Guan2006, Guan2007a}. Moreover, some recent developments for the analysis of stationary spatial point processes, which simultaneously possess both integer-valued and real-valued marks, include some graphical model approaches, and partial characteristics proposed by \cite{EckardtISR, Eckardt2018partial}. Instead of considering a spatial domain perspective, they applied a frequency domain representation to the marked points to compute different (partial) spectral density characteristics. This highlights the energy distribution of the point patterns over a range of frequencies, unlike the summary characteristics which focus on point-wise spatial interactions. Although frequency domain methods offer a highly flexible and computationally efficient way to investigate the structural interrelation of complex point processes, methodological contributions and practical applications within the spatial point process literature remain limited.

Turning back to unmarked point processes, \cite{InhomK2000} and \cite{van11} extended the second- and higher-order summary characteristics to inhomogeneous settings for particular classes of point processes, including second-order intensity-reweighted stationary and intensity-reweighted moment stationary processes; the former is a particular case of the latter. The extended versions of these summary characteristics to take integer-valued marks into account are proposed by \cite{MollerWaagepetersenBook, Cronie2016}. However, as of today, to the best of our knowledge, methods for real-valued marks still rely on the stationary assumption, and their extensions to non-stationary setting remains open and challenging. 

In an effort to study non-integer/real-valued marks, the focus is directed toward proposing novel methodologies capable of handling diverse forms of marks. More specifically, by borrowing ideas from functional data analysis \citep{ramsey2005}, extensions of Stoyan's mark correlation function \citep{StoyanStoyan1994} to function-valued marks are proposed by \citet{Comas2011, Comas2013} for stationary point processes, and \cite{Ghorbani2020} proposed a framework for functional marked point processes together with some weighted marked
reduced moment measures. Moreover,  \cite{Eckardt2023MultiFunctionMarks} and \cite{EckardtMariCMSPP} extended some summary characteristics for spatial point processes with integer-valued marks to the case of multivariate point processes with multivariate function-valued marks and constrained vector-valued quantities. These non-integer valued marks are summarized into the class of object-valued marks, where instead of a scalar mark, each point is augmented by a non-scalar mark, i.e. an object-valued attribute, that lives on a suitable mark space whose precise form depends on the object under study. Suitable choices for the mark space include the Banach/Hilbert space for function-valued marks and the simplex for constrained vector-valued quantities. As in standard mark point process investigations, these extended tools aim to explore the spatial variation/association of the specific objects, e.g. curves, for pairs of points at arbitrary distances.

In the past two decades, significant attention has been given to point processes on linear networks, where spatial locations of events are constrained. \cite{OY01} and \cite{XZY08} proposed a network-based version of Ripley's $K$-function and an intensity estimator, which replaced the Euclidean distance with the shortest-path distance without taking the geometry of the underlying network into account, leading to biased results; note that the underlying network itself often has a non-uniform distribution. As for taking the geometry of the underlying network into account, novel methodologies focusing on intensity function and summary characteristics were developed. Utilizing shortest-path distances, \cite{Ang2012} proposed geometrically-corrected second-order summary characteristics for both homogeneous and inhomogeneous point processes on linear networks. The developed framework by \cite{Ang2012} was subsequently expanded to the case of point processes on linear networks with integer-valued marks \citep{BaddeleyJammalamadakaNair2014}, and spatio-temporal point processes on linear networks \citep{MoradiMateu2020}.
Moreover, \cite{rakshit2017second} proposed the consideration of regular distances on linear networks and defined novel versions of the summary characteristics proposed by \cite{Ang2012} based on a more general class of metrics. Later, employing regular distances,  \cite{cronie2020inhomogeneous} proposed higher-order summary characteristics for point processes on linear networks. In terms of intensity functions, various techniques are proposed, which are either based on kernel functions \citep{mcswiggan2017, MFJ18} or Voronoi tessellations \citep{MoradiVor2019, MATEU2020Pseudo-separable}. Computational issues became evident on large networks,  leading to a fast kernel-based intensity estimator \cite{rakshit2019fast} and an efficient way to compute the $K$-functions for point processes on linear networks \cite{rakshit2019efficient}. Excluding the second-order summary characteristics for multitype point processes on linear networks proposed by \cite{BaddeleyJammalamadakaNair2014}, and a kernel-based smoothing approach for scalar marks on linear networks by \cite{rakshit2019fast}, the literature certainly lacks marked-based methodologies to analyze events occurring on linear networks. In particular, no specific contributions go beyond integer-valued marks.

Knowing the limitations mentioned above, we propose novel methodologies for spatial point processes on linear networks that possess either integer- or real-valued marks. More specifically, in the case of real-valued marks, we present extensions of various marked summary characteristics, e.g. Stoyan's mark correlation and mark-weighted summary characteristics to linear network settings. In the cases of multivariate/multitype point processes on linear networks, we take the higher-order summary characteristics, defined by \cite{cronie2020inhomogeneous}, and propose their cross versions, which are of great use to reveal the type of interactions between points with different marks. In Section \ref{sec:Data}, we present two motivating datasets, one on a planar state space and one on a linear network. We then, in Section \ref{sec:mpp}, provide a detailed overview of the state-of-the-art for the analysis of marked spatial point processes in $\R^2$. In particular, we start by presenting various summary characteristics for inhomogeneous multivariate/multitype point processes and then present several mark summary characteristics for stationary point processes with real-valued marks. This section further covers mark characteristics for object-valued marks as well as distinct frequency domain methods. Section \ref{sec:mlpp} starts by reviewing second-order summary characteristics for inhomogeneous multivariate/multitype point processes on linear networks together with defining their mark-weighted versions and then proposes higher-order summary characteristics for such point processes. It then extends several mark summary characteristics for point processes with real-valued marks to settings where events happen on linear networks, accompanied by a numerical evaluation highlighting the importance of considering the underlying networks. Lastly, in Section \ref{sec:realdata}, we present the outcomes obtained from the analysis of two considered datasets and close the paper with a discussion in Section  \ref{sec:final}.

\section{Data}\label{sec:Data}

\begin{figure}[!h]
    \centering
    \includegraphics[scale=0.07]{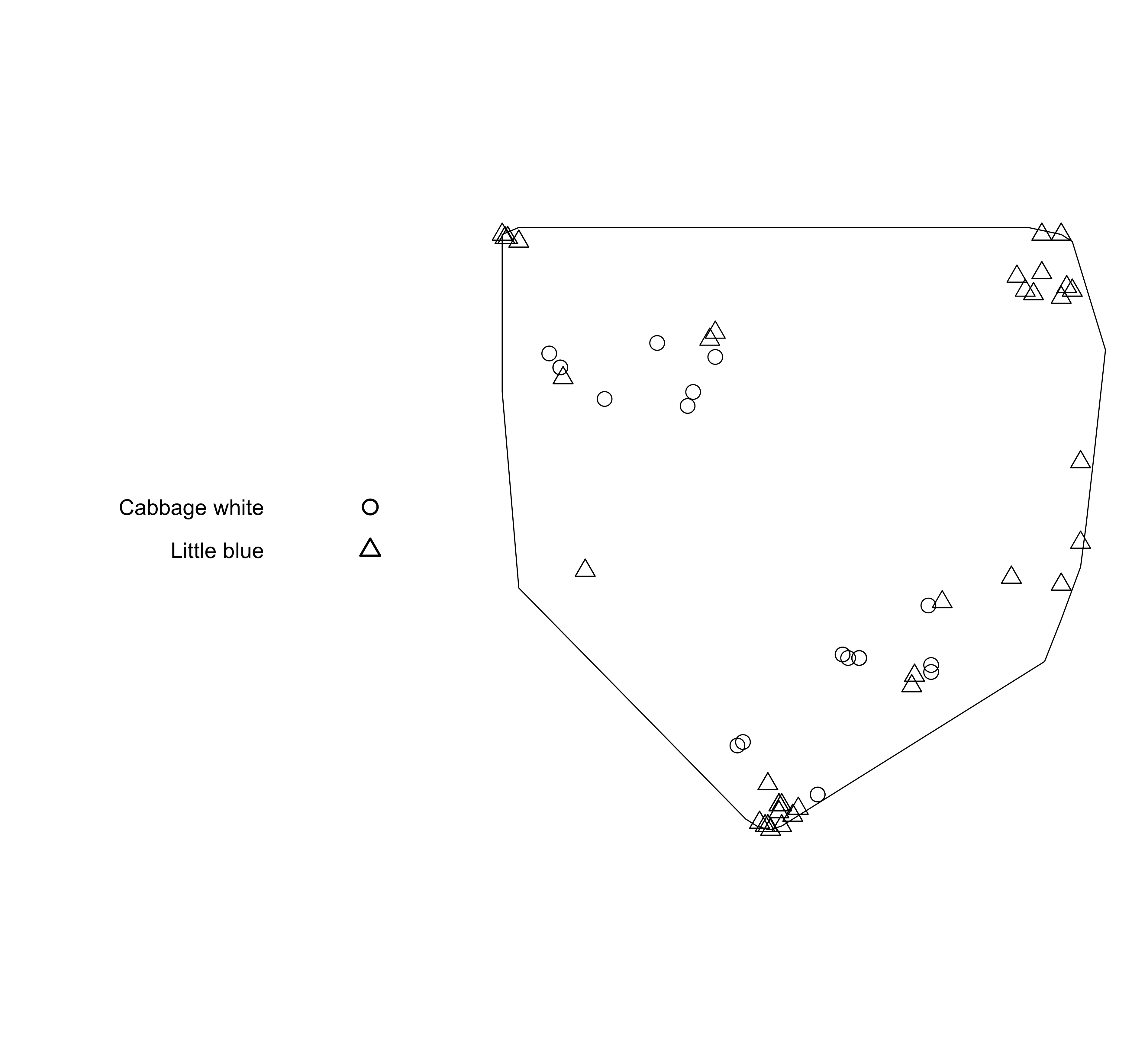}
    \caption{Butterflies in Melbourne, Australia.}
    \label{fig:butterflies}
\end{figure}

\begin{figure}[!h]
    \centering
    \includegraphics[scale=0.08]{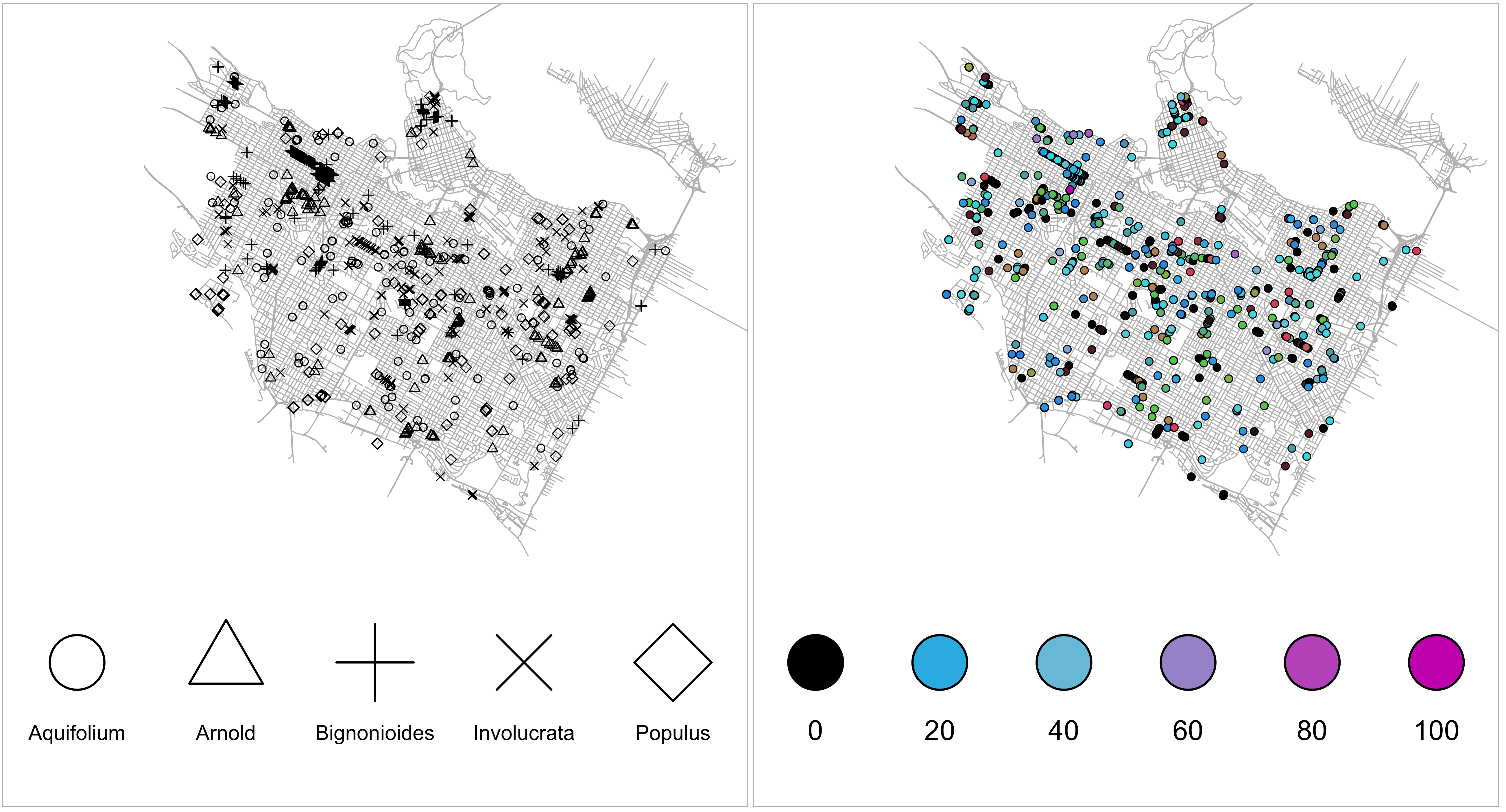}
    \caption{Public street trees in Vancouver, Canada, in 2016. Left: different shapes show species of trees; right: colors show diameter at breast height in inches. 
    }
    \label{fig:van}
\end{figure}

To illustrate the use of both the existing methodologies and our novel contributions, we consider two spatial point pattern datasets sourced from publicly accessible open data repositories. These datasets are records detailing butterfly diversity in Melbourne, Australia (Figure \ref{fig:butterflies}) and data concerning public street trees in Vancouver, Canada (Figure \ref{fig:van}).

The butterfly dataset was made accessible through the \textit{Our City's Little Gems project} (OCLG), on the \textit{Open Science Framework}, and it was sourced from the open data repository maintained by the city of Melbourne, Australia \footnote{\url{https://data.melbourne.vic.gov.au}}. Field researchers undertook data collection during January-March 2017, employing a transect tetrad study design established by the OCLG initiative \citep{KirkHolly2017OCLG}. The original data show concentrations of butterflies dispersed across multiple locations throughout the city, but we focus on some public parks around Melbourne Zoo which contain 1396 butterflies. However, after cleaning the dataset and maintaining those locations with unique species types, we obtained a point pattern of 84 points with two types of butterflies \textit{Little Blue}, with 64 points, and \textit{Cabbage White}, with 20 points. We use the convex hull of all points as the observed window. 

Figure \ref{fig:van} presents the spatial distribution of public trees, excluding park trees, located along the street network of Vancouver, Canada, in 2016. Attributes such as tree species and their corresponding diameter at breast height (dbh) in inches are attached to each location. The original data is sourced from the open data portal\footnote{\url{https://opendata.vancouver.ca}} of Vancouver, Canada, and consists of 136,574 places of public trees categorized into 282 different species. However, we only consider five species \textit{Aquifolium, Arnold, Bignonioides, Involucrata,} and \textit{Populus}, which together include 1,045 trees with a dbh varying between 2 to 94 inches. The street network comprises 49,928 vertices and 55,221 segments and spans a total length of 1,779.547 km. 

\section{Marked spatial point processes in $\R^2$}\label{sec:mpp}

Let $\x=\lbrace x_i, m(x_i)\rbrace_{i=1}^n$ be an observed finite marked point pattern within a window $W \subset \R^2$, 
where 
$x_i$ is the spatial location of $i$-th event 
and
$m(x_i)$ is
some associated 
mark living on a suitable mark space $\M$. It is assumed that $\x$ has been generated by an underlying random mechanism $X$, called a marked spatial point process, on the product space $\R^2 \times \M$, which is equipped with the spatial distance $d(u,v)=||u-v||,\ u,v \in \R^2,\ $ and the Lebesgue measure $|\cdot|$. For an unmarked version of $X$, denoted by $\X$, it is said that $\X$ is a homogeneous point process with intensity $\lambda$ if $\ee[N(\X \cap A)]=\lambda |A|,\ A \subset \R^2$, where $N$ is a count function. Otherwise, $\X$ is an inhomogeneous process, and 
\begin{eqnarray}\label{eq:lambda}
    \ee
    [N(\X \cap A)]
    =
    \int_A \lambda(u) \de u,
    \quad
    A \subset \R^2,
\end{eqnarray}
so that $\lambda(\cdot)$ governs the spatial distribution of points.  Heuristically, $\lambda(u)|\de u|, u \in \R^2,$ is the probability of finding a point of $\X$ within a vicinity of 
$u$ of size $|\de u|$. From now on, whenever we refer to the intensity function, we are specifically referring to the intensity function of an unmarked point process. 
Furthermore, according to Campbell's formulas, for any non-negative measurable function  $f:{\R^2}^m \to \R$,
\begin{eqnarray}\label{eq:Campbell}
    \ee 
    \left[
\mathop{\sum\nolimits\sp{\ne}}_{x_1,\ldots,x_m\in \X}
f(x_1,\ldots,x_m)
\right]
=
\int_{{\R^2}^m}
f(u_1,\ldots,u_m)
\lambda^{(m)} (u_1,\ldots,u_m)
\de u_1 \cdots \de u_m, 
\end{eqnarray}
where $\lambda^{(m)}(\cdot)$ is called the $m$-th order product intensity of $\X$, and $\neq$ means that the sum is taken over distinct $m$-tuples. Note that \eqref{eq:lambda} is a particular case of \eqref{eq:Campbell} when $f(x)=\1 \{ x \in A \}$ where $\1$ is an indicator function. In a similar manner, $\lambda^{(m)} (u_1,\ldots,u_m) \prod_{i=1}^m |d u_i |$ is the probability of having points of $\X$ in $m$ infinitesimal disjoint areas around $u_1,\ldots,u_m$ with sizes $|d u_1|, \ldots, |d u_m|$.

As the product intensity functions of $\X$ are not designed to study the correlation among points, if all $m$-th order product intensities exist, one may wish to employ the correlation function 
\begin{align}\label{eq:correl}
    g_m(u_1,\ldots,u_m)
    =
    \frac{
    \lambda^{(m)} (u_1,\ldots,u_m)
    }{
    \lambda(u_1) \cdots \lambda(u_m)
    }
    =
    \sum_{j=1}^m
    \sum_{D_1,\ldots,D_j}\xi_{N(D_1)}
    \left( \{ u_i:i\in D_1 \} \right)
    \cdots 
    \xi_{N(D_j)}
    \left( \{ u_i:i\in D_j \} \right)
    ,
\end{align}
where $\sum_{D_1,\ldots,D_j}$ ranges over all partitions $\{D_1,\ldots,D_j\}$ of $\{1,\ldots,m\}$ into $j$ non-empty and disjoint sets, $N(D_j)$ is the cardinality of the index set $D_j$, and the $\xi$-functions are permutation invariant \citep{van11}. If such, the point process $\X$ is called intensity-reweighted moment stationary. Note that, if $m=2$ it reduces to the so-called second-order intensity-reweighted stationary \citep{InhomK2000}, where one can obtain $g_2(u,v)-g_1(u)=(\xi_2(u,v) + 1)-1=\xi_2(u,v)$, where $g_2(u,v)$ is the so-called pair correlation function. A more stringent condition, which entails that $\X$ has an identical distribution to $\X + a = {x + a: x \in \X}$, i.e., invariance under translations, leads to the concept of stationarity. 

Next, by increasing marks' complexity, we present different summary characteristics designed to investigate the interactions between marks subject to an interpoint distance $0<r<\infty$. Under specific assumptions, such as mark independence, these summary characteristics have predetermined values, making them valuable benchmarks for indicating interactions between marks. Thus, they can be effectively utilized in the hypothesis testing framework, enabling us to draw inferences about the spatial structure of marks and identify potential deviations from mark independence. Depending on the data in question, null models include e.g. the superposition of some independent point processes and the random labeling of points generated by subsequent processes. 

\subsection{Discrete and integer-valued marks}\label{sec:discretemark}

When marks represent a set of countably many entities, which correspond to the simplest case of all potential marked point process settings, each point is assigned a single label $1\leq k \leq n$, meaning that an integer-valued quantity categorizes points. The underlying process is called a multivariate/multitype point process in this case. More specifically, the observed point pattern $\x$ will be denoted as a collection $\{ \x_1,\ldots,\x_k \},\ k\geq 2, $ where for each point pattern $\x_i, i=1,\ldots,k$,  we have $m(\cdot)=i$, meaning that each point pattern $\x_i$ consists of some points with a specific mark. Within the literature, multivariate/multitype point patterns are commonly analyzed through the so-called cross/dot-type extensions of the classical summary characteristic such as the $K$-functions and the pair correlation functions \citep{Ripley1976, InhomK2000}, 
the nearest-neighbor distance distribution function $H$, the empty space function $F$, and the $J$-function \citep{Jest1996,van11}.
The cross-type summary characteristics 
concern pairs of points with distinct marks, whereas the dot-type summary characteristics 
focus on the relationship between points with a particular mark and all the remaining points. These summary characteristics were initially developed for homogeneous marked spatial point processes \citep{Lotwick1982, HarknessIsham1983, VanLieshoutBaddeley1999}. However, since, in practice, homogeneity is rarely satisfied, we here only present the inhomogeneous versions of such summary characteristics, which reduce to their homogeneous counterparts when intensity is constant. We note that the inhomogeneous $K$-functions and pair correlation functions for multivariate/multitype point processes in $\R^2$ are defined for second-order intensity-reweighted stationary processes \citep{MollerWaagepetersenBook}, whereas the inhomogeneous nearest-neighbor distance distribution function, the inhomogeneous empty space function, and the inhomogeneous $J$-function demand intensity-reweighted moment stationarity \citep{Cronie2016}.

For a second-order intensity-reweighted stationary multivariate/multitype point process $X$ in $\R^2$, the cross-type $K$-function is given as
\begin{eqnarray}\label{eq:crossKinhom}
K_{ij}^{\mathrm{inhom}}(r)
=
\ee
\left[
\sum_{x \in X_j}
\frac{
\1\{ d(u,x)\leq r \}
}{
\lambda_j (x)
}
\Bigg|
u \in X_i
\right],
\quad 
\quad
i,j=1,\ldots,k,
\quad
r>0,
\end{eqnarray}
where $X_i,X_j$ are point processes with marks $i,j,$ respectively, and $\lambda_j(\cdot)$ is the intensity function of $X_j$. If $X_i$ does not depend on $X_j$, then $K_{ij}^{\mathrm{inhom}}(r)=\pi r^2$, which can serve as a criterion to discover independence between sub-processes of different types. If $K_{ij}^{\mathrm{inhom}}(r)>\pi r^2$, one can conclude that the expected number of points of type $j$ around points of type $i$, within an interpoint distance $r$, is more than the expected number under mark independence, indicating a tendency to occur around points of types $i$. On the contrary,  $K_{ij}^{\mathrm{inhom}}(r) < \pi r^2$ means that points of type $j$ tend to maintain an interpoint distance from points of type $j$.
Note that $K_{ii}^{\mathrm{inhom}}(\cdot)$ gives the $K$-function for $X_i$. Moreover, the dot-type $K$-function is of the form 
\begin{eqnarray}\label{eq:dotKinhom}
K_{i\bullet}^{\mathrm{inhom}}(r)
=
\ee
\left[
\sum_{\substack{x \in X \\ u \neq x}}
\frac{
\1\{ d(u,x)\leq r \}
}{
\lambda (x)
}
\Bigg|
u \in X_i
\right],
\quad 
\quad
i=1,\ldots,k,
\quad
r>0,
\end{eqnarray}
where $\lambda(\cdot)$ is the intensity function of $X$. In practice, 
\eqref{eq:dotKinhom} measures the expected number of points of any type around the points with type $i$. Furthermore, one could obtain the cross/dot-type versions of both the inhomogeneous pair correlation function and the inhomogeneous $L$-function as $\rho_{ij}^{\mathrm{inhom}}(r)=\partial K_{ij}^{\mathrm{inhom}}(r)/ 2\pi r$ and $L_{ij}^{\mathrm{inhom}}(r)=\sqrt{K_{ij}^{\mathrm{inhom}}(r)/\pi}
$; see \citet[Chapter 14]{Baddeley2015} for more details.

Apart from the above cross/dot-type second-order summary characteristics, when $X$ is further intensity-reweighted moment stationary, there exist the cross-type  inhomogeneous nearest-neighbor distance distribution function 
\begin{eqnarray}\label{eq:crossG}
    H_{ij}^{\mathrm{inhom}}(r)
    =
    1
    -
    \ee
    \left[
    \prod_{x \in X_j}
    \left(
    1
    -
    \frac{\bar{\lambda}_j}{\lambda_j(x)}
    \1 \{ d(u,x) \leq r \}
    \right)
    \Bigg|
    u \in X_i
    \right],
\end{eqnarray}
and the cross-type inhomogeneous $J$-function
\begin{eqnarray}\label{eq:crossJ}
    J_{ij}^{\mathrm{inhom}}(r)
    =
    \frac{
    1 - H_{ij}^{\mathrm{inhom}}(r)
    }{
    1 - F_{j}^{\mathrm{inhom}}(r)
    }, \quad F_{j}^{\mathrm{inhom}}(r) \neq 1,
\end{eqnarray}
where $\bar{\lambda}_j=\inf_{u \in X_j} \lambda_j(u)$,
and 
\begin{eqnarray}
    F_{j}^{\mathrm{inhom}}(r)
    =
    1
    -
    \ee
    \left[
    \prod_{x \in X_j}
    \left(
    1 -
    \frac{\bar{\lambda}_j}{\lambda_j(x)}
     \1 \{ d(u,x) \leq r \}
    \right)
    \right],
    \quad
    u \in \R^2,
\end{eqnarray}
which is the inhomogeneous empty space function of $X_j$ and does not depend on the choice $u$ \citep{Cronie2016}. Moreover, if $X_i$ and $X_j$ are independent, then $H_{ij}^{\mathrm{inhom}}(r)=F_{j}^{\mathrm{inhom}}(r)$ and consequently $J_{ij}^{\mathrm{inhom}}(r)=1$, which can be used as a benchmark to detect independence between points of different types. Note that these functions are not symmetric in $i$ and $j$ meaning that e.g., $H_{ij}^{\mathrm{inhom}}(r)\neq H_{ji}^{\mathrm{inhom}}(r)$, and by setting $j=i$ in \eqref{eq:crossG} and \eqref{eq:crossJ} one obtains the nearest-neighbor distance distribution function and the $J$-function of $X_i$. Furthermore,
\begin{align}\label{eq:dotG}
H_{i\bullet}^{\mathrm{inhom}}(r)
    &=
    1
    -
    \ee
    \left[
    \prod_{\substack{x \in X \\ u \neq x}}
    \left(
    1
    -
    \frac{\bar{\lambda}}{\lambda (x)}
    \1 \{ d(u,x) \leq r \}
    \right)
    \Bigg|
    u \in X_i
    \right],
    \\
    J_{i\bullet}^{\mathrm{inhom}}(r)
    &=
    \frac{
    1 - H_{i\bullet}^{\mathrm{inhom}}(r)
    }{
    1 - F^{\mathrm{inhom}}(r)
    },
    \quad F^{\mathrm{inhom}}(r) \neq 1,
\end{align}
where $F^{\mathrm{inhom}}(r)$ is the inhomogeneous empty space function of $X$. Due to the properties of $J$-functions under superposition of stationary and independent processes, \cite{VanLieshoutBaddeley1999} introduced the $I$-function, which its analogous for inhomogeneous processes becomes $I^{\mathrm{inhom}}(r) = \sum_{i=1}^{k} p_i J^{\mathrm{inhom}}_{i}(r) - J^{\mathrm{inhom}}(r)$ where $p_i$ is the probability of type $i$. Under independence of marks, the $I$-function becomes zero, whereas deviations from zero suggest either positive or negative associations among the components.

If point processes $X_i$ and $X_j$ are independent, then one expects to have $J^{\mathrm{inhom}}_{ij}(r)=1$ at least for small distances, while $J^{\mathrm{inhom}}_{ij}(r)>1$ points to a tendency for point of type $j$ to happen near points of type $i$, and $J^{\mathrm{inhom}}_{ij}(r)<1$ means that points of type $j$ prefer not to occur close to points of type $i$ showing an inhibition between the two types \citep{van11, Cronie2016}. In general terms, cross-type summary characteristics are employed to examine whether the presence of a specific type can influence the occurrence of points belonging to another type.


\subsubsection{Mark connection and  mingling functions}

Within the context of discrete/integer-valued marks, a variety of more accessible empirical summary characteristics are available for stationary point processes.
For instance, the mark connection functions $p_{ij}(r)=p_ip_j\rho_{ij}(r)/\rho(r)$, where $p_i$ is the probability of type $i$ and $\rho(\cdot)$ is the pair correlation function, can be intuitively interpreted as the conditional probability that two points at a distance $r>0$ from each other possess marks $i$ and $j$ provided that these points belong to the point process $X$; under the independence of $X_i$ and $X_j$, we have $p_{ij}(r)=p_ip_j$ \citep{StoyanStoyan1994, Illian2008,baddeley2010multivariate}.
For $p_{ij}(r)>p_ip_j$, the mark connection function indicates a tendency of aggregation of components $i$ and $j$ around each other 
while  $p_{ij}(r)<p_ip_j$ suggests repulsion. 
Moreover, the normalized 
mark mingling function is given by
\begin{eqnarray}\label{eq:mingleR2}
\nu(r)
=
\frac{1}{c}
\ee
\left[
\sum_{x,y \in X}^{\neq}
\1 \{
m(x)\neq m(y)
\}
\1\{d(x,y)\leq r\}
\right],
\end{eqnarray}
where $c
=\sum^k_{i=1} (n_i(n-n_i))/(n(n-1))$ is a normalizing constant, $n_i$ is the number of points of type $i$, and $\sum^{\neq}$ means that points are distinct \citep{Lewandowski1997,20113358596, HUI2014125}. The mark mingling function $\nu(r)$ may be used to disclose aggregation of points with (dis)similar marks so that under the independence of marks we have $\nu(r)\approx 1$, while $\nu(r)>1$ indicates heterospecific attraction, and $\nu(r)<1$ suggests conspecific attraction. Note that the main difference between the mingling function and the cross/dot-type summary characteristics is that the mingling function quantifies the closeness of points with distinct types.

\subsection{Real-valued marks}\label{sec:reals}

Unlike discrete/integer-valued marks, which are investigated by reformulating a marked point process into $k>1$ components, real-valued marks are commonly analyzed through test functions $\tf_f: \mathds{M}\times\mathds{M} \to \R^+$, which is usually a function of the marks of two points located at a distance of $r>0$ from each other \citep{PenttinenandStoyan1989, Schlather2001}. Moreover, in contrast to the presented summary characteristics in Section \ref{sec:discretemark}, which address the aggregation/repulsion between points of different types, test functions $\tf_f$ analyze numerical differences between the marks of distinct points as a function of distance. For instance, neighboring points may exhibit (dis)similar mark values, and specific points might exert dominance over the marks of neighboring points, having large marks while other points in their vicinity possess smaller marks \citep[Chapter 5]{Illian2008}.

For stationary point processes, the $\tf_f$-correlation function $\kappa_{\tf_f}(r)$ is given as
\begin{eqnarray}\label{eq:tfcorr}
\kappa_{\tf_f}(r)
    =
    \frac{
    \ee \left[
    \tf_f \left(
    m(x),m(y)
    \right) \big| x,y \in X
    \right]
    }{
    c_{\tf_f}
    },
    \quad  
    d(x,y)=r,
\end{eqnarray}
where $c_{\tf_f}$ is a normalizing factor; note that the numerator in \eqref{eq:tfcorr} is a conditional expectation
 with respect to the joint distribution of marks.
Thus, one can see how marks of the points, located at a distance $r$ from each other, interact in comparison to the average behavior of marks \citep{PenttinenandStoyan1989,baddeley2010multivariate}. Indeed, the explicit form of these mark characteristics relies on the choice of test functions, which often account for the spatial variation of marks as well as their pairwise relationships. 
Within the literature, denoting the average and the variance of all marks as $\mu_m$ and $\sigma^2_m$, different test functions and normalizing factors are proposed, which are given in Table \ref{tab:testfuns:scalars}. We note that \cite{StoyanStoyan1994} proposed the so-called nearest-neighbor indexes, which are similarly constructed through different test functions but only accounting for the marks at a point $x \in X$ and its nearest neighbor, rather than considering all potential pairs of points within a distance $r$ from $x$.

\begin{table}[!h]
\caption{
Common test functions for point processes with real-valued marks. For Schlather's $I$, $\mu_m(r)$ is the mean of the marks for all points 
located at a distance $r$ from $x$,
and for the mark differentiation function
$c_{t_f}^{\Delta}
=
1 - (2/n(n-1))\sum_{i=1}^n R_i/m(x_{(i)})$ where $m(x_{(i)})$ is the $i$-th increasingly-ordered mark, $R_1=0$ and $R_i=\sum_{j=1}^{i-1}m(x_{(j)})$, for $2\leq i \leq
 n$.
}
\begin{center}
\begin{tabular}{l | ccc}
\hline
Name of the function & Symbol & Test function ($\tf_f$) & Normalizing \\
                     &   &  & factor $(c_{\tf_f})$ \\
\hline
Mark variogram & $\gamma_{mm}(r)$ & $0.5
    [m(x)-m(y)]^2 $ &  $\sigma^2_m$
\\
\hline
Stoyan's mark correlation function & $\kappa_{mm}(r)$ & $m(x)m(y)$ &  $\mu_m^2$
\\
\hline
$\mathbf{r}$-mark correlation function & $\kappa_{m\bullet}(r)$ & $m(x)$ &  $\mu_m$
 \\
 \hline
$\mathbf{r}$-mark correlation function& $\kappa_{\bullet m}(r)$ & $ m(y) $ &  $\mu_m$
 \\
\hline
Beisbart/Kerscher's  $\kappa_{mm}(r)$& $\kappa_{mm}^{\mathrm{Bei}}(r)$ & $m(x) + m(y)$ &  $2\mu_m$
\\
\hline
Isham's  $\kappa_{mm}(r)$& $\kappa^{\mathrm{Ish}}_{mm}(r)$ & $m(x)m(y)$ $-\mu_m^2$&  $\sigma_m^2$
\\
\hline
Stoyan's covariance function & $\cov_{mm}(r)$ & $m(x)m(y)-\mu_m^2$ &  1
\\
\hline
Schlather's $I$ & $I_{mm}(r)$ & $(m(x) - \mu_m(r)) (m(y) - \mu_m(r))$ &  $\sigma^2_m$
\\
\hline

Shimanti's $I$ &  $I^{\mathrm{Shi}}_{mm}(r)$ & $(m(x) - \mu_m) (m(y) - \mu_m)$ &  $\sigma^2_m$
\\
\hline
Mark differentiation function & $\Delta_{mm}(r)$ & $1 - 
    \frac{ \min \left( m(x),m(y) \right)
    }{
    \max \left( m(x),m(y) \right)
    }$ &  $c_{t_f}^{\Delta}$
\\
\hline
\end{tabular}
\label{tab:testfuns:scalars}
\end{center}
\end{table}

The mark variogram $\gamma_{mm}(r)$ \citep{cressie93,markvar, Stoyan2000} measures the local half-squared differences among marks of a pair of points while Stoyan's mark correlation function $\kappa_{mm}(r)$ \citep{StoyanStoyan1994} focuses on the mean of the local product of such marks. In other terms, the objective of the mark variogram $\gamma_{mm}(r)$ is to assess the pairwise differences between the marks associated with a pair of points separated by a distance $r$, which can provide valuable insights into the spatial relationship and variability of marks within a specified distance range. Concerning Stoyan's mark correlation function $\kappa_{mm}(r)$, the pairwise product is expected to coincide with the global squared mean of the marks $\mu_m^2$ under the independence assumption such that $\kappa_{mm}(r)=1$. If nearby points have smaller (larger) marks, their average product will also be small (large) and deviate from $1$. We note that Stoyan's mark covariance function $\cov_{mm}(r)$ \citep{DBLP:journals/eik/Stoyan84} is indeed a linear transformation of his mark correlation function such that both $\cov_{mm}(r)$ and $\kappa_{mm}(r)$  essentially convey the same message \citep{Schlather2001}.
Regarding the use of $\mathbf{r}$-mark correlation functions, $\kappa_{m\bullet}(r)$ and $\kappa_{\bullet m}(r)$, both reflect the average of the mark for 
either of the two considered points
 with respect to a distance $r$. 
  Except for mark independence, the averaged $\mathbf{r}$-mark correlation will not be equal to the overall mark mean since either the mark of the first or second point is chosen. 
Shedding some light on Beisbart's and Kerscher's version of $\kappa_{mm}(r)$, to have a large contribution in \eqref{eq:tfcorr}, only one of the points is required to have a large mark. However, under mark Independence, this mark correlation function coincides with twice the mark mean yielding $\kappa^{\mathrm{Bei}}_{mm}(r)=1$. 
In contrast to $\kappa_{mm}(r)$ and $\kappa^{\mathrm{Bei}}_{mm}(r)$,
Isham's function $\kappa^{\mathrm{Ish}}_{mm}(r)$ reveals a 
Pearson-type correlation for marks.
Schlather's function $I_{mm}(r)$ gives insight into how marks, for a pair of points with an interpoint distance $r$, are related to each other \citep{Schlather2004}. Basically, this function centers each mark by the conditional mean mark $\mu_m(r)$, i.e., the $\mathbf{r}$-mark correlation function, and normalizes the product by the mark variance $\sigma^2_m$. Shimanti's function $I_{mm}(r)$ is constructed similarly but uses the global mark mean instead of the conditional mark mean \citep{Shimatani:MoranI}. Finally, the mark differentiation function $\Delta_{mm}(r)$ shows how the ratio of marks varies with respect to a distance $r$. In the case of mark independence, the minimum and the maximum of marks are assumed to be, on average, similar such that the proportion becomes $1$, i.e., $\Delta_{mm}(r)=0$ \citep{20113358596, HUI2014125}.

Aiming at defining a $K$-function for stationary point processes with real-valued marks and a constant intensity $\lambda$, which accounts for the correlation between points as well as between marks,  \cite{pettinen1992forest} proposed a mark-weighted $K$-function  as 
\begin{equation}\label{eq:Kmm}
 K_{{\tf}_f}(r)
 =
 \frac{1}{\lambda c_{{\tf}_{f}}}
 \ee
 \left[
 \sum_{\substack{x \in X \\ u \neq x} 
} 
 \tf_{f} \left(
    m(u),m(x)
    \right)
 \1\{ d(u,x)\leq r \}
 \Bigg|
    u \in X
 \right],
\end{equation}
where  $c_{{\tf}_{f}}$ is the expected value of $\tf_{f}(\cdot)$ under mark independence  for which $K_{{\tf}_f}(r)$ reduces to the so-called Ripley’s $K$-function \citep[Chapter 15]{Baddeley2015}; \eqref{eq:Kmm} can be extended to inhomogeneous cases. \cite{pettinen1992forest} initially considered the Stoyan's mark correlation function in \eqref{eq:Kmm}, i.e., $\tf_{f}(\cdot)=\kappa_{mm}(\cdot)$, and commonly denoted $K_{{\tf}_f}(\cdot)$ as $K_{mm}(\cdot)$, however, one may use any of the test functions presented in Table \ref{tab:testfuns:scalars}. Note that, in the case of $K_{mm}(\cdot)$, by taking the pairwise product of marks as weight into the $K$-function, the estimated curves become scaled versions of the original $K$-function, except under mark independence. 
More specifically, 
for pairwise positively correlated marks, i.e. if the pairwise product of marks exceeds the overall expected value, the empirical curves are up-scaled versions of $K$-function. Otherwise, they are down-scaled. To better understand the impact of marks, it is advisable to compare the original $K$-functions with their mark-weighted counterparts. 
Further, normalizing  $K_{\tf_f}(r)$ by its unmarked counterpart, i.e. the Ripley’s $K$-function, one can obtain a cumulative mark-correlation function
corresponding to the mean value of the employed test function for a pair of points with an interpoint distance $r$ \citep{WiegandMoloney2013}. We note that \cite{doi:10.1080/10618600.2023.2206441} recently discussed local mark-weighted cumulative summary statistics, which account for the contribution of each marked point to the global second-order summary characteristics.


A further summary statistics for  a stationary point process $X$, with an intensity function $\lambda$ and the pair correlation $\rho(\cdot)$, is  
\begin{equation}\label{eq:ustat}
U(r)
=
\lambda^2
\rho(r) \kappa_{\tf_f}(r)
|\de x| |\de  y|, 
\end{equation}
where $|\de x|, |\de  y|$ are sizes of two infinitesimal small areas around $x$ and $y$ separated by a distance $r$. In a similar manner as $K_{\tf_f}(r)$ in \eqref{eq:Kmm}, $U(\cdot)$ considers the correlation between points and between marks. Since under complete spatial randomness  $\rho(\cdot)=1$, and for independent marks $\kappa_{\tf_f}(\cdot)=1$, thus, in the combination of those cases, $U(\cdot)$ reduces to the second-order product density of $X$ \citep{Capobianco1998,Renshaw2002}. 
If at least one of these two does not vanish, then $U(\cdot)$ is a version of the second-order product density of $X$ weighted by the spatial interaction between the points via $\rho(\cdot)$ and/or association between the marks via $\kappa_{\tf_f}(\cdot)$.
We note that further
summary characteristics,
considered within the literature, include the (non)cumulative density correlation functions \citep{Fedriani:Wiegand:2015}.

Turning to the case where each data point has two distinct marks $m_i(\cdot),\ i=1,2,$ by considering a bivariate test function $\tf_f(m_1(x),m_2(y))=m_1(x)m_2(y),\ x,y \in X$, \cite{Stoyan1987} proposed a bivariate mark correlation function 
with a normalizing constant
$\mu_1\mu_2$ where $\mu_i,\ i=1,2,$ is the mean of mark $m_i(\cdot),\ i=1,2$; see also \citet{Raven2011, WiegandMoloney2013}. Note that such an idea can be used to construct other bivariate and multivariate (nearest-neighbor-based) test functions \citep{PommereningBook, Eckardt2023MultiFunctionMarks}. In addition, the cross/dot-type pair correlation function and the bivariate/multivariate mark summary characteristics can also be used to derive cross/dot-type and multivariate versions of $U(r)$ \citep{EckardtISR, EckardtMateuGonzales2020}. Such extended versions not only help to investigate the interrelations between different types of points but also allow to understand the association between different real-valued quantities over space.  

\subsection{Frequency domain approaches}

 Next, we discuss characterizations of marked spatial point processes through frequency domain methods, all of which could be translated
 into their corresponding distance-based characteristics, e.g. $U(\cdot)$, given in \eqref{eq:ustat}, by using the inverse Fourier theorem. This close relationship of the frequency and distance-based approaches allows to derive planar partial mark characteristics which reflect the interrelation between two components $X_i, X_j, i,j=1,\ldots,k,\ $ conditional on all the remaining types.

\subsubsection{Frequency domain approaches for discrete and integer-valued marks}

In contrast to previous marked summary characteristics, frequency domain characteristics for marked spatial point processes have received limited attention.
Bartlett's complete covariance density function \citep{Bartlett1964} is given as $\mathfrak{K}_{ij}(u,v)=\lambda_i (u) \delta_{ij}(u-v)+\vartheta_{ij}(u,v), i,j=1,\ldots,k,\ $ where 
$\vartheta_{ij}(\cdot)$
is a covariance density function \citep[Equation 2]{Mugglestone1996a} and  $\delta_{ij}(\cdot)$ is a two-dimensional Dirac delta-function. 
Similar to the covariance density function, $\mathfrak{K}_{ij}(u,v)$ may be used to describe the cross second-order behavior of points within the corresponding space. However, unlike the covariance function, it controls for multiple coincident points via $\delta_{ij}$. 
For the components $X_i, X_j$,\ $i,j=1,\ldots,k$, and at frequencies $\boldsymbol{\omega}=(\omega_1,\omega_2)$, using a discrete Fourier transform representation 
\cite{Mugglestone1996a}
introduced the cross-spectral density function $f_{ij}(\boldsymbol{\omega})$ as 
\begin{eqnarray*}
f_{ij}(\boldsymbol{\omega})
=
\int^{\infty}_{-\infty} \mathfrak{K}_{ij}(r)\exp(-\imath \boldsymbol{\omega}^\top r)\de r,
\quad \quad
\mathfrak{K}_{ij}(r)
= \lambda_i\delta_{ij}(r)+\vartheta_{ij}(r)
,
\end{eqnarray*}
where $r=d(u,v)$, $u,v \in \R^2$,  $\boldsymbol{\omega}^\top$ is the transpose of $\boldsymbol{\omega}$, and $\imath=\sqrt{-1}$. Note that, in this case,  $\mathfrak{K}_{ij}(\cdot)$ only depends on distances between points.
Two components are said to be independent if the cross-spectral density function equals zero for all frequencies. 
We note that by applying Theorem 8.3.1 of \cite{Brillinger1981}, the above spectrum could be translated into a partial version $f_{ij|\boldsymbol{X}\setminus\lbrace i,j \rbrace}(\boldsymbol{\omega})$, i.e. computing the Schur complement of the cross-spectral densities \citep{EckardtISR}.

The cross-spectral function $f_{ij}(\cdot)$ further leads to the definition of some  interesting spectral functions, including the spectral coherence function $R^2_{ij}(\boldsymbol{\omega})
=
f_{ij}^2(\boldsymbol{\omega})/
\left(
f_{ii}(\boldsymbol{\omega})f_{jj}(\boldsymbol{\omega})
\right)$,
the cross-amplitude spectrum $\zeta_{ij}(\boldsymbol{\omega})=\mod\lbrace f_{ij}(\boldsymbol{\omega})\rbrace$,
the cross-phase spectrum $\wp_{ij}(\boldsymbol{\omega})=\arg(f_{ij}(\boldsymbol{\omega}))$, the gain function $\mathfrak{G}_{i|j}(\boldsymbol{\omega})=\sqrt{(f_{ij}(\boldsymbol{\omega})R_{ij}(\boldsymbol{\omega}))/f_{i}(\boldsymbol{\omega})}$, and also their corresponding partial versions \citep{Mugglestone1996a, Eckardt2018partial}. 
To offer some insight into these functions,  the cross-phase spectrum $\wp(\boldsymbol{\omega})$ measures the similarity between two point patterns to see if the spectrum of one pattern is a linear shift of that of the other one, 
while the amplitude spectrum $\zeta_{ij}(\boldsymbol{\omega})$ encodes the relative magnitude of
frequencies for two point patterns. Thus, both functions are useful tools to investigate the characteristics of the empirical spectrum with respect to the frequencies $\boldsymbol{\omega}$. The gain spectrum $\mathfrak{G}_{i|j}(\boldsymbol{\omega})$, in contrast, can be interpreted as a regression coefficient in a linear regression at frequency $\boldsymbol{\omega}$, for two corresponding components.
Additionally, one could make directional inferences by transforming the spectral density functions into the polar form; see e.g. 
\citet{Renshaw1983,Renshaw1984}.

\subsubsection{Frequency domain approaches for real-valued marks}

The accessibility of frequency domain approaches for real-valued marks is more constrained than that of discrete marks. The auto/cross-spectral density functions for real-valued marks, at frequencies $\boldsymbol{\omega}=(\omega_1,\omega_2)$, can be obtained from a discrete Fourier transformation of the corresponding $U(r)$ functions,  as
\begin{align*}
f^{\mathrm{m}}_{ii}(\boldsymbol{\omega})
=
\int U_{ii}(r)\exp(-\imath \boldsymbol{\omega}^\top r)\de r,
\quad \quad
f^{\mathrm{m}}_{ij}(\boldsymbol{\omega})
=
\int U_{ij}(r)\exp(-\imath \boldsymbol{\omega}^\top r)\de r,
\end{align*}
where $U_{ii}(r)$ and $U_{ij}(r)$ are the auto/cross-type versions of \eqref{eq:ustat}, respectively; see 
\cite{Renshaw2002, EckardtISR} for more details. 
Analogous to the multivariate spatial point processes,  the marked spectra can be transformed into partial versions using the results of Theorem 8.3.1 of \cite{Brillinger1981}.

As of some recent development,   \cite{EckardtISR}, using a re-scaled version of the partial spectral coherence, defined a spatial dependence graph model for multivariate point processes with, potentially, real-valued marks, according to which different components are represented as nodes and the conditional independence structure among the $k$ components is reflected by missing edges.

\subsection{Object-valued marks}\label{sec:objects}



We now discuss mark summary characteristics for an observed point pattern  $\x=\lbrace x_i, \mo(x_i)\rbrace^n_{i=1}$ in which each point $x_i$ is augmented by a non-scalar quantity $\mo_i$ living on some suitable mark space $\mathds{M}$, which 
depends on the specificity of marks in question. Apart from the function-valued mark setting, where $\mathds{M}$ is the Hilbert/$\mathcal{L}_2$ space \citep{comas2008METMA, Comas2011, Comas2013, Ghorbani2020, Eckardt2023MultiFunctionMarks},  this newly introduced class of marked spatial point processes also includes the cases where marks are constrained arrays or inherently structured quantities  \citep{EckardtMariCMSPP}.  

Focusing explicitly on function-valued mark scenarios, where for each point $x_i$ the corresponding mark $\mo(x_i)$ is a function-valued quantity $g(x_i)$ on $\mathds{F}(\mathcal{T})$ with $\mathcal{T}=(a,b), -\infty\leq a\leq b\leq \infty$ and $g(x_i): \mathcal{T}\subseteq \R \mapsto \R$, different mark summary characteristics, similarly to Section \ref{sec:reals}, can be defined through an extended test function $\tf_f: \mathds{F}\times\mathds{F}\rightarrow \R^+$. Starting with a pointwise specification, \cite{Eckardt2023MultiFunctionMarks} proposed a generalized version of the ${\tf_f}$-correlation function in \eqref{eq:tfcorr}, given as
\begin{eqnarray}\label{eq:tfcorr:ftc}
    \kappa_{\tf_f}(r,t)
    =
    \frac{
    \ee \left[
    \tf_f \left(
    m_x(t),m_y(t)
    \right) \big| x,y \in X
    \right]
    }{
    c_{\tf_f}(t)
    },
    \quad  
    d(x,y)=r,
\end{eqnarray}
where $m_x(t)$ is the mark of $x$ at $t\in\mathcal{T}$, and $c_{\tf_f}(t)$ is, for a fixed $t\in\mathcal{T}$, a pointwise normalizing constant corresponding to the expectation of the test function when $r$ tends to infinity.  Denoting the mean and variance of all marks at $t\in \mathcal{T}$ by  $\mu_g(t)$ and $\sigma_g^2(t)$, a summary of the extended test functions is presented in Table \ref{tab:testfuns:objects}. 
As of example, normalizing $\ee
\left[
0.5
\left(
g_x(t)- g_y(t) \right)^2
\big| x,y \in X
\right]$ by $\sigma^2_g(t)$ gives the pointwise mark variogram $\gamma_{gg}(r,t)$, and, to obtain a pointwise version of Stoyan's mark correlation function, the expectation $\ee
\left[
g_x(t)g_y(t)
\big| x,y \in X
\right]$ needs to divided by $\mu_g(t)$  \citep{Eckardt2023MultiFunctionMarks}. 
 
\begin{table}[!h]
\caption{
Pointwise test functions for point processes with function-valued marks.
}
\begin{center}
\begin{tabular}{l | ccc}
\hline
Name of the function & Symbol & Test function ($\tf_f$) & Normalizing \\
                     &   &  & factor $(c_{\tf_f})$ \\
\hline
Mark variogram & $\gamma_{gg}(r,t)$ & $0.5
    [g_x(t)-g_y(t)]^2 $ &  $\sigma^2_{g}(t)$
\\
\hline
Stoyan's mark correlation function & $\kappa_{gg}(r,t)$ & $g_x(t)g_y(t)$ &  $\mu_{g}^2(t)$
\\
\hline
$\mathbf{r}$-mark correlation function & $\kappa_{g\bullet}(r,t)$ & $g_x(t)$ &  $\mu_{g}(t)$
 \\
 \hline
$\mathbf{r}$-mark correlation function & $\kappa_{\bullet g}(r,t)$ & $ g_y(t) $ &  $\mu_{g}(t)$
 \\
\hline
Beisbart and Kerscher's  $\kappa_{gg}(r,t)$& $\kappa_{gg}^{\mathrm{Bei}}(r,t)$ & $g_x(t) + g_y(t)$ &  $2\mu_{g}(t)$
\\
\hline
Isham's  $\kappa_{gg}(r,t)$& $\kappa^{\mathrm{Ish}}_{gg}(r,t)$ & $g_x(t)g_y(t)$ $-\mu_{g}(t)^2$&  $\sigma_{g}^2(t)$
\\
\hline
Stoyan's covariance function & $\cov_{gg}(r,t)$ & $g_x(t)g_y(t)-\mu_{g}^2(t)$ &  1
\\
\hline
Schlather's $I$ & $I_{gg}(r,t)$ & $(g_x(t) - \mu_{g}(r,t)) (g_y(t) - \mu_{g}(r,t))$ &  $\sigma^2_{g}(t)$
\\
\hline

Shimanti's $I$ &  $I^{\mathrm{Shi}}_{gg}(r,t)$ & $(g_x(t) - \mu_{g}(t)) (g_y(t) - \mu_{g}(t))$ &  $\sigma^2_{g}(t)$
\\
\hline
\end{tabular}
\label{tab:testfuns:objects}
\end{center}
\end{table}

While all the test functions presented in Table \ref{tab:testfuns:objects}, with respect to the argument $t$, have similar interpretations as those in Table \ref{tab:testfuns:scalars}, they do not convey any information on the overall pairwise interrelation between the function-valued quantities under study. The desired global mark characteristics, however, can be constructed from their pointwise versions by the integration of the normalized expectation of $\tf_f(g_x(t),g_y(t))$  over $\mathcal{T}$. In this respect, the pointwise mark variogram and mark correlation functions  translate into global versions
\begin{align*}
\gamma_{gg}(r)
&=
\int_{\mathcal{T}}\ee
\left[
0.5
\left(
g_x(t)-g_y(t)
\right)^2
\Big| x,y \in X
\right]
\de t \eqaldot \int_{\mathcal{T}}
\gamma_{gg}(r,t)
\de t,
\\
\kappa_{gg}(r)
&=
\int_{\mathcal{T}}
\ee
\left[
g_x(t) g_y(t)
\Big| x,y \in X
\right]
\de t 
\eqaldot \int_{\mathcal{T}}
\kappa_{gg}(r,t)
\de t. 
\end{align*}
Then, any such global characteristic allows, in a $\mathcal{L}_2$ sense, for a similar interpretation as discussed for Table \ref{tab:testfuns:scalars}. In other words, the global mark variogram $\gamma_{gg}(r)$ gives the average variability over distinct pairs of functions for two points with an interpoint distance $r$. Under mark independence, the average differences among the curves are expected to coincide with the functional variance such that $\gamma_{gg}(r)=1$. Likewise, $\kappa_{gg}(r)$ assesses the product of any two function-valued quantities with respect to the interpoint distance of points $x,y \in X$, where $\kappa_{gg}(r)=1$ for independent marks. Unlike $\kappa_{gg}(r)$,  both $\mathbf{r}$-mark correlation functions $\kappa_{g\bullet}(r)$ and $\kappa_{\bullet g}(r)$ concern the average of the function-valued point attribute of either the first or the second point subject to a distance $r$. This usually differs from $\mu_g$ except under mark independence settings. Similarly to Beisbart and Kerscher's original version, $\kappa^{\mathrm{Bei}}_{gg}(r)$ reflects the average pairwise sum, i.e. the perturbation, of any two functions for a pair of points at a distance $r$ from each other, which is expected to coincide with twice the functional mean $\mu_g$ if marks are independent. If two functions of two nearby locations vary strongly from the functional average, the sum of both would also be, on average, different from the expected case leading to $\kappa_{gg}^{\mathrm{Bei}}(r)\neq 1$. Lastly, in both function-valued versions of Schlather's and Shimanti's $I$-functions, marks are centered by the conditional and unconditional functional means squared, respectively, and scaled by the functional variance. We note that the same ideas can also be applied to the mark-weighted $K$-function $K_{\tf_f}(\cdot)$, given in \eqref{eq:Kmm}, and the function $U(r)$, given in \eqref{eq:ustat}, leading to  
$K_{\tf_f}(t) = \int K_{\tf_f}(r,t)\de t$ and $U(r) =  \int U(r,t)\de t$. 

The extensions of the above summary characteristics to multivariate spatial point processes, including \textit{cross-type}, \textit{cross-function} and \textit{cross-type cross-function} mark summary characteristics are discussed by \cite{Eckardt2023MultiFunctionMarks}.
While the cross-type characteristics reveal the interrelation of mark $g(\cdot)$ for points $x\in X_i$ and $y \in X_j$, the cross-function versions are defined for function-valued marks $g_1(\cdot)\neq g_2(\cdot)$ on $\mathds{F}^2$ at locations $x,y\in X$. These ideas are also extended to \textit{multi-function} versions for the sets $\{x_i, \mathbf{m}(x_i)\}^n_{i=1}$, with $\mathbf{m}(x_i)\in\mathds{F}^s, ~s>2$, see \cite{Eckardt2023MultiFunctionMarks} for details.



\section{Marked spatial point processes on linear networks}\label{sec:mlpp}

In certain point process applications, the spatial distribution of events may become confined within an underlying structure. Instances of these applications include the study of street crimes, traffic accidents, trees positioned alongside roads, and various other scenarios where spatial locations are restricted to a linear network. 

A linear network $\LL \subset \R^2$ is considered as a finite union of some line segments, i.e., $\LL=\bigcup_{i=1}^k l_i$ where $l_i=[u_i,v_i]=\{tu_i + (1-t)v_i:0\leq t\leq 1\}$, $u_i\neq v_i\in\R^2$. The linear network $\LL$ is equipped with a regular distance metric $\dL$, with the shortest-path distance being an example of it \citep{rakshit2017second}. Moreover, no over/under-pass is assumed within $\LL$, meaning that each intersection of segments is a node. The total length of $\LL$ is denoted by $|\LL|=\sum_{i=1}^k |l_i|$ where $|l_i|=\dL(u_i,v_i)=d(u_i,v_i)$. We denote a marked point process on $\LL$ by $X^{\LL}$ for which an observed point pattern is denoted by $\x^{\LL}$, and similarly to \eqref{eq:lambda}, if we let $\X^{\LL}$ be the unmarked version of $X^{\LL}$, then 
\begin{eqnarray}\label{eq:lambdanet}
    \ee
    [N(\X^{\LL} \cap A)]
    =
    \int_A \lambda^{\LL}(u) \de_1 u,
    \quad
    A \subset \LL,
\end{eqnarray}
where $\lambda^{\LL}(\cdot)$ is the intensity function of $\X^{\LL}$, and $\de_1$ stands for integration with respect to arc length on the network. In this case, the intensity function $\lambda^{\LL}(u)$ 
gives the expected number of points per unit length of the network in the vicinity of a location $u \in \LL$  \citep{BADDELEY2021100435}. Similarly to \eqref{eq:Campbell} and \eqref{eq:correl}, for any non-negative 
measurable function $f(\cdot)$ on the product space $\LL^m$,
\begin{align}\label{eq:product}
\ee
\left[
\mathop{\sum\nolimits\sp{\ne}}_{x_1,\ldots,x_m\in \X^{\LL}}f
(x_1,\ldots,x_m)
\right] 
=
\int_{\LL^m}
f(u_1,\ldots,u_m) 
\lambda^{(m)}_{\LL} (u_1,\ldots,u_m)
\de_1u_1 \cdots \de_1u_m ,
\end{align}
and
\begin{align}\label{eq:correlnetwork}
g_m^{\LL}(u_1,\ldots,u_m)
    =
    \frac{
    \lambda^{(m)}_{\LL} (u_1,\ldots,u_m)
    }{
    \lambda^{\LL}(u_1) \cdots \lambda^{\LL}(u_m)
    }
    =
    \sum_{j=1}^m
    \sum_{D_1,\ldots,D_j}\xi_{N(D_1)}^{\LL}
    \left(\{u_i:i\in D_1\}\right)
     \cdots \xi_{N(D_j)}^{\LL}
     \left(\{u_i:i\in D_j\}
     \right)
    ,
\end{align}
where $\lambda^{(m)}_{\LL}(\cdot)$  is the $m$-order product intensity of $X^{\LL}$, $\sum_{D_1,\ldots,D_j}$ ranges over all partitions $\{D_1,\ldots,D_j\}$ of $\{1,\ldots,m\}$ into $j$ non-empty and disjoint sets, and $N(D_j)$ is the cardinality of the index set $D_j$ \citep{cronie2020inhomogeneous}. In order to propose marked summary characteristics for point processes on linear networks, similarly to Section \ref{sec:mpp}, we need some form of stationarity. However, addressing this challenge has proven to be quite complex \citep{baddeley2017stationary}, as, currently, there is no transformation that can transform points on a linear network with the guarantee that the shifted point will remain on the same network. \cite{cronie2020inhomogeneous} proposed different notions of stationarity on linear networks, according to which whenever the product intensities $\lambda_{\LL}^{(m)}(\cdot)$, $1\leq m\leq k$, exist, $\bar\lambda_{\LL}=\inf_{u\in \LL}\lambda_{\LL}(u)>0$, and for any $m\in\{2,\ldots,k\}$ the correlation function $g_m^{\LL}(\cdot)$ satisfies
\begin{align}
    \label{IRMPSstrong}
g_m^{\LL}(u_1,\ldots,u_m)
=
\bar g_m^{\LL}(d_\LL(u,u_1),\ldots,d_\LL(u,u_m)),
\end{align}
for any $u\in \LL$ and some function
$\bar g_m^{\LL}:[0,\infty)^m\to[0,\infty)$, 
they said that $\X^{\LL}$ is $k$-th order intensity reweighted pseudostationary (with respect to a regular distance $d_{\LL}$);  $\X^{\LL}$ is called intensity reweighted moment pseudostationary (IRMPS) when this holds for any order $k\geq2$. Moreover, a homogeneous point process $\X^{\LL}$, which is also $k$-th order intensity reweighted pseudostationary, is called  $k$-th order pseudostationary. 
Finally, a moment pseudostationary point process $\X^{\LL}$ is considered (strongly)  pseudostationary if its moments completely and uniquely characterize its distribution. 

\subsection{Discrete and integer-valued marks}\label{sec:lppmethods}

We now consider the case where $X^{\LL}$ can be decomposed into $1 < k \leq n$ components according to a discrete/integer-valued mark. In this case, $X^{\LL}$ is called a multivariate/multitype spatial point process on a linear network $\LL$, and $\xl$  is denoted as a collection of $k$ distinct point patterns $\{ \xl_1,\ldots,\xl_k \}, k\geq 2$. Marked summary characteristics for multivariate/multitype point processes on linear networks can be naturally proposed by 
mimicking the underlying principles presented in Section \ref{sec:discretemark}.
However, despite the growing availability of multivariate/multitype spatial network-constrained point processes,  there exist only a few of such extensions \citep{Spooner2004, BaddeleyJammalamadakaNair2014, EckardtJCGS, Eckardt2020}. In particular, most attention is paid to the pair correlation function and the $K$-function, for which, by following \citet{Ang2012, BaddeleyJammalamadakaNair2014,rakshit2017second}, for a second-order pseudostationary ($d_\LL$-correlated in the language of \cite{rakshit2017second}) marked point process $X^\LL$, we have 
\begin{eqnarray}\label{eq:crossKinhom:linnet}
K_{ij}^{\LL,\mathrm{inhom}}(u, \rL)
=
\ee
\left.
\left[
\sum_{x \in X_j^{\LL}}
\frac{
\1\{\dL(u,x)\leq \rL \}  \nabla\{u, \dL(u,x)  \}
}{
\lambda_j^{\LL} (x)
}
\right|
u \in X_i^{\LL}
\right],
\quad 
\quad
i,j=1,\ldots,k,
\quad
\rL>0,
\end{eqnarray}
and 
\begin{eqnarray}\label{eq:dotKinhom:linnet}
K_{i\bullet}^{\LL,\mathrm{inhom}}(u, \rL)
=
\ee
\left.
\left[
\sum_{\substack{x \in X^{\LL} \\ u \neq x}}
\frac{
\1\{ \dL(u,x)\leq \rL \}
\nabla\{u, \dL(u,x)  \}
}{
\lambda^{\LL} (x)
}
\right|
u \in X_i^{\LL}
\right],
\quad 
\quad
i=1,\ldots,k,
\quad
\rL>0,
\end{eqnarray}
where $\nabla(\cdot)$ acts as an edge correction factor, and $\lambda^{\LL}_j(\cdot)$ is the intensity function of the $j$-th component. If the employed metric is the shortest-path distance, then $\nabla\{u, \dL(u,x)  \}$ becomes the reciprocal of the number of points lying exactly $\dL(u,x)$ units from $u\in X_i$. In a similar manner, one could obtain the cross/dot-type versions of the inhomogeneous pair correlation function. 

\subsubsection{Mark connection and mingling function}

For homogeneous point processes on linear networks, 
\cite{BaddeleyJammalamadakaNair2014} presented the network-based versions of the mark connection and mark equality functions as
\begin{eqnarray}
p_{ij}^{\LL}(\rL)
=
\frac{\lambda_i^{\LL} \lambda_j^{\LL} \rho^{\LL}_{ij}(\rL)}{[{\lambda^{\LL}]}^2
\rho^{\LL}(\rL)},
\quad \quad
p^{\LL}(\rL)
=
\sum^{k}_{i=1}p^{\LL}_{ii}(\rL),
\end{eqnarray}
where $\rho^{\LL}_{ij}(\cdot)$ is the cross pair correlation function for the processes with types $i,j$ and $\rho^{\LL}(\cdot)$ is the pair correlation of the full process. The mark connection function $p_{ij}^{\LL}(\rL)$ is, given the presence of a pair of points separated by a distance $\rL$, intuitively the conditional probability that those points are of types $i,j$. Similarly, $p^{\LL}(\rL)$ is the conditional probability that those two points, separated by a distance $\rL$, have the same type. 

Next, we propose the counterpart of the normalized mingling function \eqref{eq:mingleR2} for homogeneous point processes on linear networks as
\begin{eqnarray}\label{eq:mingleln}
\nu_{\LL}(\rL)
=
\frac{1}{c}
\ee
\left[
\sum_{x,y \in X^{\LL}}^{\neq}
\1 \{
m(x)\neq m(y)
\}
\1\{d_{\LL}(x,y)\leq \rL \}
\right],
\end{eqnarray}
where $c$ is the same normalizing factor as in  \eqref{eq:mingleR2}. Similarly, $\nu_{\LL}(\cdot)$ may be used to discover aggregation/repulsion between points of different types subject to an interpoint distance $\rL$. Note that, unlike the cross/dot-type summary characteristics, the mingling functions \eqref{eq:mingleR2} and \eqref{eq:mingleln} quantify the aggregation and/or repulsion tendencies between any given types, rather than focusing on the interaction between two specific types or comparing a single type against all other types.

\subsubsection{Higher-order marked inhomogeneous summary characteristics}

Next, for IRMPS point processes, we extend the inhomogeneous higher-order summary characteristics of \cite{cronie2020inhomogeneous} to multivariate/multitype settings. In particular, we have the following cross-type summary characteristics 
\begin{align}
    H^{\LL, \mathrm{inhom}}_{ij}(u,\rL)
    &= 
    1
    -
    \ee
    \left[
    \prod_{x \in X_j^{\LL}}
    \left(
    1
    -
    \frac{
\bar{\lambda}^{\LL}_j \1 \{ d_{\LL}(u,x) \leq r \}
}{\lambda_j^{\LL}(x)} 
\nabla\{u, \dL(u,x)  \} 
    \right)
    \Bigg|
    u \in X_i^{\LL}
    \right],
    \\
    J^{\LL, \mathrm{inhom}}_{ij}(u,\rL)
    &= 
\frac{1-H^{\LL, \mathrm{inhom}}_{ij}(u,\rL)}{1-F^{\LL, \mathrm{inhom}}_{j}(u,\rL)}, 
\quad
F^{\LL, \mathrm{inhom}}_{j}(u,\rL) \neq 1,
\\
F^{\LL, \mathrm{inhom}}_{j}(u,\rL)
&=
    1- 
        \ee
    \left[
    \prod_{x \in X_j^{\LL}} 
    \left(
1 -
\frac{\bar\lambda^{\LL}_j
\1
\{
d_{\LL}(u,x) \leq r
\}
}{\lambda_j^{\LL}(x)}
\nabla\{u, \dL(u,x)  \}
\right)
\right],
\end{align}
and the dot-type ones 
\begin{align}
    H^{\LL, \mathrm{inhom}}_{i\bullet}(u,\rL)
    &= 
    1
    -
    \ee
    \left[
    \prod_
    {\substack{x \in X^{\LL} \\ u \neq x}}
    \left(
    1
    -
    \frac{
\bar{\lambda}^{\LL} \1 \{ d_{\LL}(u,x) \leq r \}
}{\lambda^{\LL}(x)} 
\nabla\{u, \dL(u,x)  \} 
    \right)
    \Bigg|
    u \in X_i^{\LL}
    \right],
    \\
    J^{\LL, \mathrm{inhom}}_{i\bullet}(u,\rL)
    &= 
\frac{1-H^{\LL, \mathrm{inhom}}_{i\bullet}(u,\rL)}{1-F^{\LL, \mathrm{inhom}}(u,\rL)},
\quad 
F^{\LL, \mathrm{inhom}}(u,\rL) \neq 1,
\end{align}
where $F^{\LL, \mathrm{inhom}}(u,\rL)$ and $F^{\LL, \mathrm{inhom}}_{j}(u,\rL)$ are the inhomogeneous empty space functions of $X^{\LL}$ and $X_j^{\LL}$, respectively. For any two fixed types $i, j$, $J^{\LL, \mathrm{inhom}}_{ij}(\cdot)=1$ means points of type $j$ are randomly distributed around points of type $i$, $J^{\LL, \mathrm{inhom}}_{ij}(\cdot)>1$ means that points of type $j$ tend to happen around points of type $i$, and $J^{\LL, \mathrm{inhom}}_{ij}(\cdot)<1$ indicates that points of type $j$ prefer to maintain a distance from points of type $i$. Turning to the dot-type $J$-function, if $J^{\LL, \mathrm{inhom}}{i\bullet}(\cdot)=1$, it implies that points of different types than $i$ exhibit no interaction with points of type $i$. When $J^{\LL, \mathrm{inhom}}{i\bullet}(\cdot)>1$, it signifies a propensity for other types to appear in proximity to points of type $i$. Conversely, $J^{\LL, \mathrm{inhom}}_{i\bullet}(\cdot)<1$ suggests a preference for other types to hold a distance from points of type $i$.
Note that these higher-order marked summary characteristics are expected to provide a deeper understanding of the spatial patterns and clustering/repulsion behavior among points of different types since they go beyond pairwise interactions providing information about interactions involving more than two points.  


\subsection{Real-valued marks}\label{sec:lppnew}

We now turn to a situation where points are labeled by real-valued marks,  
taking the presented methods in Section \ref{sec:reals}, and present novel mark summary characteristics that take the geometry of the underlying network into account.



\subsubsection{Mark correlation functions and mark-weighted summary characteristics}\label{sec:lppnew1}

Generalizing the $\tf_f$-correlation functions in \eqref{eq:tfcorr} to the present setting, the ${\tf_f^{\LL}}$-correlation function for point processes on linear networks is of the form
\begin{eqnarray}\label{eq:tfcorrNet}
\kappa^{\LL}_{\tf_f^{\LL}}(\rL)
=
\frac{
\ee \left[
    \tf_f^{\LL} 
    \left(
    m(x),m(y)
    \right) 
    \big| x,y \in X^{\LL}
    \right]
    }{
    c_{\tf_f^{\LL}}
    },
    \quad  
    \dL(x,y)=\rL,
\end{eqnarray}
where $c_{\tf_f^{\LL}}$ is a normalizing factor; the superscript $^{\LL}$ emphasizes that the ${\tf_f}$-correlation function will be evaluated over the linear network ${\LL}$.
The numerator in \eqref{eq:tfcorrNet} is the conditional expectation of a given test function evaluated over a pair of marks whose corresponding spatial locations are $r_\LL$ distances away, given that the two points belong to the point process $X^{\LL}$. The denominator $c_{\tf_f^{\LL}}$ is the expected value of the considered test function under the mark independence assumption.
Applying the same principles as in Section \ref{sec:reals}, the specific linear-network-based mark characteristic depends on the explicit formulation of the test function. A summary of potential test functions for point processes on linear networks and their notations is given in Table \ref{tab:testfuns:nets}.

\begin{table}[!h]
\caption{Test functions for point processes on linear networks with real-valued marks. The average and variance of all marks are denoted by $\mu_{m,{\LL}}$ and $\sigma^2_{m,{\LL}}$, and $\mu_{m,{\LL}}(r_{\LL})$ is the conditional mean of the marks for points with an interpoint distance $\rL$.
}
\begin{center}
\begin{tabular}{l | ccc}
\hline
Name of the function & Symbol & Test function ($\tf_f$) & Normalizing \\
                     &   &  & factor $(c_{\tf_f})$ \\
\hline
Mark variogram& $\gamma^{\LL}_{mm}(\rL)$ & $0.5
    [m(x)-m(y)]^2 $ &  $\sigma^2_{m,{\LL}}$
\\
\hline
Stoyan's mark correlation function & $\kappa^{\LL}_{mm}(\rL)$ & $ m(x)m(y)$ &  $\mu_{m,{\LL}}^2$
\\
\hline
$\mathbf{r}^{{\LL}}$-mark correlation function & $\mathbf{r}^{{\LL}}_{m\bullet}(\rL)$ & $m(x)$ &  $\mu_{m,{\LL}}$
 \\
 \hline
$\mathbf{r}^{{\LL}}$-mark correlation function & $\mathbf{r}^{{\LL}}_{\bullet m}(\rL)$ & $ m(y) $ &  $\mu_{m,{\LL}}$
 \\
\hline
Beisbart and Kerscher's  $\kappa^{\LL}_{mm}(\rL)$& $\kappa_{mm}^{\mathrm{Bei},\LL}(\rL)$ & $m(x) + m(y)$ &  $2\mu_{m,{\LL}}$
\\
\hline
Isham's  $\kappa^{\LL}_{mm}(\rL)$& $\kappa^{\mathrm{Ish},\LL}_{mm}(\rL)$ & $m(x)m(y)$ $-(\mu_{m,\LL})^2$&  $\sigma_{m,{\LL}}^2$
\\
\hline
Stoyan's covariance function & $\cov^{\LL}_{mm}(\rL)$ & $m(x)m(y)-(\mu_{m,\LL})^2$ &  1
\\
\hline
Schlather's $I$ & $I^{\LL}_{mm}(\rL)$ & $(m(x) - \mu_{m,\LL}(r_{\LL})) (m(y) - \mu_{m,\LL}(r_{\LL}))$ &  $\sigma^2_{m,{\LL}}$
\\
\hline

Shimanti's $I$ &  $I^{\mathrm{Shi},\LL}_{mm}(\rL)$ & $(m(x) - \mu_{m,\LL}) (m(y) - \mu_{m,\LL})$ &  $\sigma^2_{m,{\LL}}$
\\
\hline
\end{tabular}
\label{tab:testfuns:nets}
\end{center}
\end{table}
Taking the geometry of linear networks into account and employing a regular metric \citep{rakshit2017second}, the mark variogram $\gamma_{mm}^{\LL}(\rL)$ quantifies the variability of marks for pairs of points with an interpoint distance $\rL$.
Note that, 
in contrast to the classic mark variogram $\gamma_{mm}(r)$, given in Table \ref{tab:testfuns:scalars}, which disregards the underlying geometry of the linear network  $\LL$ and, thus, may yield biased outcomes and potentially mislead interpretations (see Section \ref{sec:numerical}), the geometrically corrected mark variogram $\gamma_{mm}^{\LL}(\rL)$ takes the specific properties of the linear network into account and provide more accurate outcomes. The large/small local differences, i.e. high/low local variation, between marks give rise to large/small values for the mark variogram function, deviating from the mark independence assumption.
Similarly, the geometrically corrected version of Stoyan's mark correlation $\kappa_{mm}^{\LL}(\rL)$ quantifies the average pairwise product of marks for points separated by a distance $\rL$ along the linear network $\LL$. If marks are independent, it coincides with the mark mean squared $\mu_{m,\LL}^2$ providing that $\kappa_{mm}^{\LL}(\rL)=1$. If marks for nearby points on a linear network differ from the expected mark under independence, their average would also be different from $\mu^2_{m,\LL}$ and thus $\kappa_{mm}^{\LL}(\rL)\neq 1$. Similarly, the linear network $\mathbf{r}^{{\LL}}$-mark correlation functions would only coincide with the mark mean if the marks are independent subject to a distance $\rL$. Consequently, any deviations of $\mathbf{r}^{{\LL}}$-mark correlation functions from one indicate the existence of a structure in the distribution of marks. 
The geometrically corrected version of Beisbart and Kerscher's mark correlation  $\kappa^{\mathrm{Bei},\LL}_{mm}(\rL)$ expects that the average pairwise sum of marks is expected to be equal to twice the mark mean under independence. Thus, deviations of  $\kappa^{\mathrm{Bei},\LL}_{mm}$ from unity highlight local dependencies and the existence of a particular structure among marks. In a similar manner, one may describe other mark characteristics presented in Table \ref{tab:testfuns:nets}. 

In addition to the above test functions, we further, for a homogeneous point process $X^{\LL}$ with constant intensity $\lambda^{\LL}$, extend the mark-weighted $K$-functions $K_{\tf_f}(r)$, given in \eqref{eq:Kmm}, to the linear network settings, and propose its counterpart as 
\begin{eqnarray}\label{eq:linnetMarked:K}
K^{\LL}_{\tf_f^{\LL}}(\rL)
=
\ee 
\left[
\frac{1}{c_{\tf_f^{\LL}}\lambda^{\LL}}
\sum_{\substack{x \in X^{\LL} \\ u \neq x}}
\tf_f^{\LL}(m(x),m(u))
\1 
\lbrace \dL(x,u) \leq \rL \rbrace
\nabla(u, d_{\LL}(u,x))
\Big| u\in X^{\LL} 
\right], 
\end{eqnarray}
which becomes 
\begin{eqnarray}\label{eq:linnetMarked:Kinhom}
K^{\LL,\mathrm{inhom}}_{\tf_f^{\LL}}(\rL)
=
\ee\left[
\frac{1}{c_{\tf_f^{\LL}}}
\sum_{\substack{x \in X^{\LL} \\ u \neq x} 
}
\tf_f^{\LL}(m(x),m(u))
\frac{
\1 
\lbrace \dL(x,u) \leq \rL \rbrace
\nabla(u, d_{\LL}(u,x))
}{\lambda^{\LL}(x)
}
\Big| u\in X^{\LL}
\right]
\end{eqnarray}
for inhomogeneous point processes.
Implementing the test function $\tf_f^{\LL}(\cdot)$ as a weight in the geometrically corrected $K$-function for spatial point processes on linear networks as in \eqref{eq:linnetMarked:K} and \eqref{eq:linnetMarked:Kinhom}, gives rise to a scaled version of the original $K$-function. In particular, since the implemented weight becomes one if marks are independent, the up/down-scaled version of the original $K$-function highlights some interaction between marks. Under the assumption that points are generated from a Poisson process with independent marks, the above mark-weighted $K$-functions possess a known value denoted by $r_{\LL}$. However, due to the presence of two sources for re-scaling the $K$-function, namely pairwise interaction between points and between marks, it is advisable to compute both the weighted and unweighted versions for a more comprehensive understanding. Note that the weighting idea is applicable to other summary characteristics such as the inhomogeneous empty space function, the inhomogeneous nearest neighbor distance distribution function, and the inhomogeneous $J$-function.


Next, for a homogeneous point process $X^{\LL}$, on a linear network $\LL$, with the pair correlation $\rho^{\LL}(\cdot)$, we propose an extension of \eqref{eq:ustat} as
\begin{eqnarray}\label{eq:linnetMarked:U}
U^{\LL}(\rL)
=
[{\lambda^{\LL}}]^{2}
\rho^{\LL}(\rL)
\kappa^{\LL}_{\tf_f^{\LL}}(\rL)
|\de_1 x| |\de_1  y|,
\end{eqnarray}
where $|\de_1 x|$ and $ |\de_1  y|$ are the sizes of two infinitesimal segments around the points $x, y \in \LL$; similar interpretation, as the planar case, holds for $U^{\LL}(\rL)$.


\subsubsection{Numerical evaluation}\label{sec:numerical}

\begin{figure}[!h]
    \centering
    \includegraphics[scale=0.12]{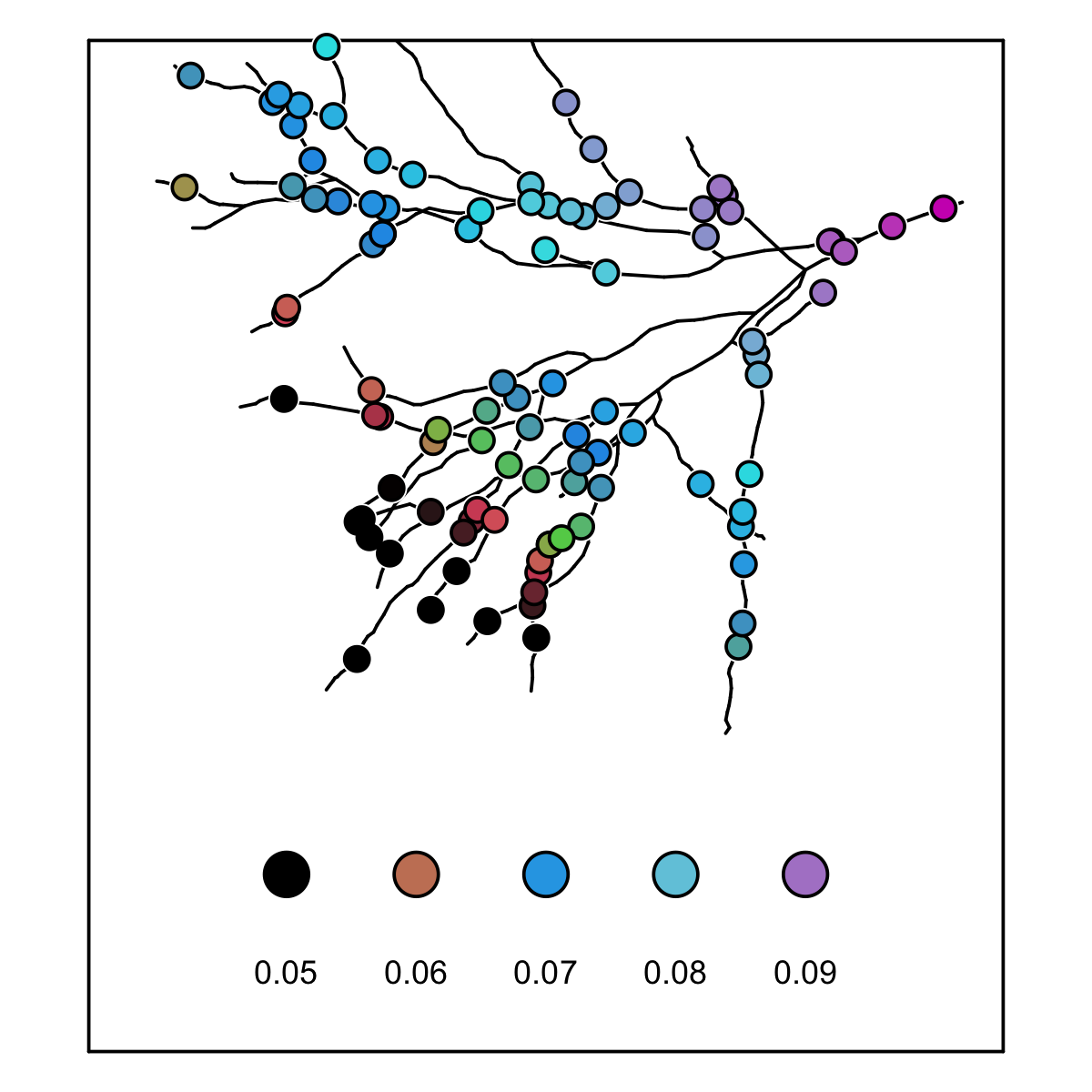}
    \includegraphics[scale=0.12]{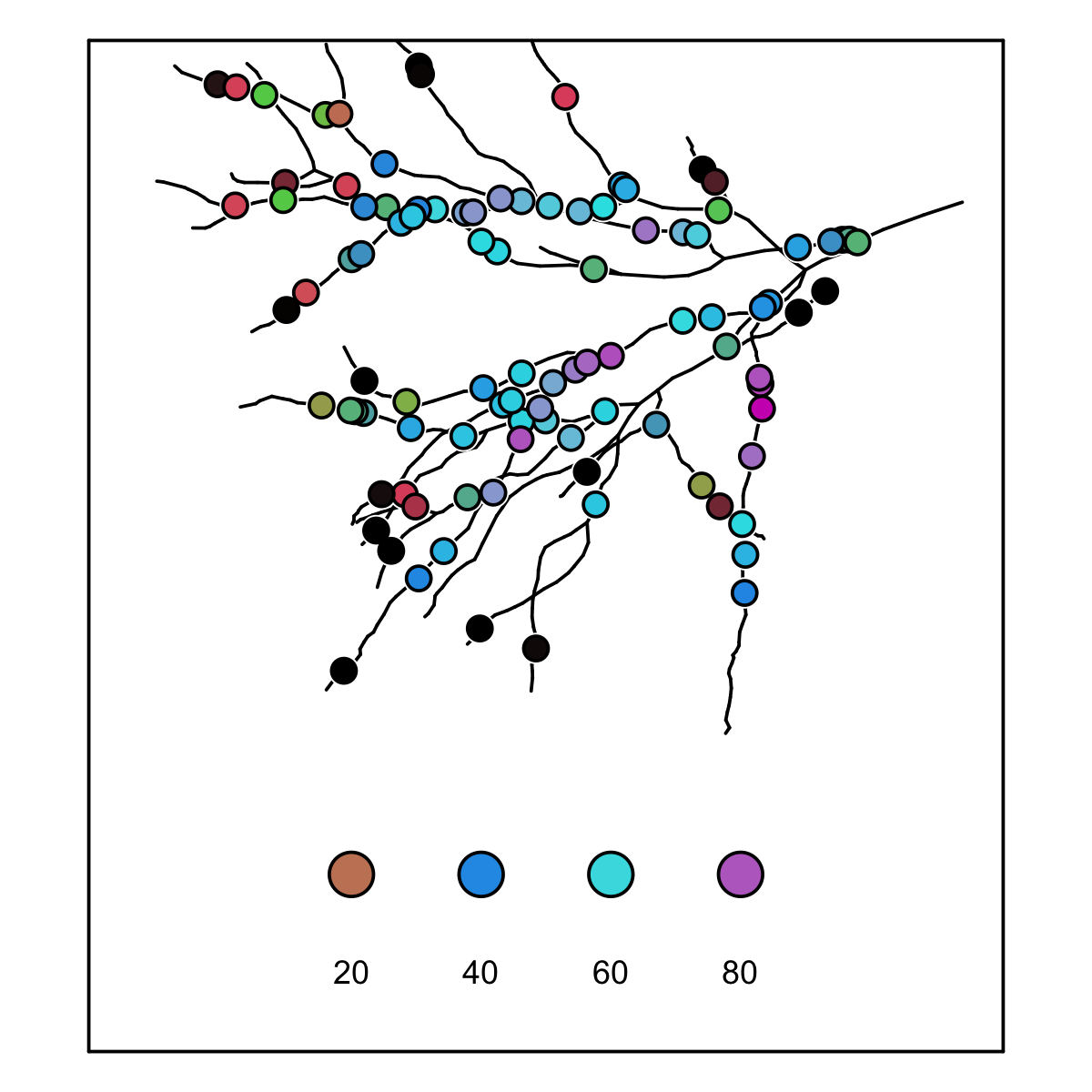}
    \includegraphics[scale=0.12]{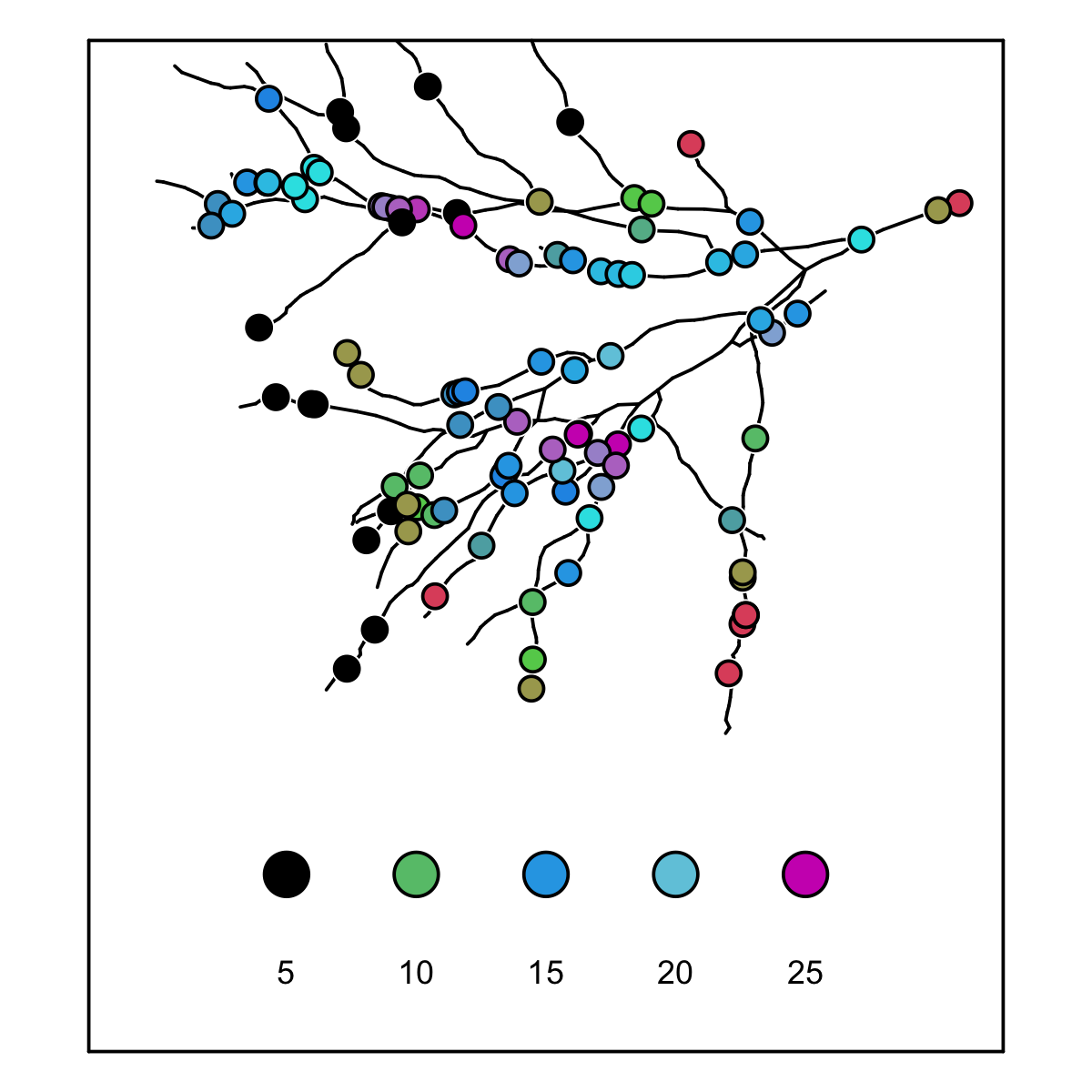}
    \caption{
    Examples of simulated data. From left to right: Models I, II, and III. Colors show mark values.
    }
    \label{fig:datasim}
\end{figure}

This section is devoted to evaluating the performance of the proposed linear-network-based mark characteristics and investigating potential variations between the classic planar mark correlation functions and those that take the underlying linear network into account. In terms of the spatial distribution of points, following the assumptions concerning the definition of mark correlation functions, we consider homogeneous point patterns, with 100 uniformly distributed points, on a dendrite network, which is previously used in \citet{jammalamadaka2013statistical, BaddeleyJammalamadakaNair2014}. 
Next, to convert the simulated point patterns into marked point patterns, we generate real-valued quantities according to three distinct scenarios, called models I, II, and III, and assign the simulated marks to each point location. In model I, marks follow the function $f(x,y)=(x+y)/5000$, having a trend from bottom-left to top-right of the network; in model II, marks are the shortest-path distance from each point to the dendrite's border, and in model III marks are the number of points which have a distance less than 80 to a target point, i.e., the number of nearest neighbor points. Figure \ref{fig:datasim} shows examples of the considered models. Note that within all three considered models, there is a clear structure for the marks on the dendrite network. 


\begin{figure}[!h]
    \centering
    \includegraphics[scale=0.12]{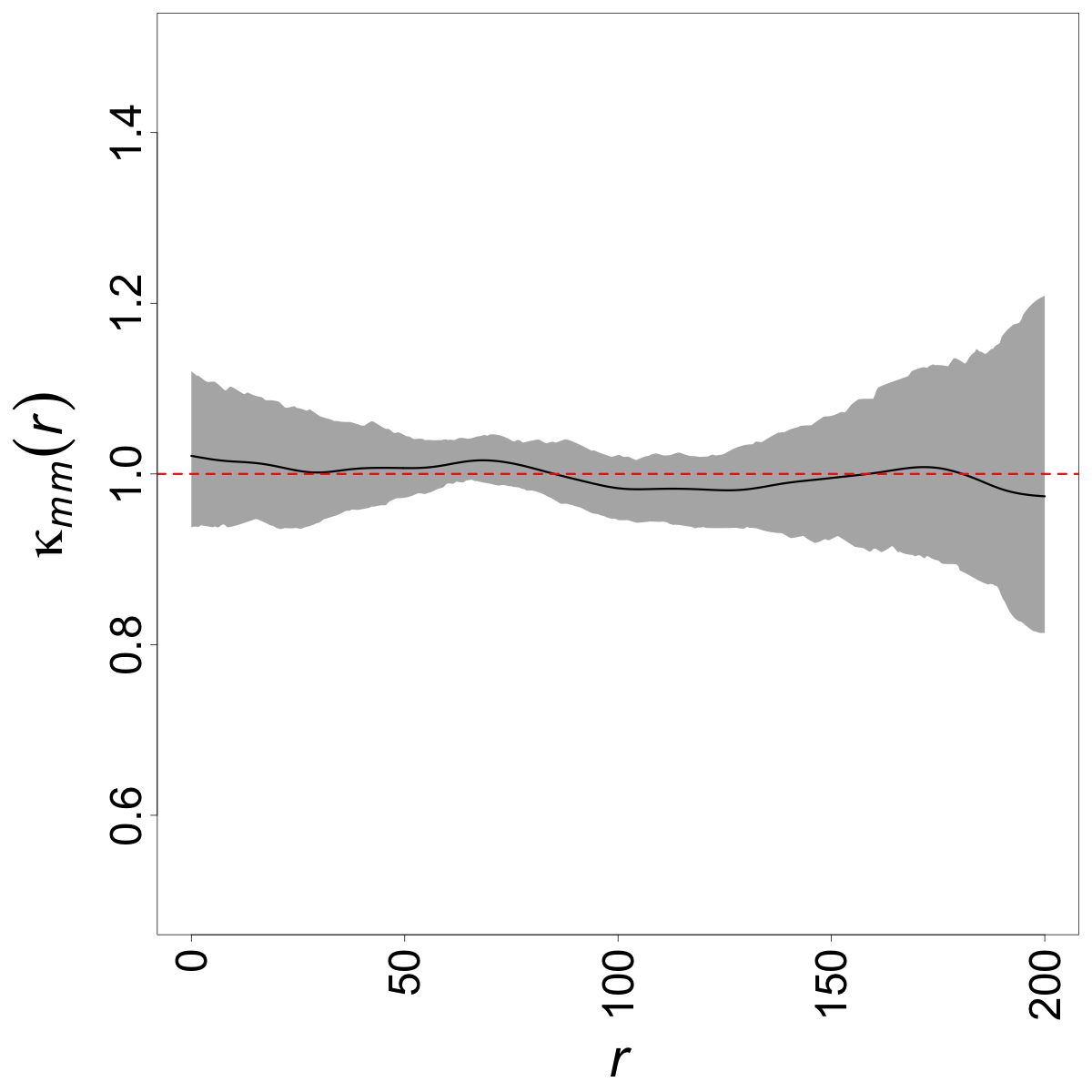}
    \includegraphics[scale=0.12]{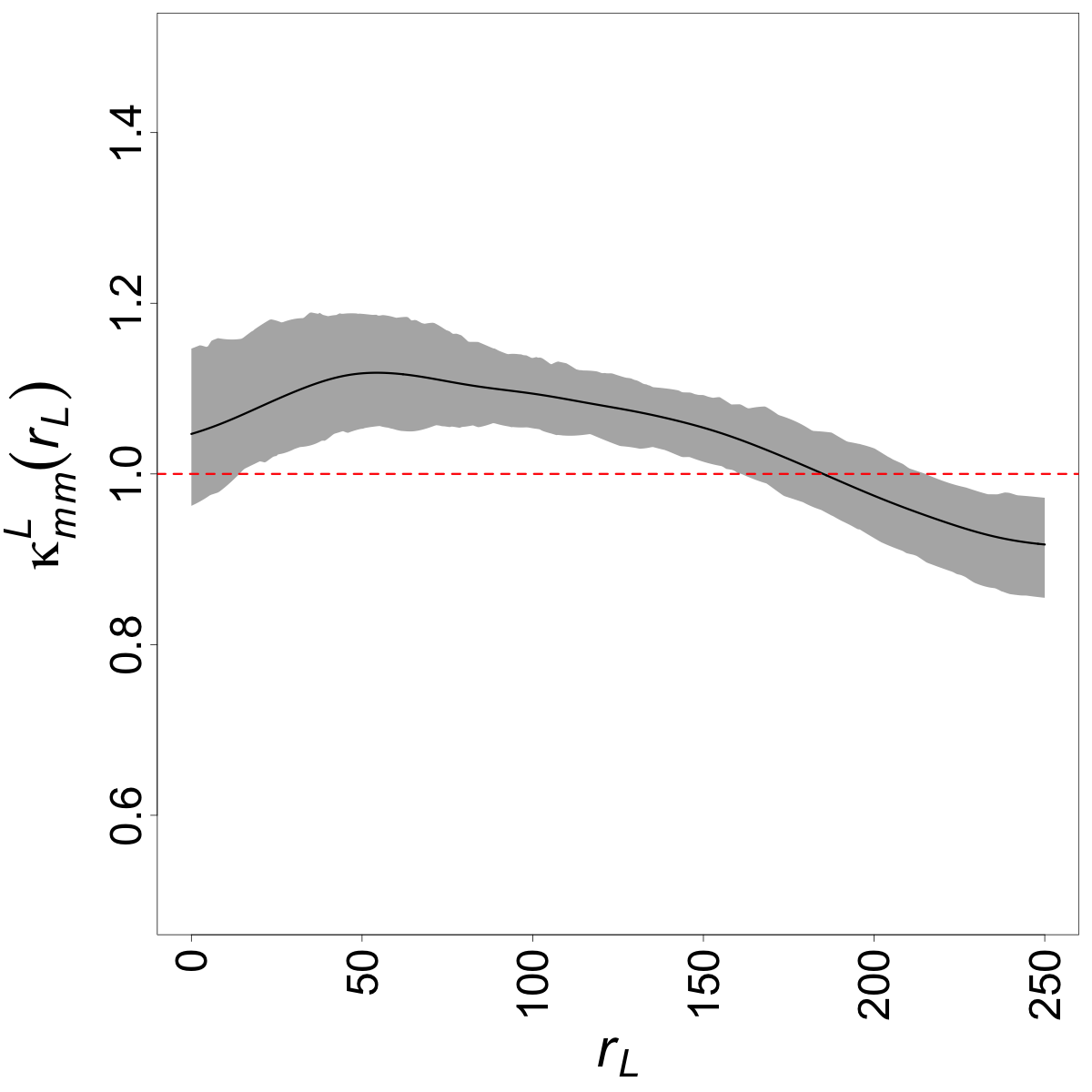}
    \includegraphics[scale=0.12]{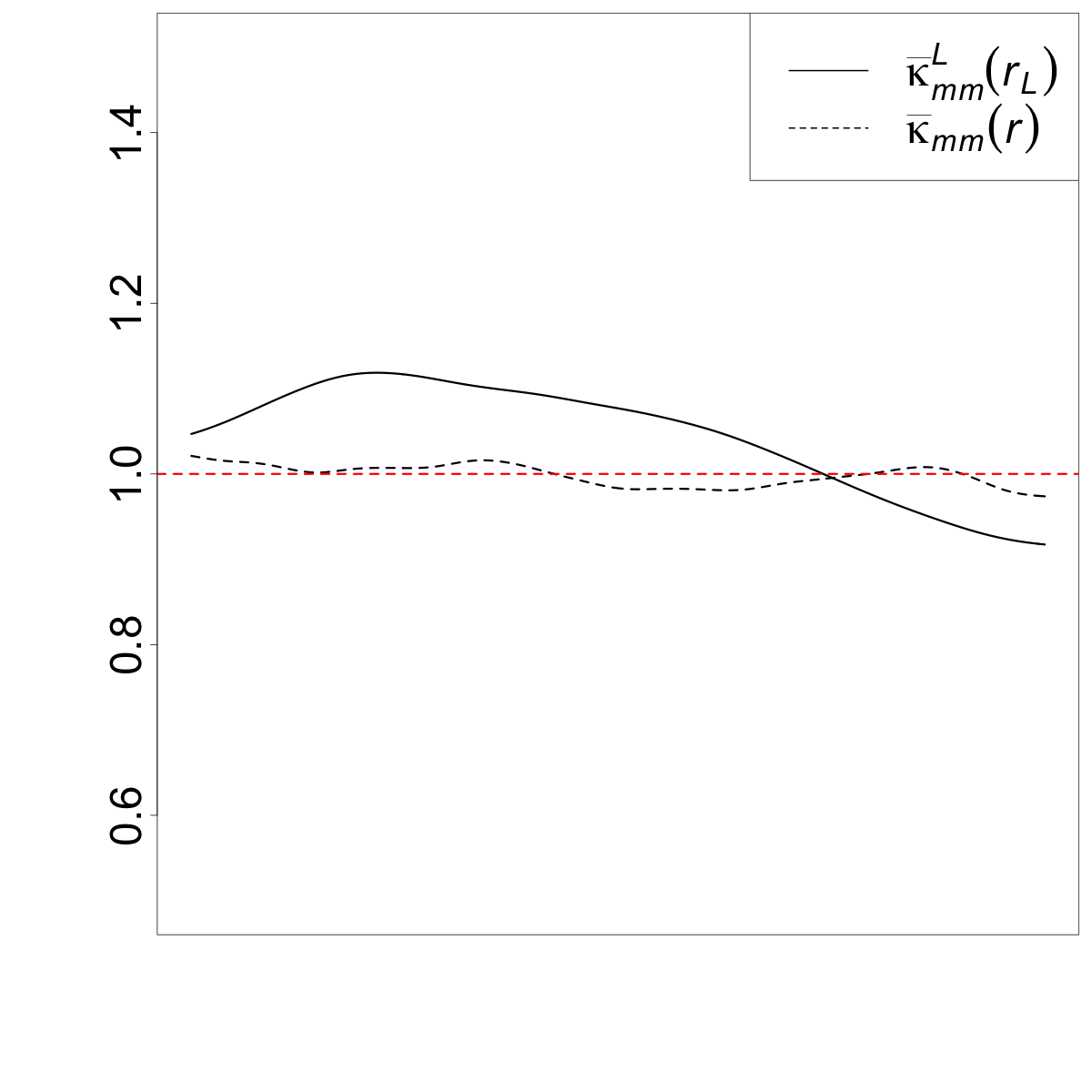}
    
    \includegraphics[scale=0.12]{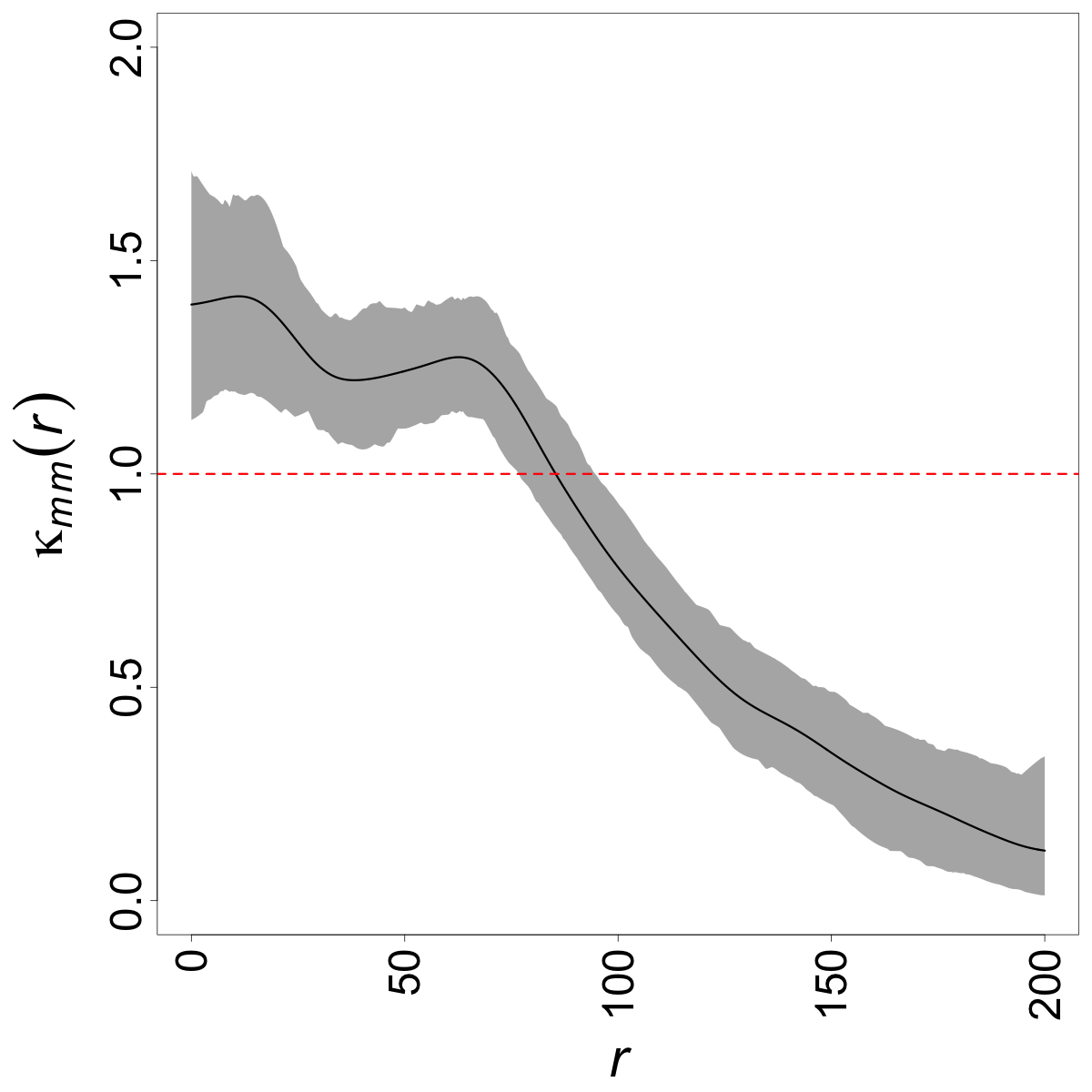}
    \includegraphics[scale=0.12]{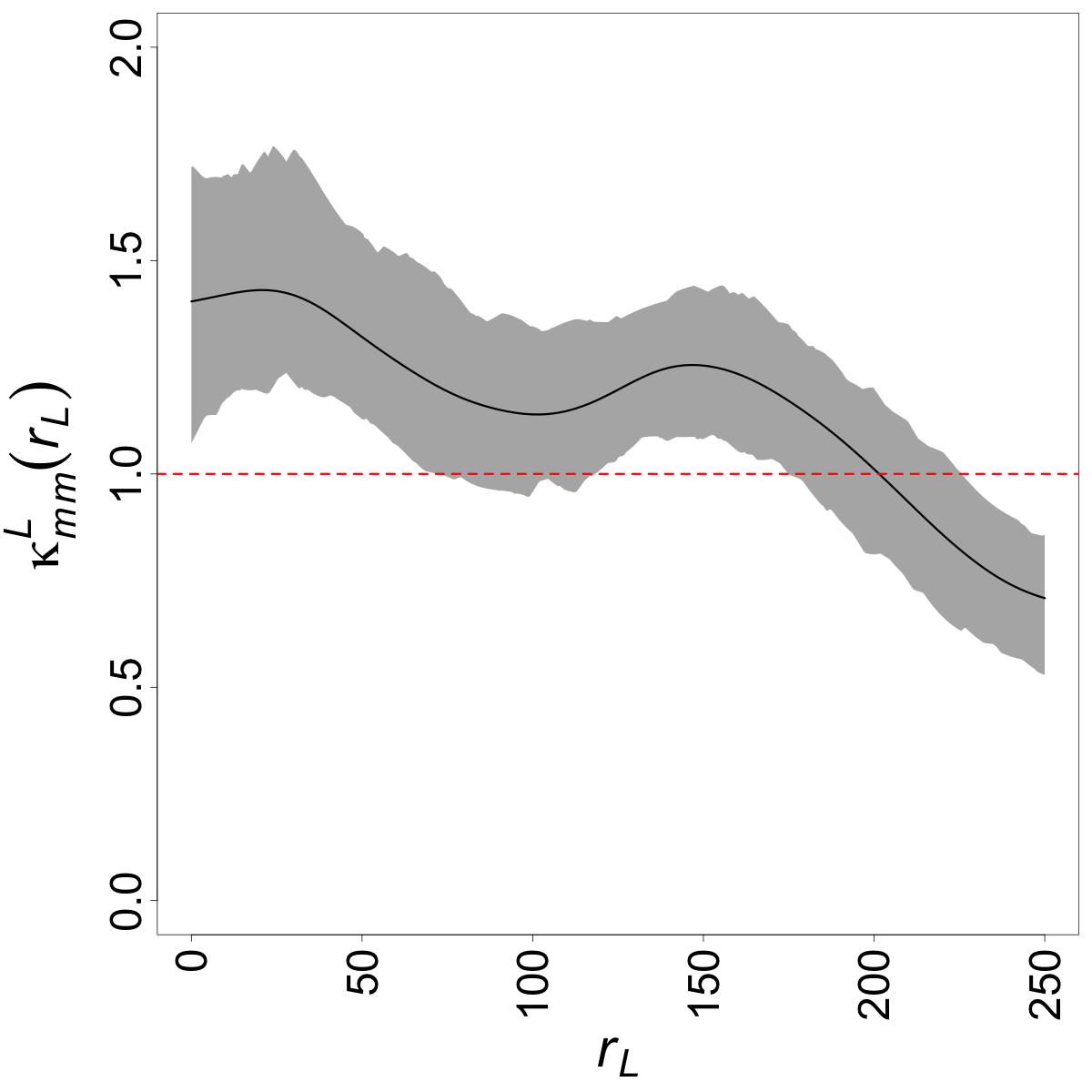}
    \includegraphics[scale=0.12]{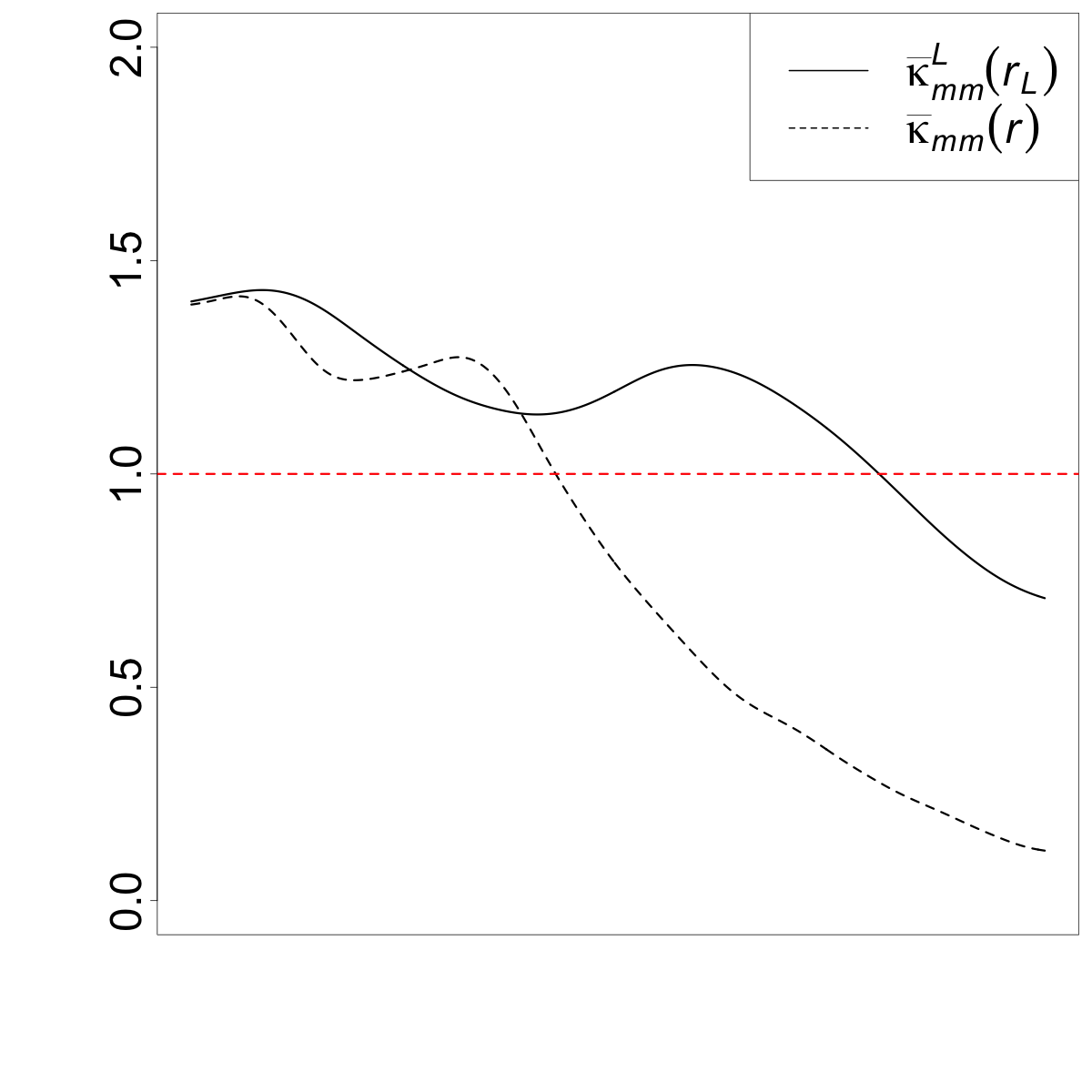}

    \includegraphics[scale=0.12]{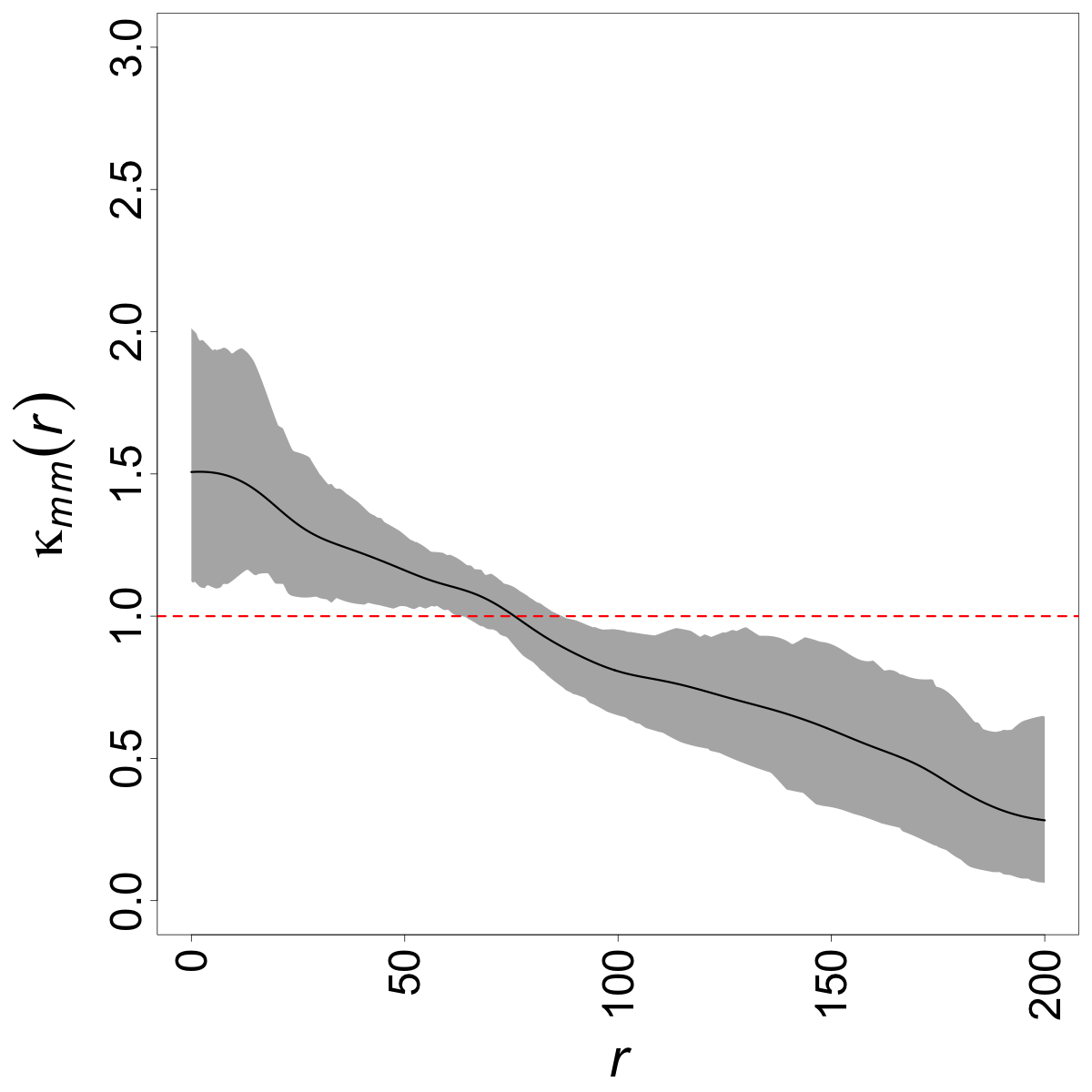}
    \includegraphics[scale=0.12]{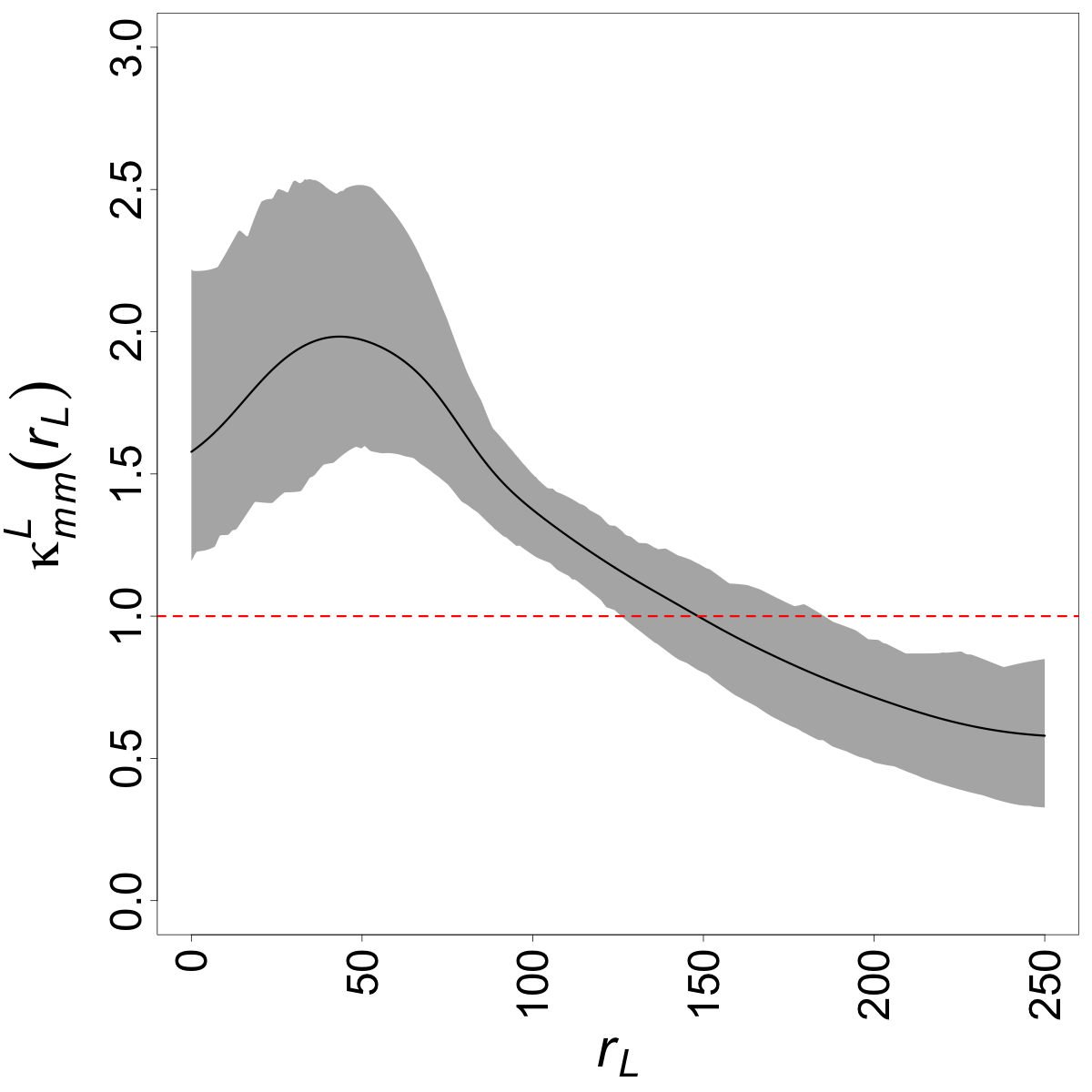}
    \includegraphics[scale=0.12]{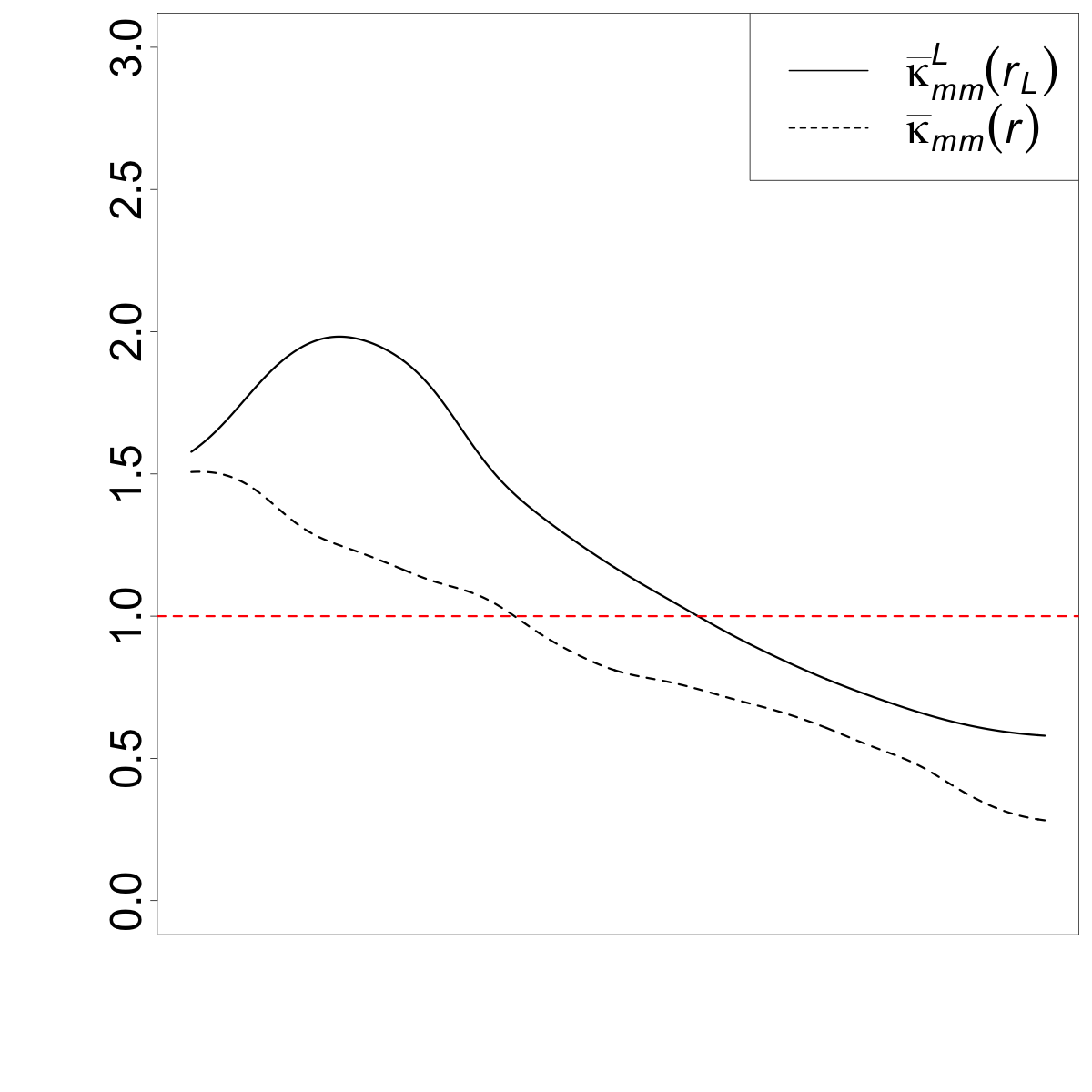}
    \caption{
    $95\%$ pointwise envelopes for  Stoyan's mark correlation functions based on 199 simulated patterns from Models I, II, and III. From left to right: $\kappa_{mm}(r)$, $\kappa^{\LL}_{mm}(\rL)$, and their average. From top to bottom: Model I, Model II, and Model III.
    }
    \label{fig:markcorrs}
\end{figure}

From each model, we simulate 199 patterns, and then, for each simulated point pattern, we calculate the Stoyan's mark correlation functions $\kappa_{mm}(r)$ and $\kappa^{\LL}_{mm}(\rL)$; we let $d_{\LL}$ be the shortest-path distance. We also let $r_{\LL}=1.25r$ \citep{rakshit2019fast}, leading to $\kappa_{mm}(r)$ and $\kappa^{\LL}_{mm}(\rL)$ being evaluated over distances $r \in [0,200]$ and $r_{\LL} \in [0,250]$, respectively. Based on the 199 estimated mark correlation functions, we obtain $95\%$ pointwise envelopes to compare the general behaviors of  $\kappa_{mm}(r)$ and $\kappa^{\LL}_{mm}(\rL)$ when the data are generated on a dendrite network. Figure \ref{fig:markcorrs} shows the estimated Stoyan's mark correlation functions for all models together with their $95\%$ pointwise envelopes and average behavior.
Looking at the top row of Figure \ref{fig:markcorrs}, showing the results for the model I, while the structure of the marks is not revealed by Stoyan's mark correlation function $\kappa_{mm}(r)$ (left panel), indicating mark independence, our proposed linear-network-based Stoyan's mark correlation functions $\kappa^{\LL}_{mm}(\rL)$ (central panel) clearly reveals a positive mark correlation for shortest path distances up to around 175 units. For larger distances, the mark correlation becomes less than one, which means that the average product of marks for points that are apart by a distance higher than 175 units is smaller than the squared mark mean value. To better compare the average behavior of $\kappa_{mm}(r)$ and $\kappa^{\LL}_{mm}(\rL)$, the averaged mark correlation functions are displayed in the right panel. 
Similarly, model II (middle row) and model III (bottom row) show apparent dissimilarities between the planar and the linear network mark correlation functions. 
Looking at the middle row and the presented results for model II, one can see a similar tendency of having a high mark correlation for smaller distances, which turns out to be lower than the correlation value under mark independence for larger distances. However, the clear difference between the two is that $\kappa_{mm}(r)$ turns from positive to negative quickly, around $r=100$, whereas  $\kappa^{\LL}_{mm}(\rL)$ maintain values higher than the expected value under mark independence up to $r_{\LL}=200$.
Note that, as the shortest path distances from the points at the border to the central point on the dendrite network increase in value, central points are surrounded mainly by higher mark values. In contrast, very small marks appeared only very close to the dendrite's border. Even though both $\kappa_{mm}(r)$ and $\kappa^{\LL}_{mm}(\rL)$ may show a positive correlation for smaller distances and a negative correlation for larger distances, their main difference is in the degree of positiveness/negativeness and further on the turning point from positive to negative.
This impression also holds for the obtained results for model III displayed in the bottom row. While the $\kappa^{\LL}_{mm}(\rL)$ clearly show the structure of the marks for small distances, increasing until $r_{\LL}=50$ and then decreasing, $\kappa_{mm}(r)$ entirely show a decreasing trend from small to large distances. Note also the difference between the turning points.

\begin{figure}[!h]
    \centering
    \includegraphics[scale=0.12]{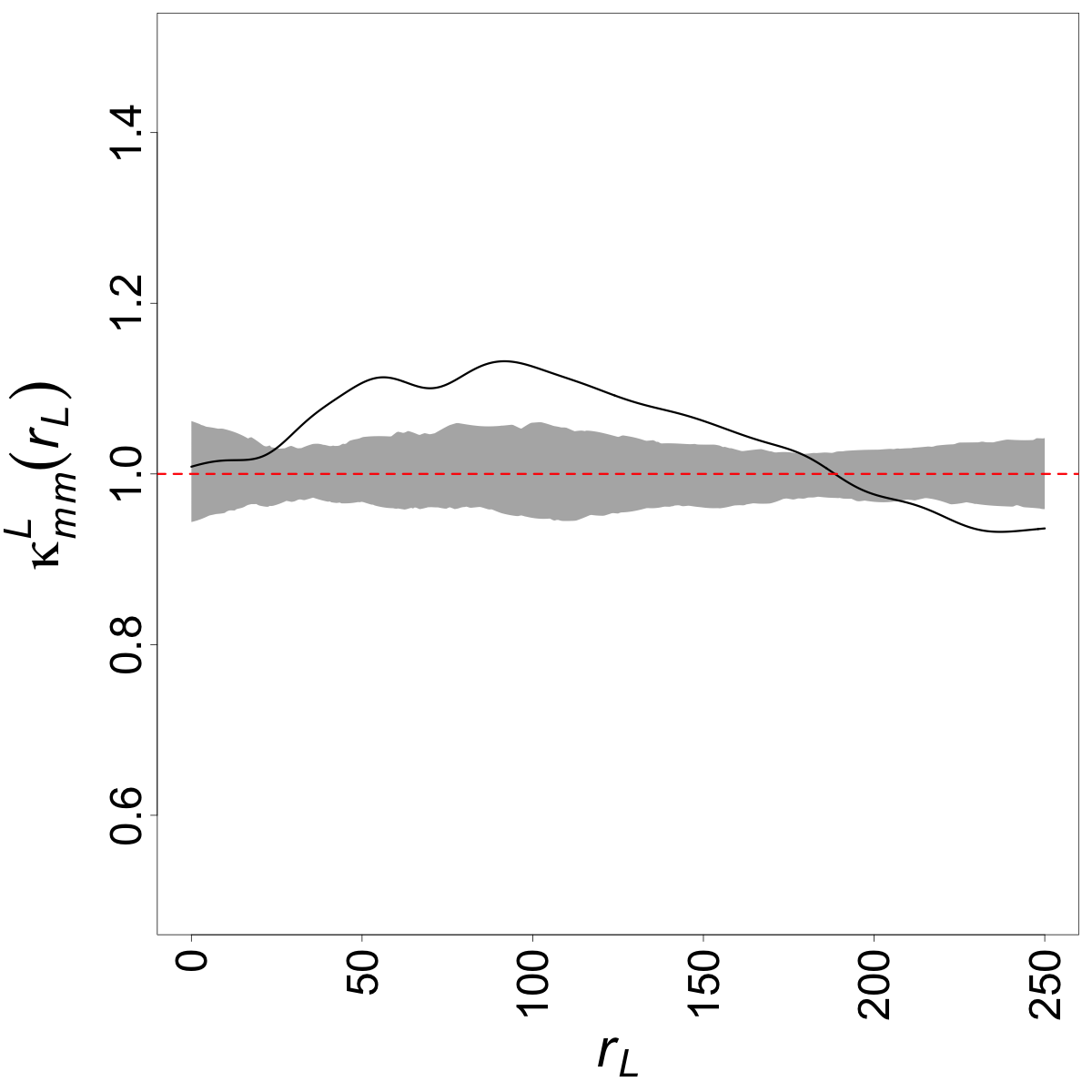}
    \includegraphics[scale=0.12]{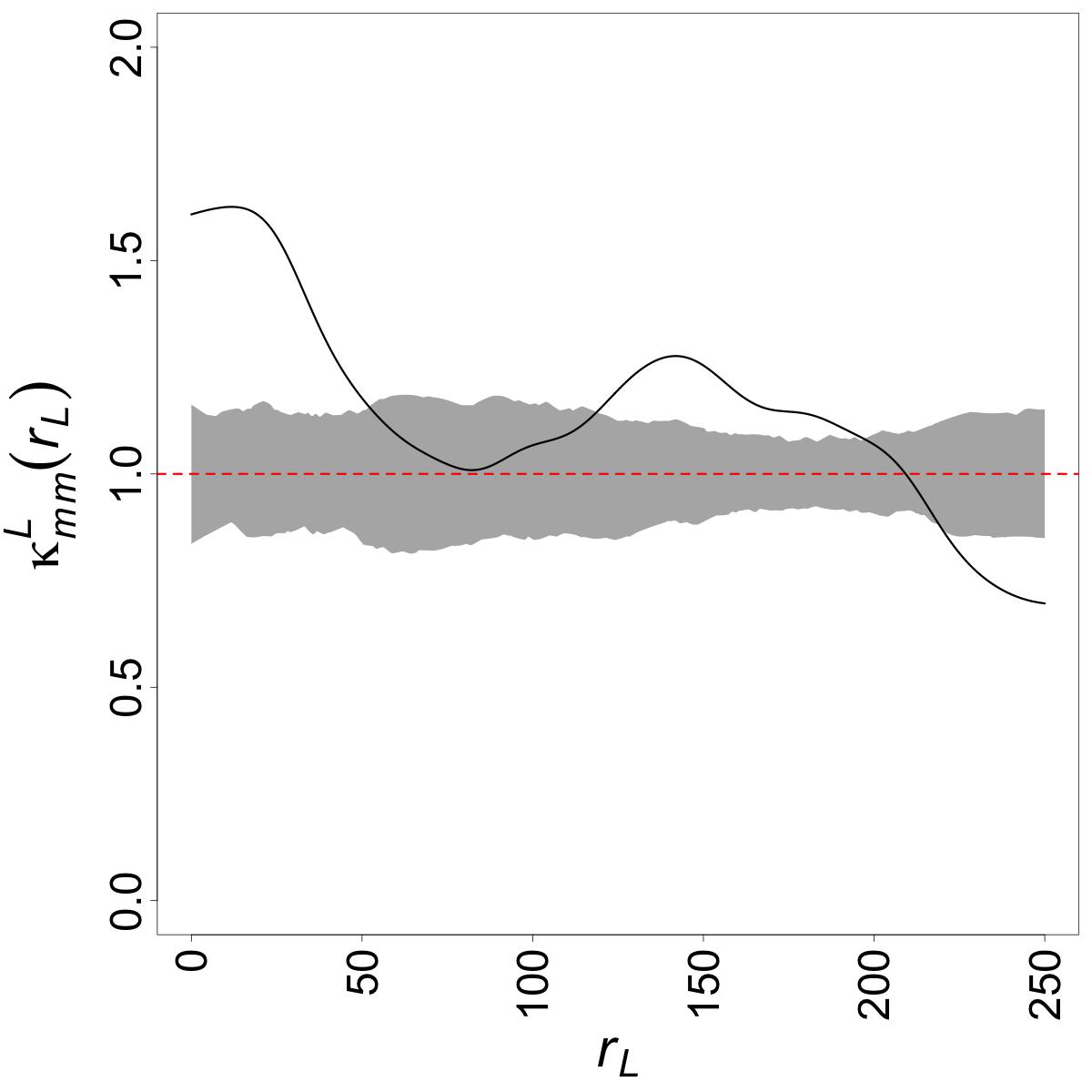}
    \includegraphics[scale=0.12]{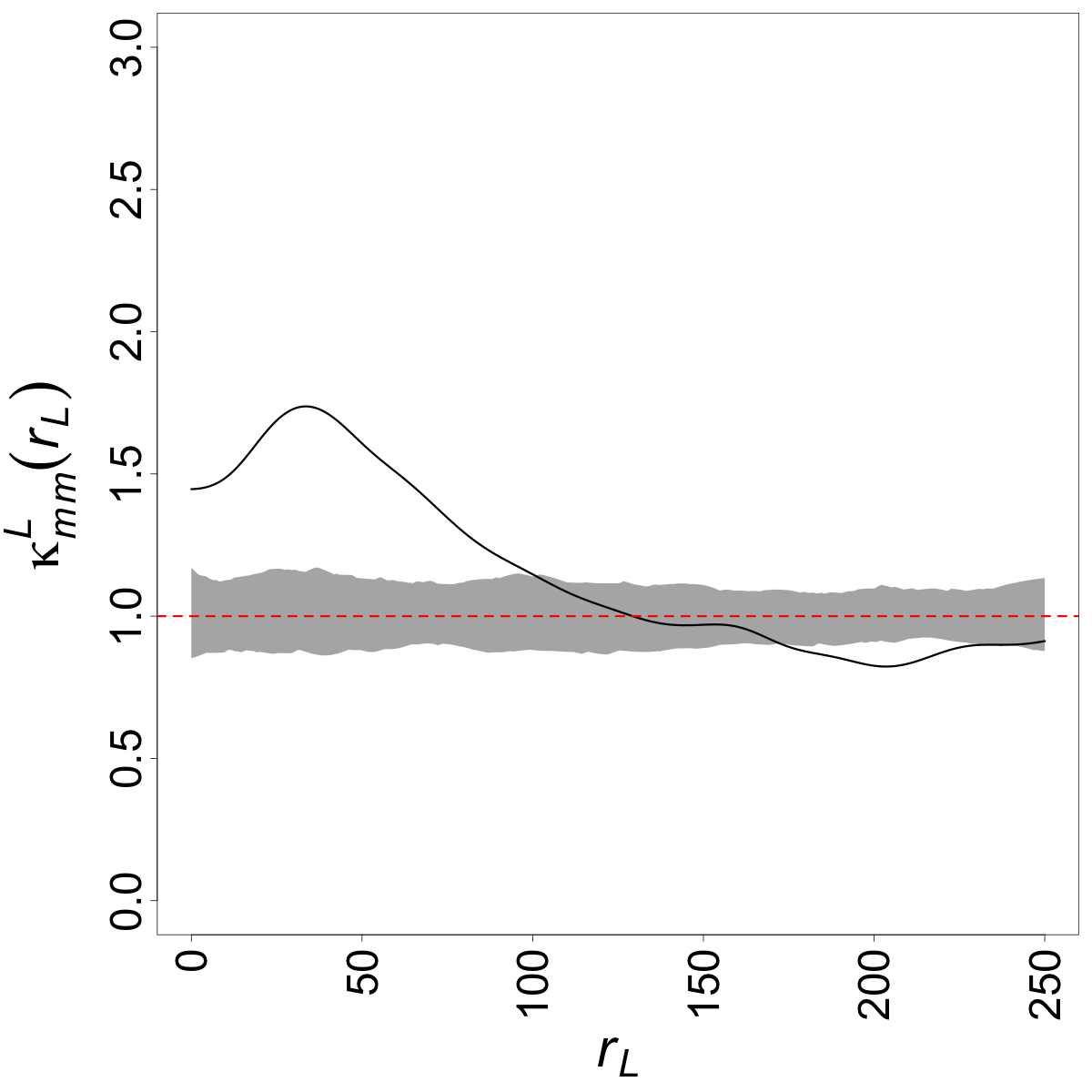}
    \includegraphics[scale=0.12]{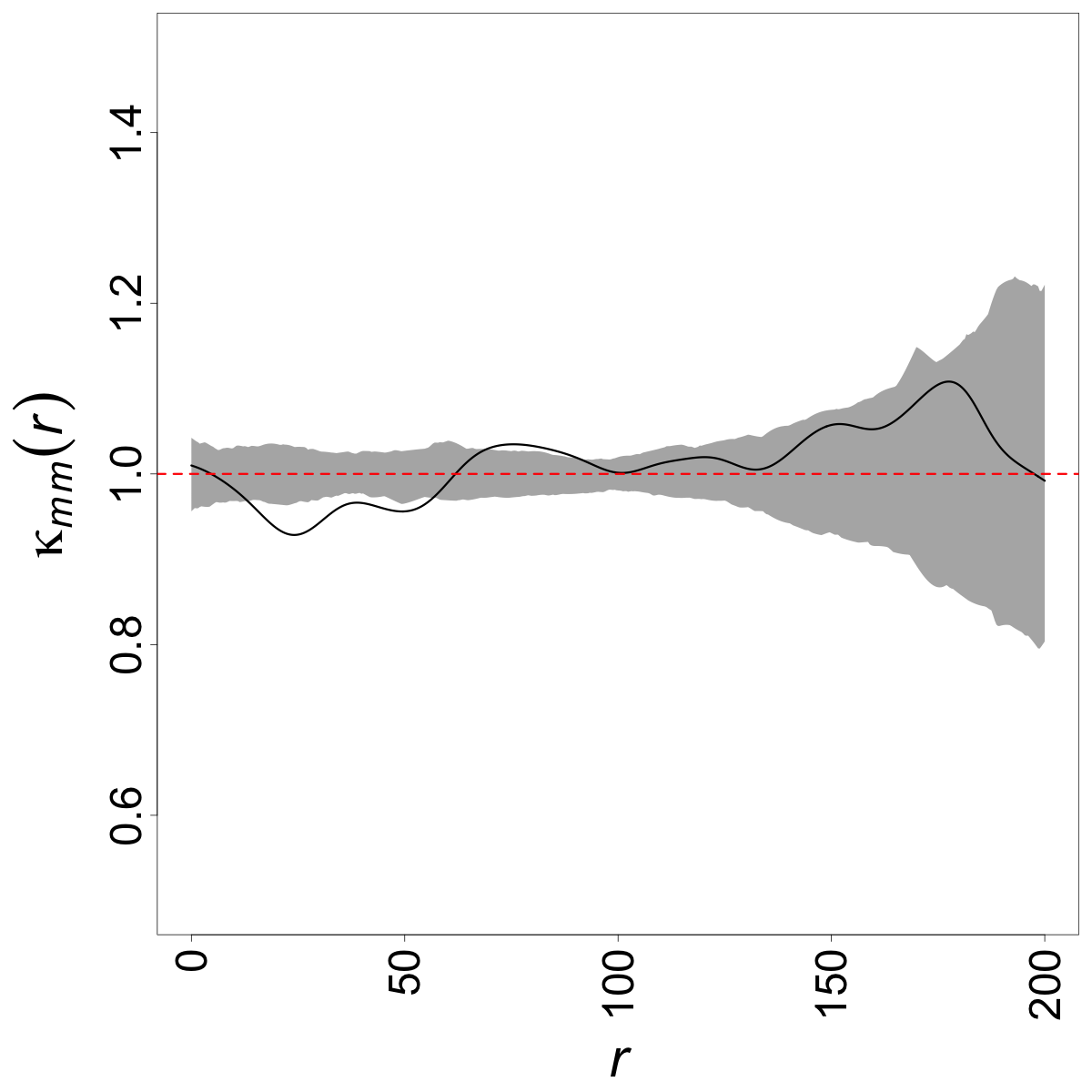}
    \includegraphics[scale=0.12]{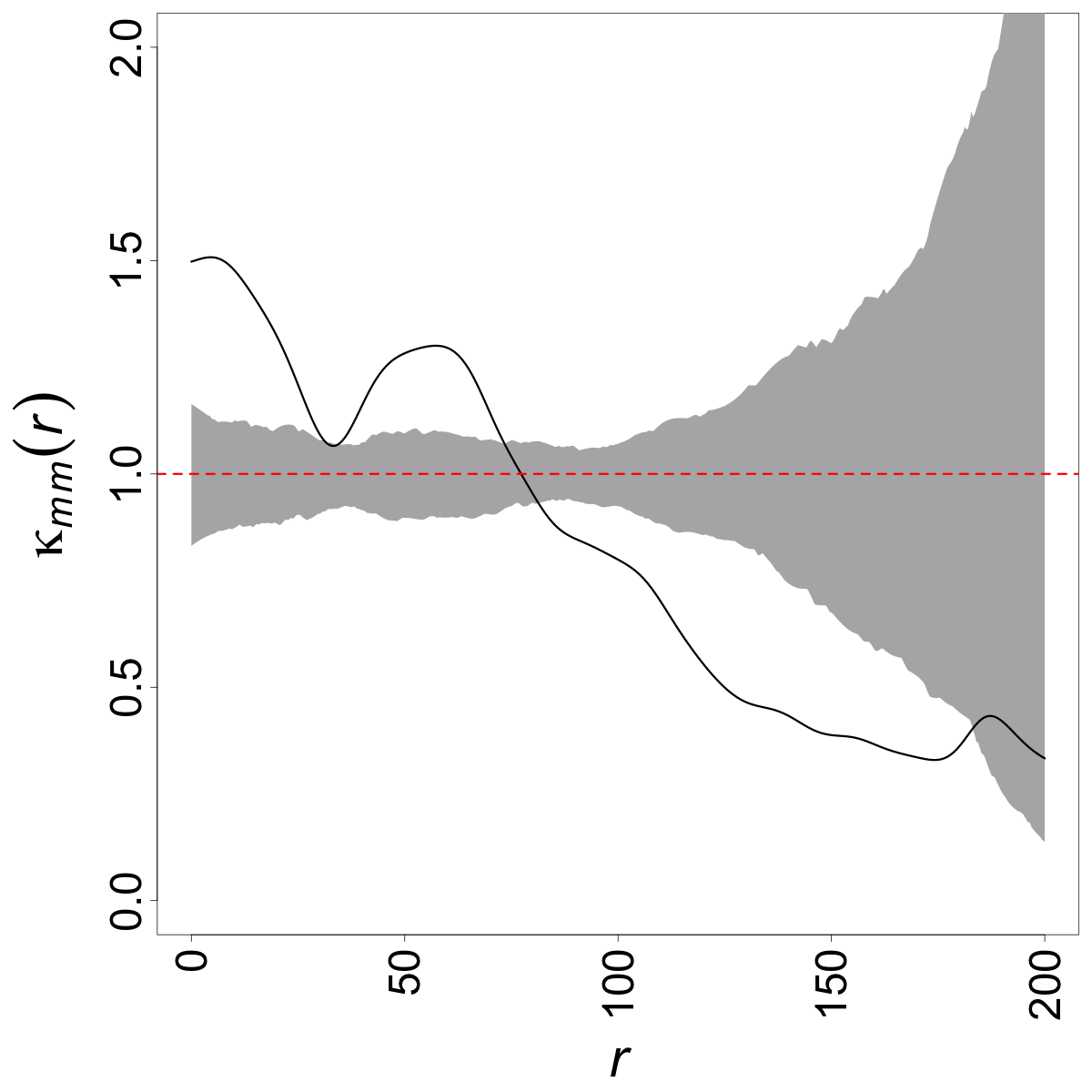}
    \includegraphics[scale=0.12]{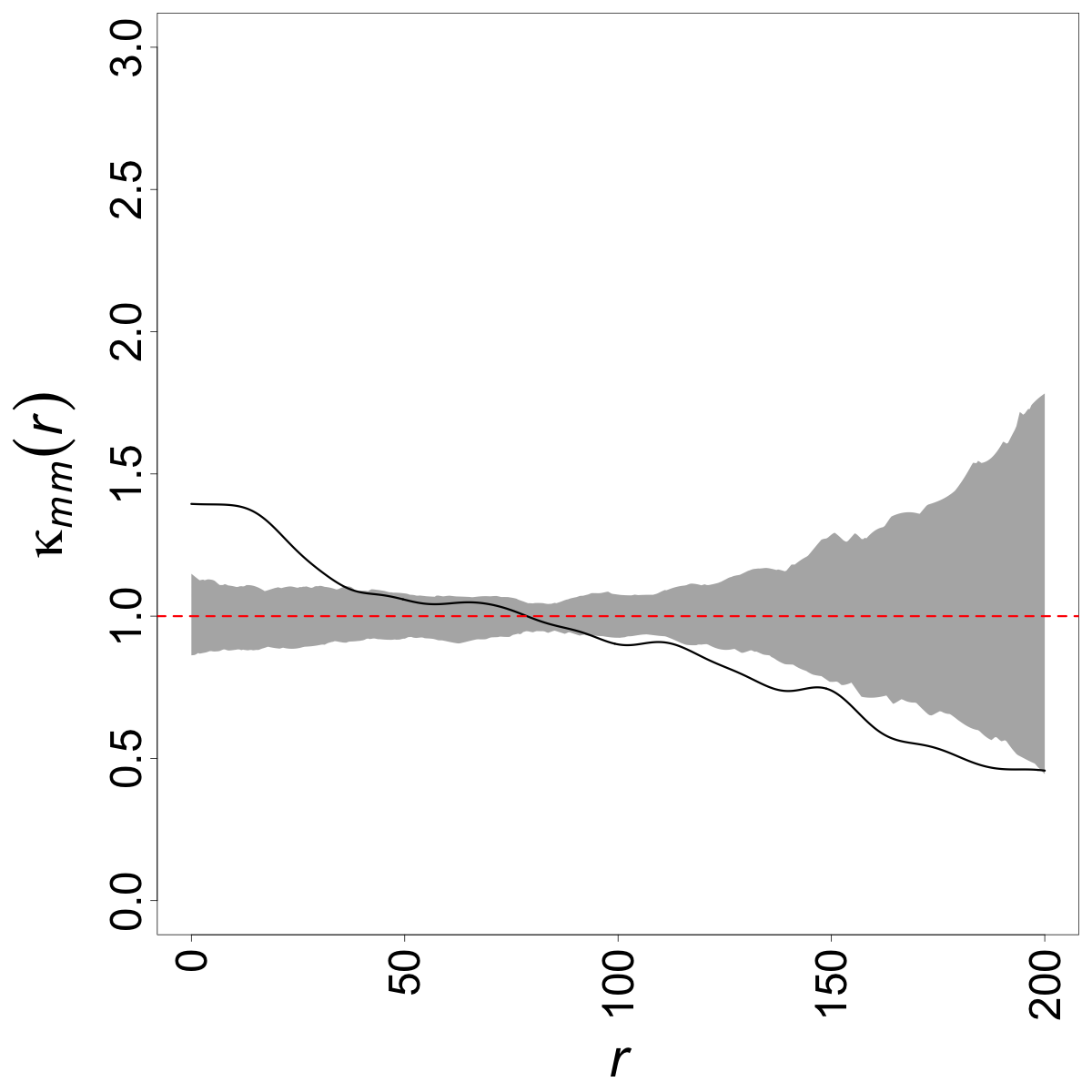}
    \caption{
    $95\%$ envelopes for Stoyan's mark correlation functions, under mark independence, based on 199 simulated patterns. The solid curves show Stoyan's mark correlation functions for one point pattern per each model I, II, and III (left to right).
    }
    \label{fig:markcorrtests}
\end{figure}

Upon further comparison between $\kappa_{mm}(r)$ and $\kappa^{\LL}_{mm}(\rL)$, we proceed by selecting a single realization from each model. We then compare the Stoyan's mark correlation function of these realizations to a $95\%$ pointwise envelope under mark independence. This envelope is constructed using 199 point patterns with randomly allocated marks derived from the original pattern. Looking at Figure \ref{fig:markcorrtests}, a clear difference between the envelopes for $\kappa_{mm}(r)$ and $\kappa^{\LL}_{mm}(\rL)$ is that $\kappa_{mm}(r)$ shows a quite higher variation for larger distances compared to $\kappa^{\LL}_{mm}(\rL)$, especially for model II displayed in the middle column. Another significant distinction is the overall fluctuation pattern of $\kappa^{\LL}_{mm}(\rL)$ compared to $\kappa_{mm}(r)$, particularly with respect to their containment within the envelopes, indicating different conclusions. By joining the findings from both Figures \ref{fig:markcorrs} and \ref{fig:markcorrtests}, it becomes evident that disregarding the impact of underlying linear networks could potentially lead to inaccurate conclusions. 

\section{Real data analysis}\label{sec:realdata}

We now study the mark structure for two real data presented in Section \ref{sec:Data}. For the case of butterflies in Melbourne, Australia, we further let the data points be labeled by the intensities so that we can make use of both cross-type and mark correlation functions in our analysis. Intensities, separately for each type, are estimated via the non-parametric Jones-Diggle kernel-based estimators \citep{jones1993simple} employing Scott’s rule of thumb for bandwidth selection \citep{scott2015multivariate}. In the case of public street trees in Vancouver, Canada, we only make use of our proposed version of Stoyan's mark correlation function for point processes on linear networks and make an additional comparison with its counterpart for point processes in $\R^2$ in the case of real datasets. We point out that both real datasets exhibit inhomogeneity, which is expected when dealing with real phenomena. However, we believe mark correlation functions can serve as good indicators for discovering interactions between marks, although they are essentially developed for stationary and homogeneous point processes.

\subsection{Butterflies in Melbourne, Australia}

\begin{figure}[!h]
    \centering
    \includegraphics[scale=0.09]{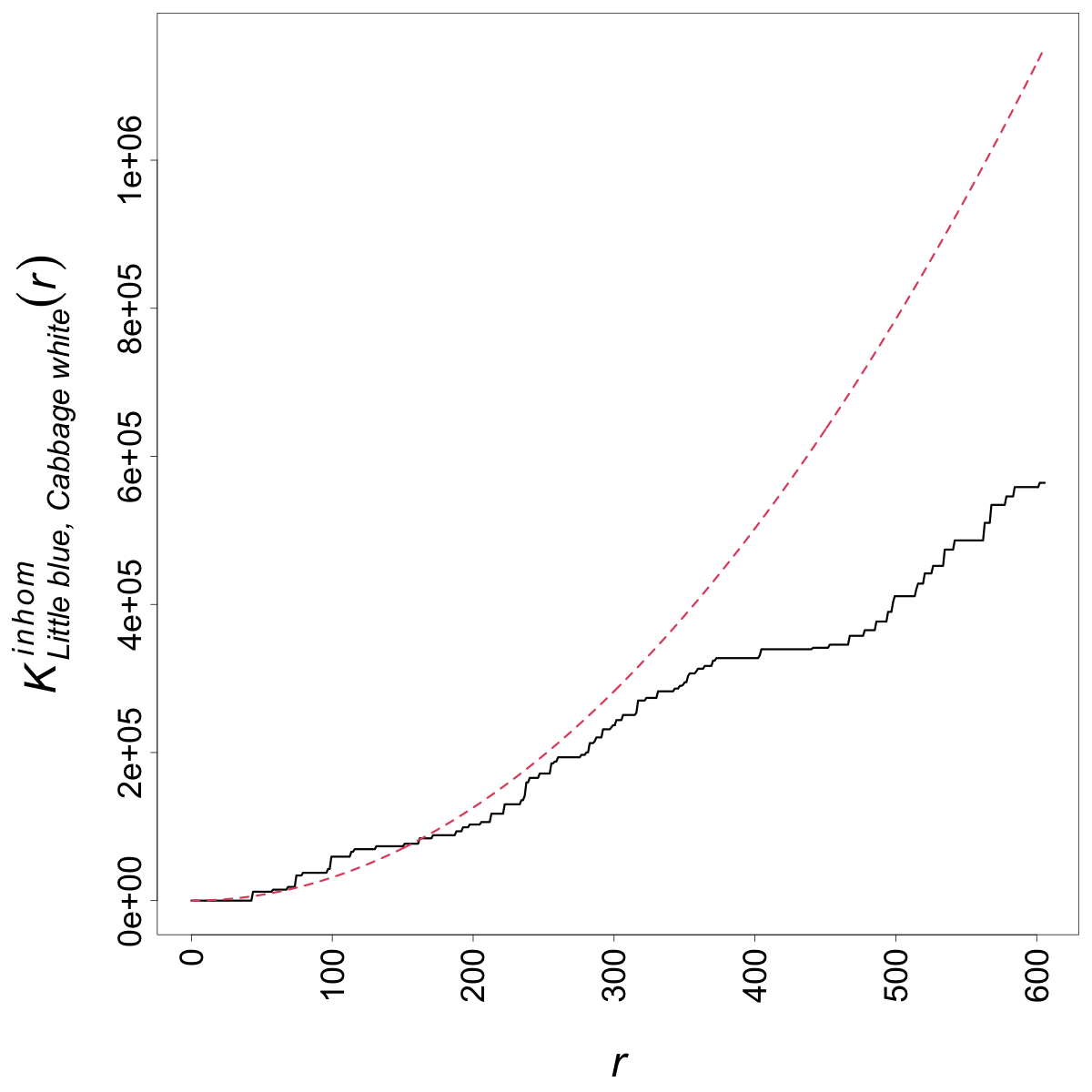}
    \includegraphics[scale=0.09]{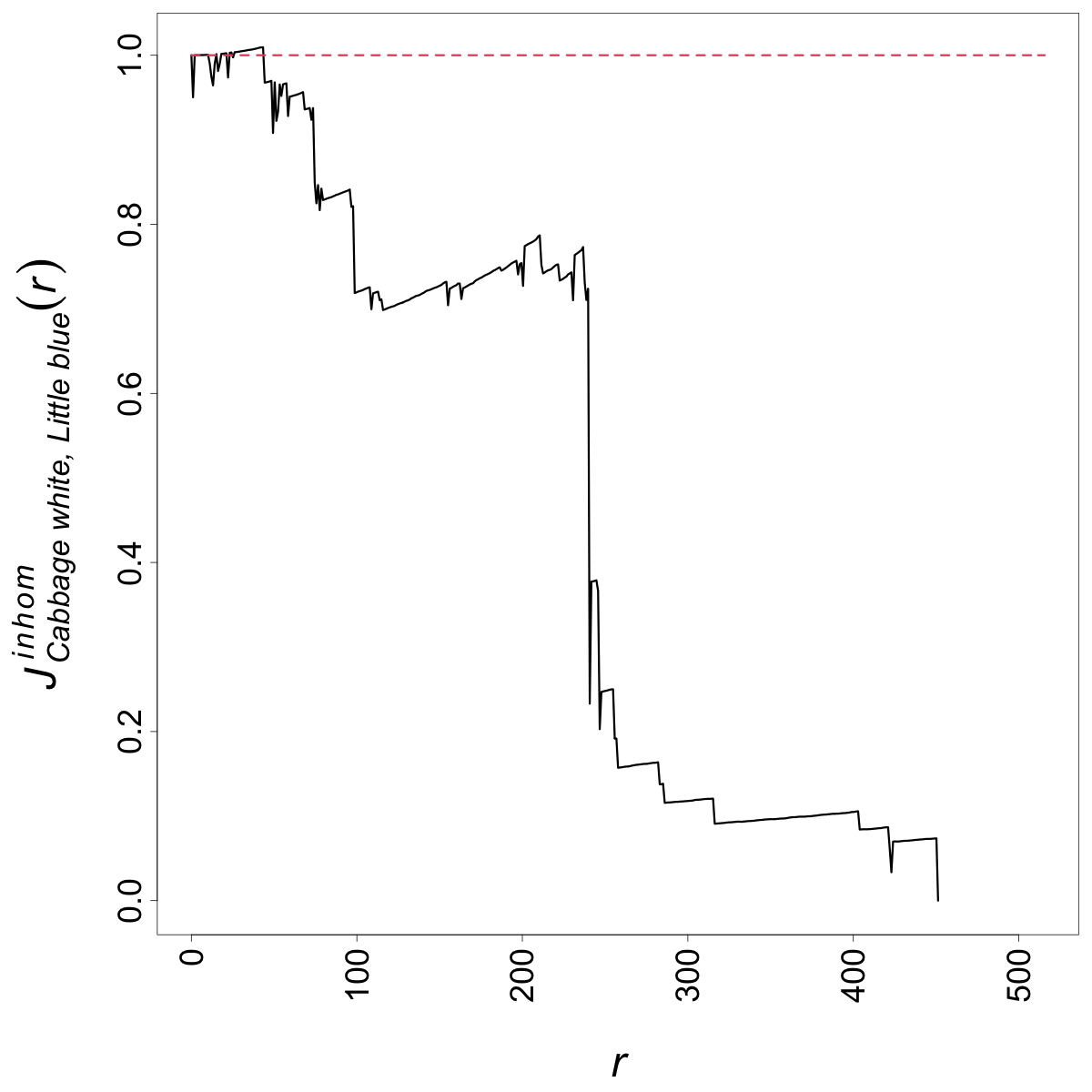}
    \includegraphics[scale=0.09]{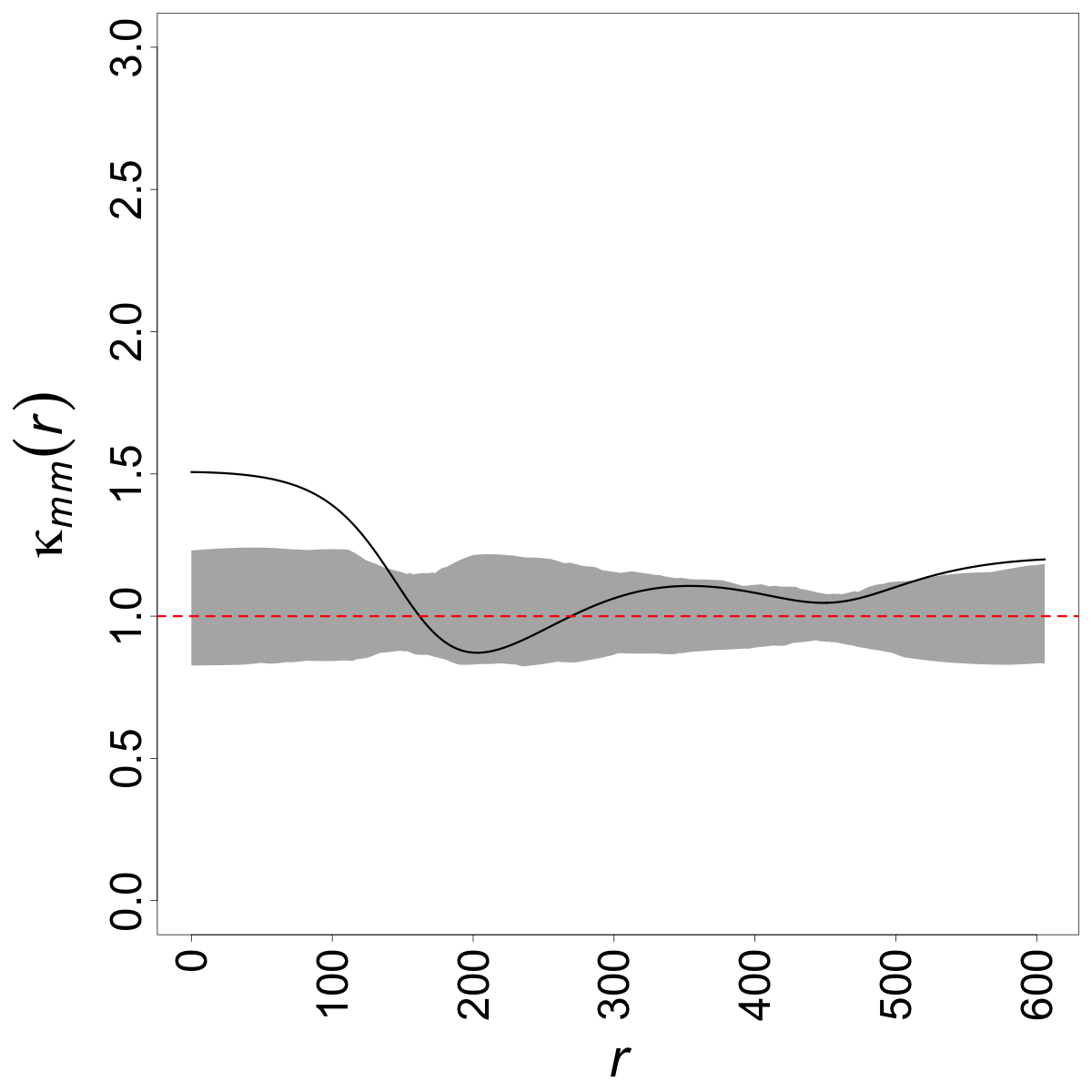}
    \includegraphics[scale=0.09]{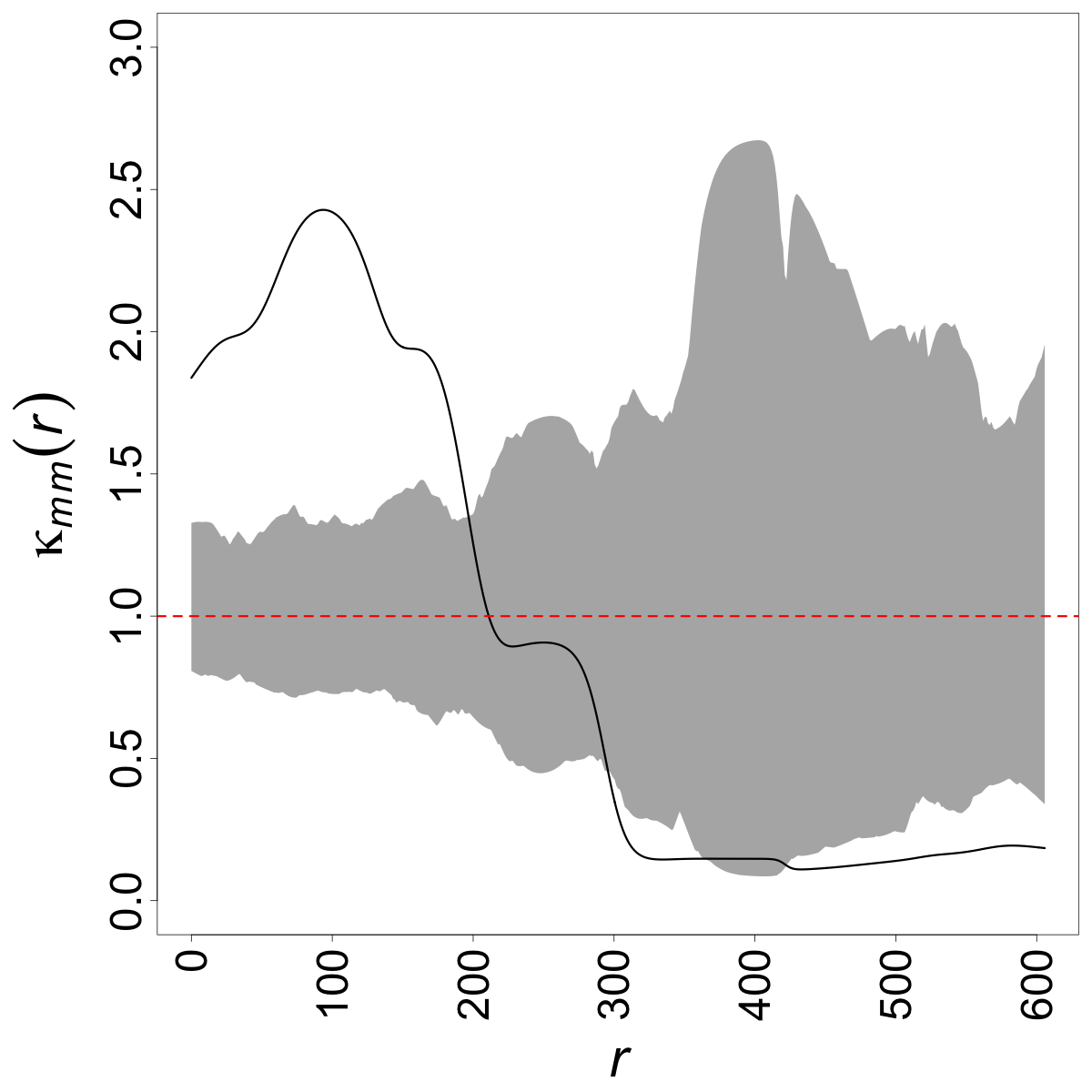}
    \caption{
    Mark summary characteristics for butterfly data in Melbourne, Australia. From left to right, cross-type inhomogeneous $K$-function, cross-type inhomogeneous $J$-function, Stoyan's mark correlation function for the Cabbage White type, and Stoyan's mark correlation function for the Little Blue type.
    }
    \label{fig:butterflyresults}
\end{figure}

Figure \ref{fig:butterflyresults} shows the cross-type summary characteristics and Stoyan's mark correlation function for butterflies in Melbourne, Australia. In the cross-type inhomogeneous $K$-function case, we only use its estimator with an isotropic edge correction. For small distances ($r \leq 200$), the cross-type inhomogeneous $K$-function moves around its theoretical value for Poisson processes, meaning that the butterflies of type Cabbage White are randomly distributed around butterflies of type Little Blue. However, for larger distances, the cross-type inhomogeneous $K$-function lies below its theoretical value for Poisson processes, indicating that, at those distances, the number of butterflies of type Cabbage White around butterflies of type Little Blue is smaller than that of complete spatial randomness. A similar message is conveyed by the cross-type inhomogeneous $J$-function.

Turning to Stoyan's mark correlation function $\kappa_{mm}(r)$, we first obtain 199 point patterns for each type of butterfly by randomly allocating marks. Then, we compare Stoyan's mark correlation function $\kappa_{mm}(r)$ for each type of butterfly with $95\%$ pointwise envelopes obtained from the 199 patterns under mark independence. We can see that, for distances smaller than $r=150$, $\kappa_{mm}(r)$ for butterflies of type Cabbage White stays above the envelope, indicating that the average product of densities is higher than the expected average under mark independence; for interpoint distances smaller than $r=150$ points have higher intensities. For large distances, it generally stays within the envelope. In the case of butterflies of type Little Blue, $\kappa_{mm}(r)$ similarly remains above the envelope for small distances. The obtained envelope, however, shows a higher variation, compared to that of Cabbage White, which might be due to the fact that the number of butterflies of type Little Blue is relatively lower than those of type Cabbage White.

\subsection{Public street trees in Vancouver, Canada}

\begin{figure}[!h]
    \centering
    \includegraphics[scale=0.07]{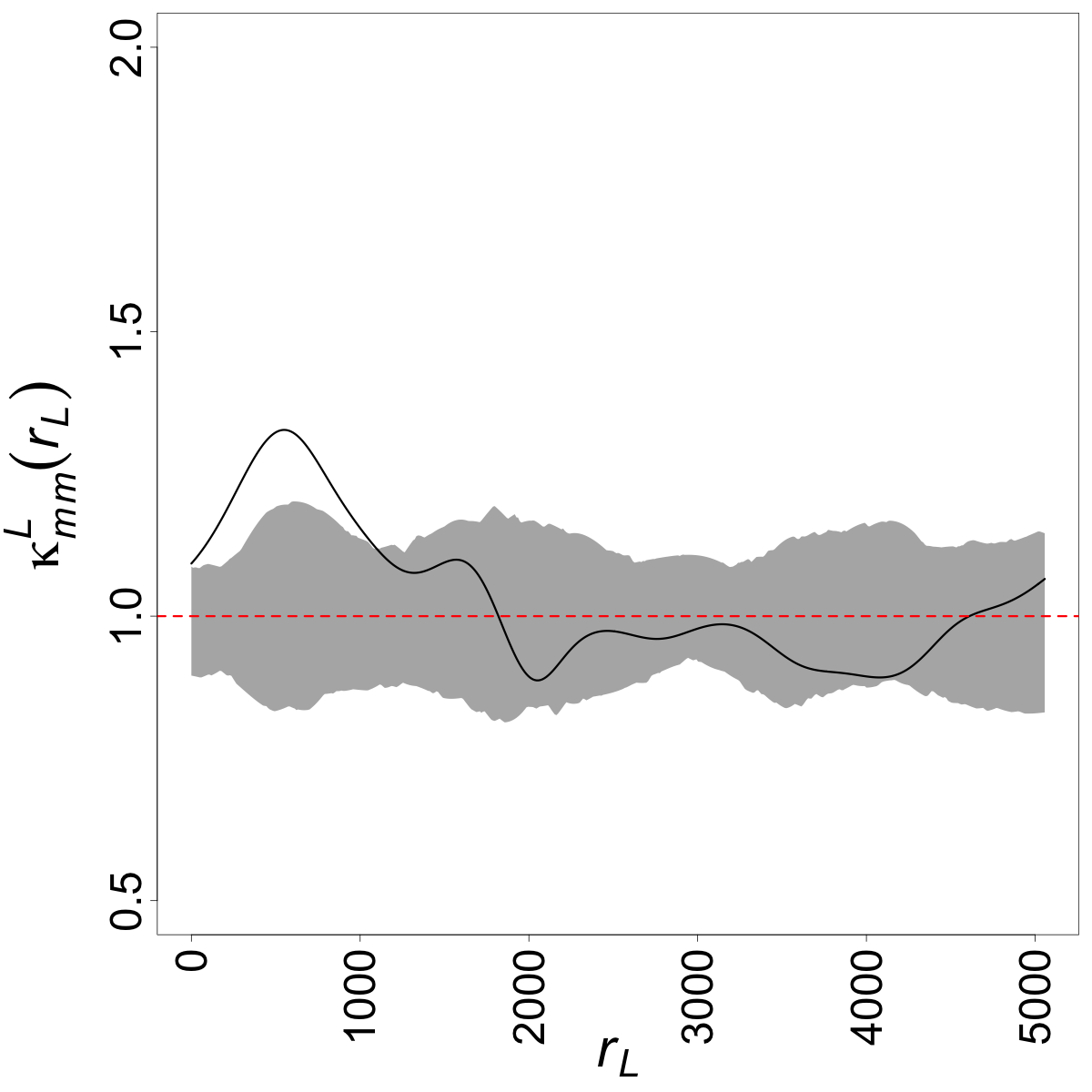}
    \includegraphics[scale=0.07]{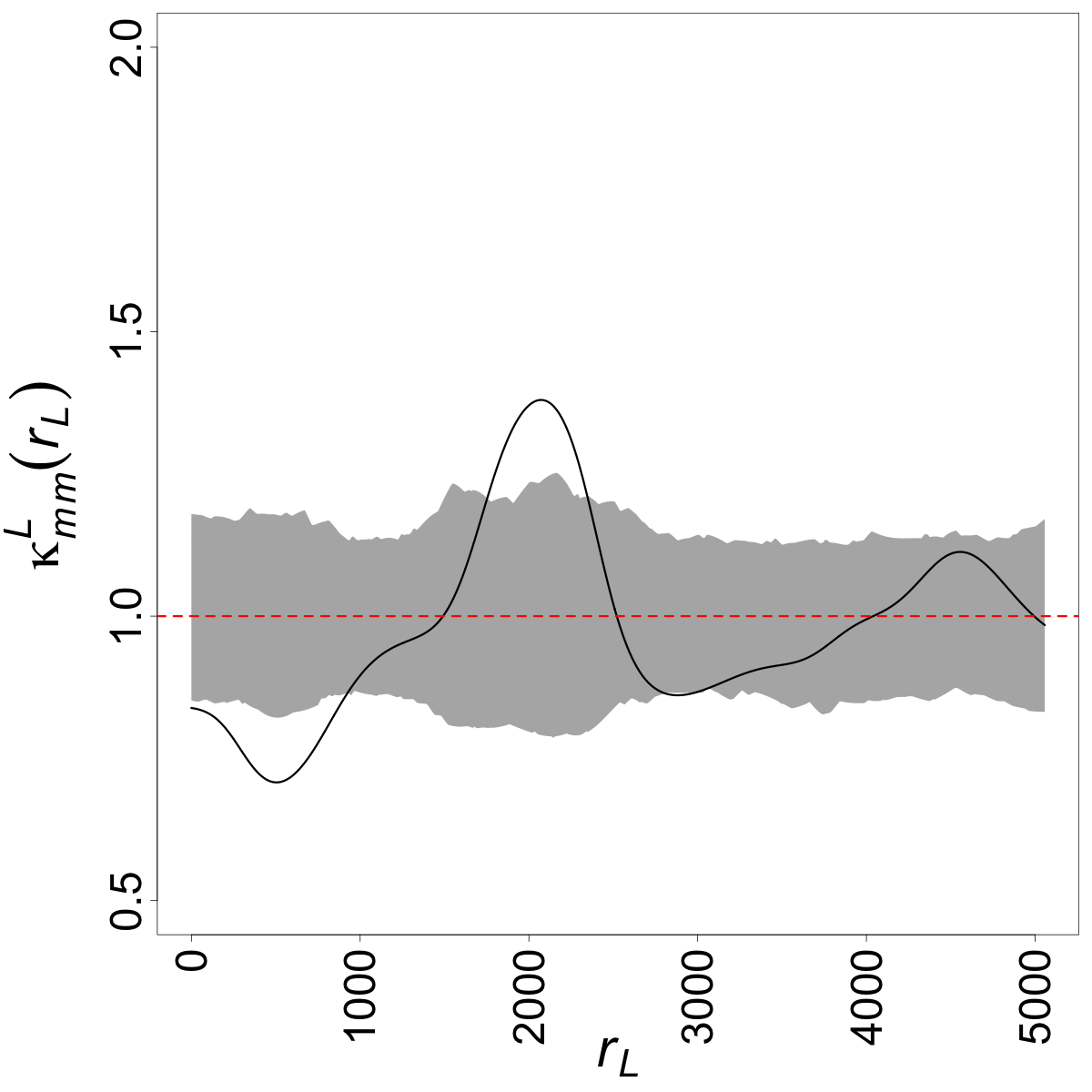}
    \includegraphics[scale=0.07]{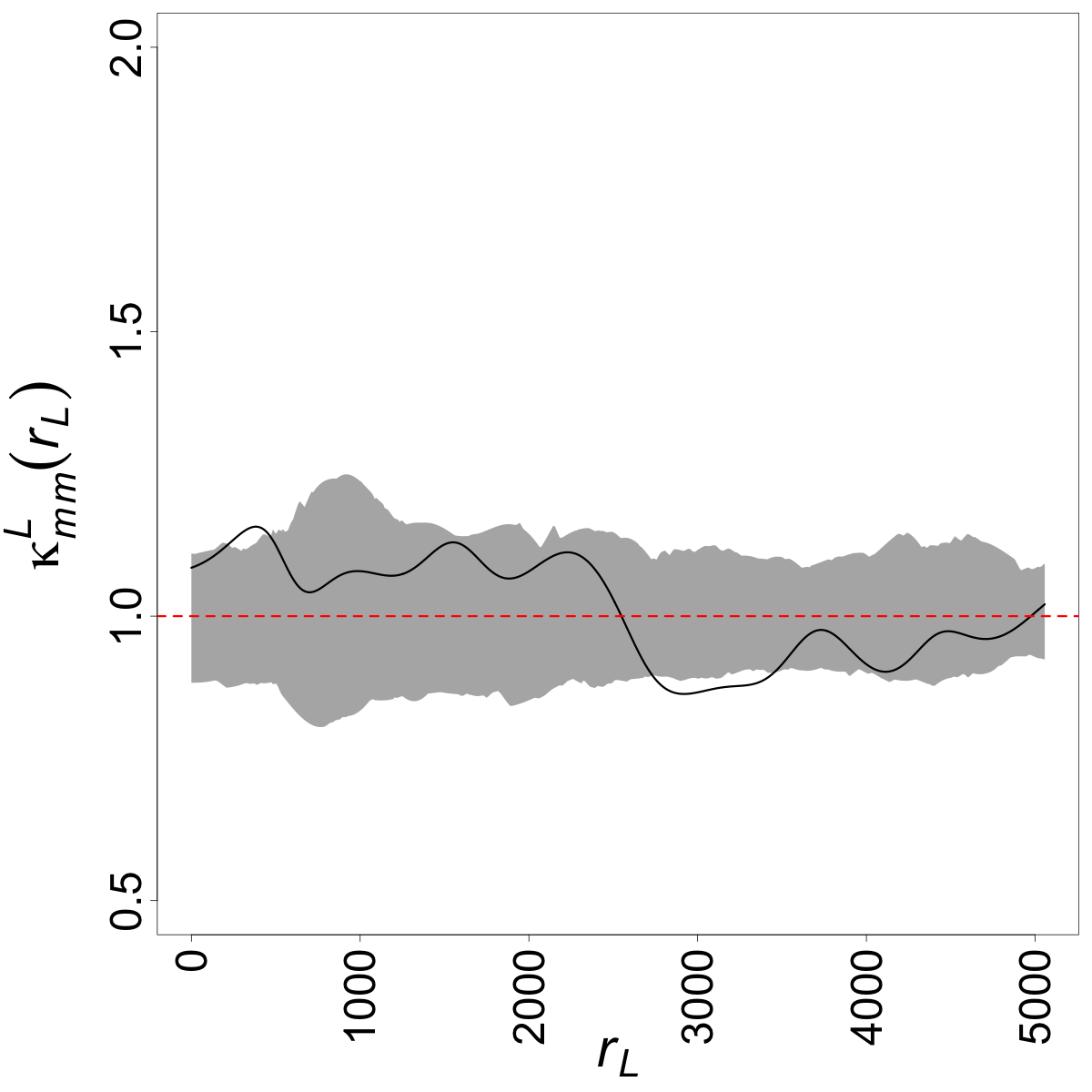}
    \includegraphics[scale=0.07]{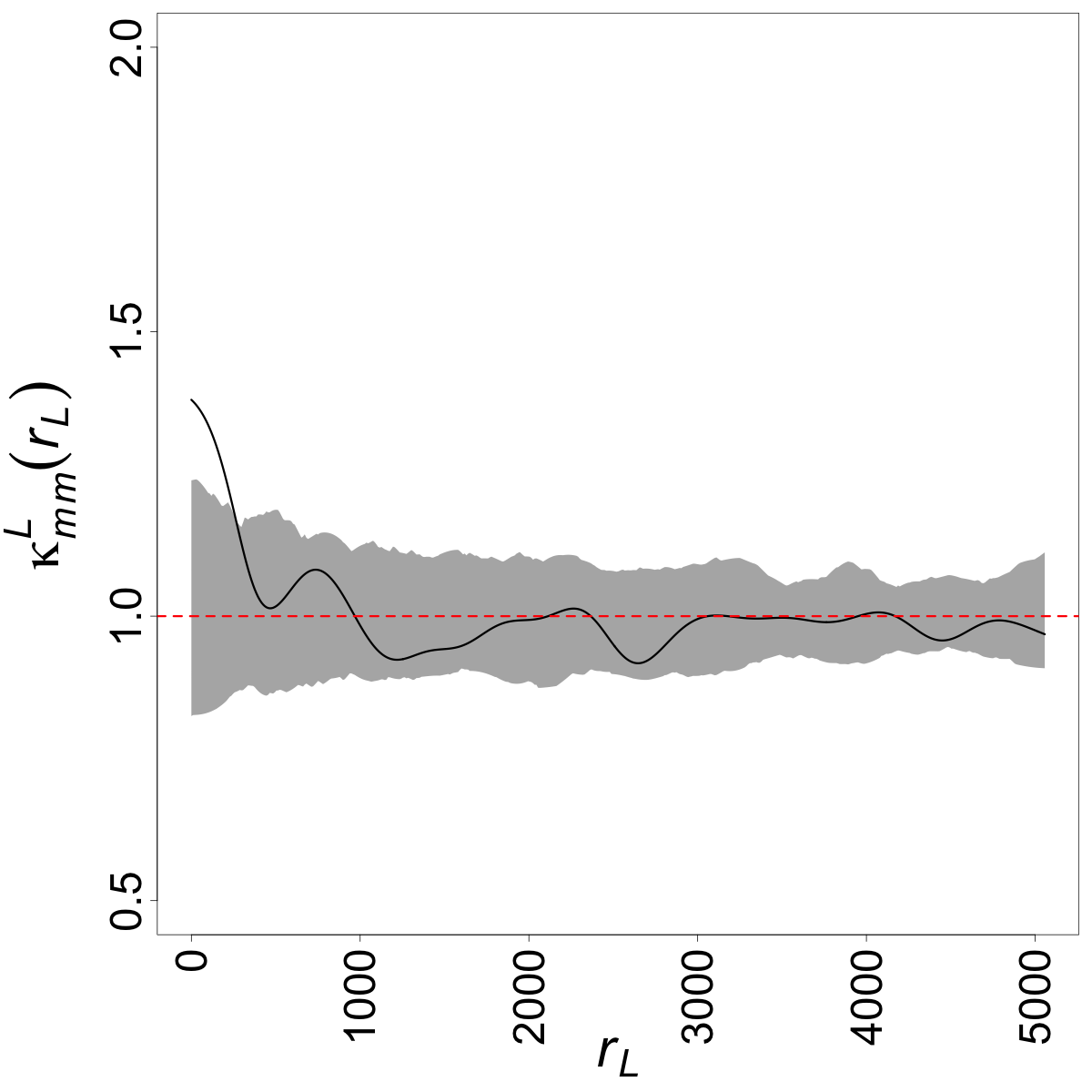}
    \includegraphics[scale=0.07]{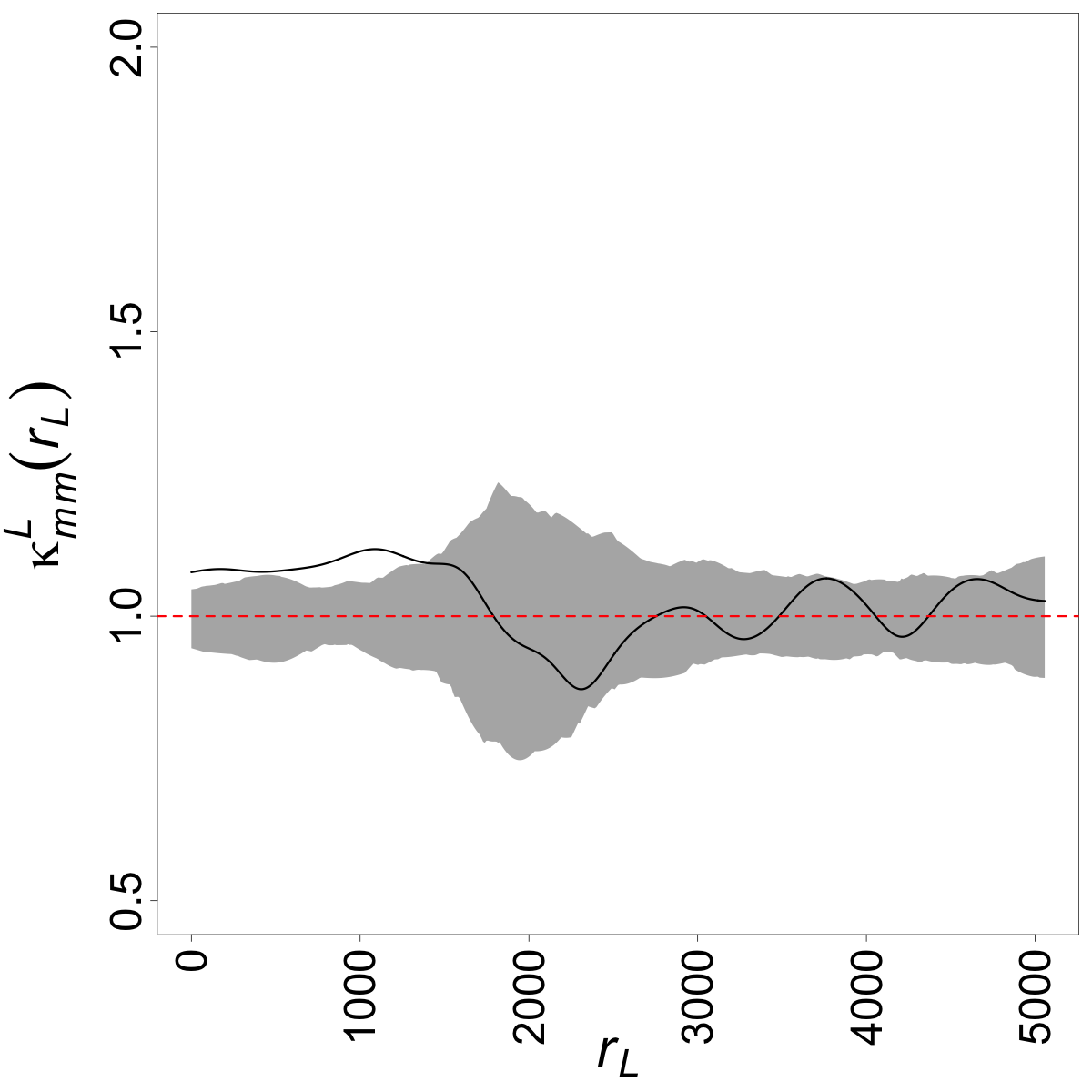}
    \includegraphics[scale=0.07]{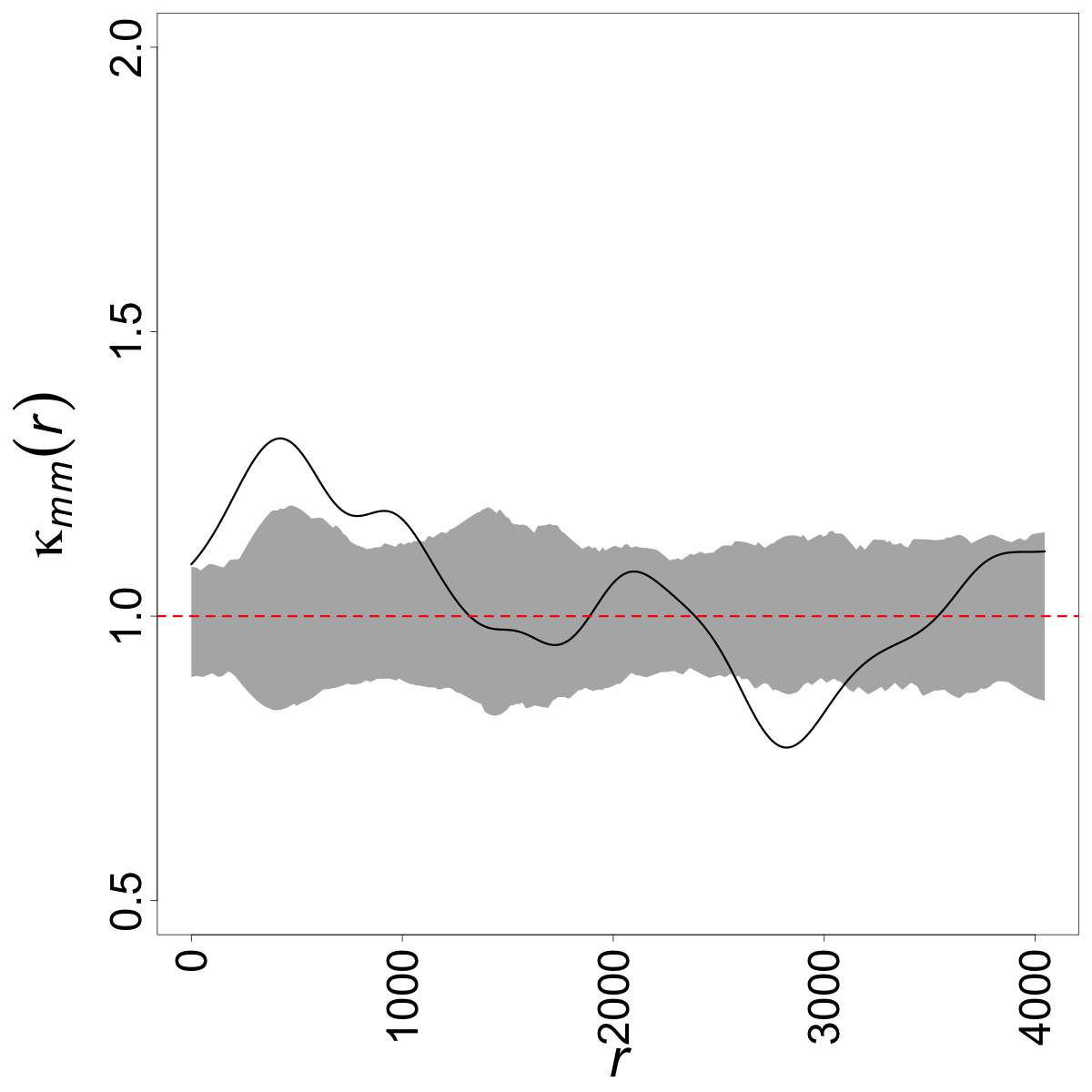}
    \includegraphics[scale=0.07]{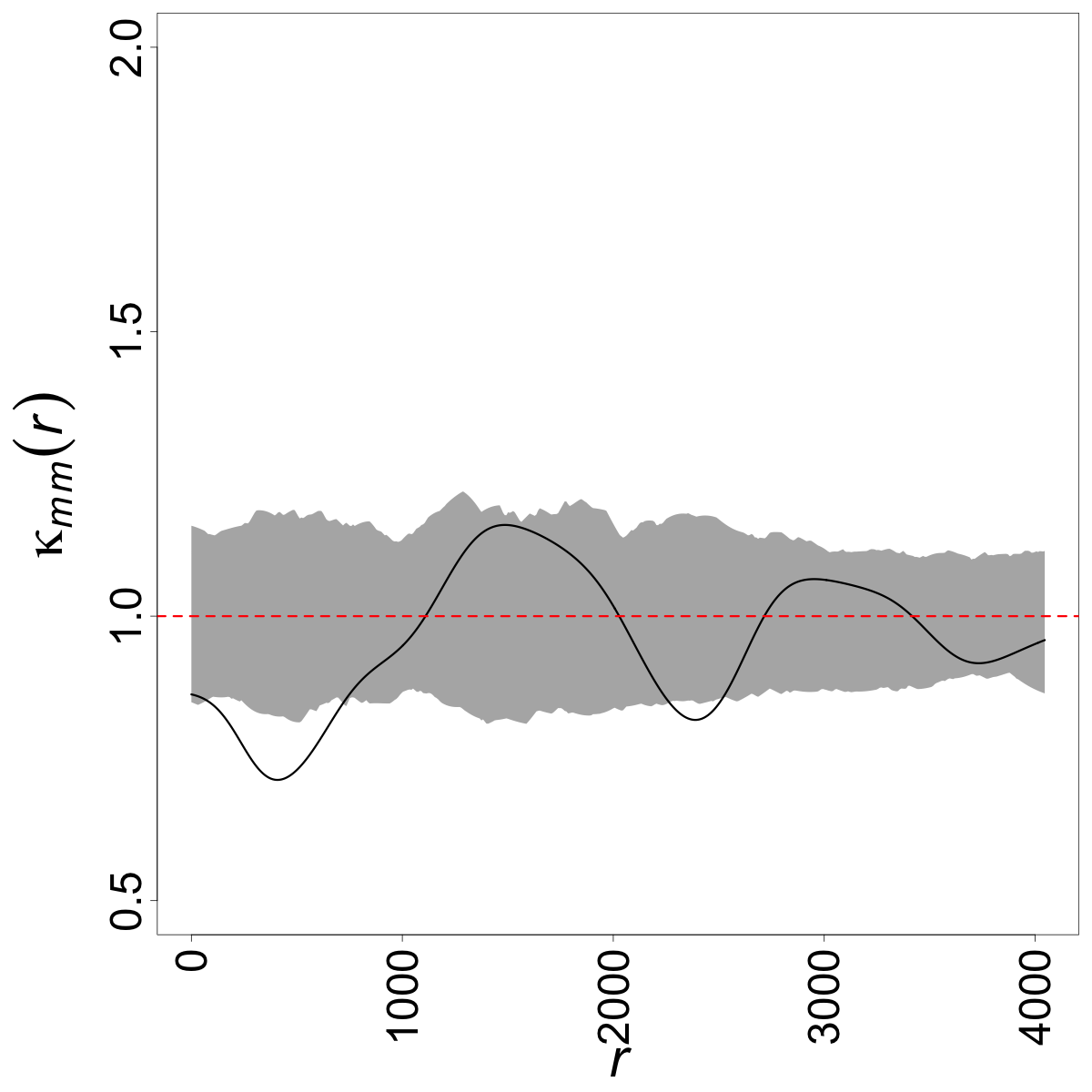}
    \includegraphics[scale=0.07]{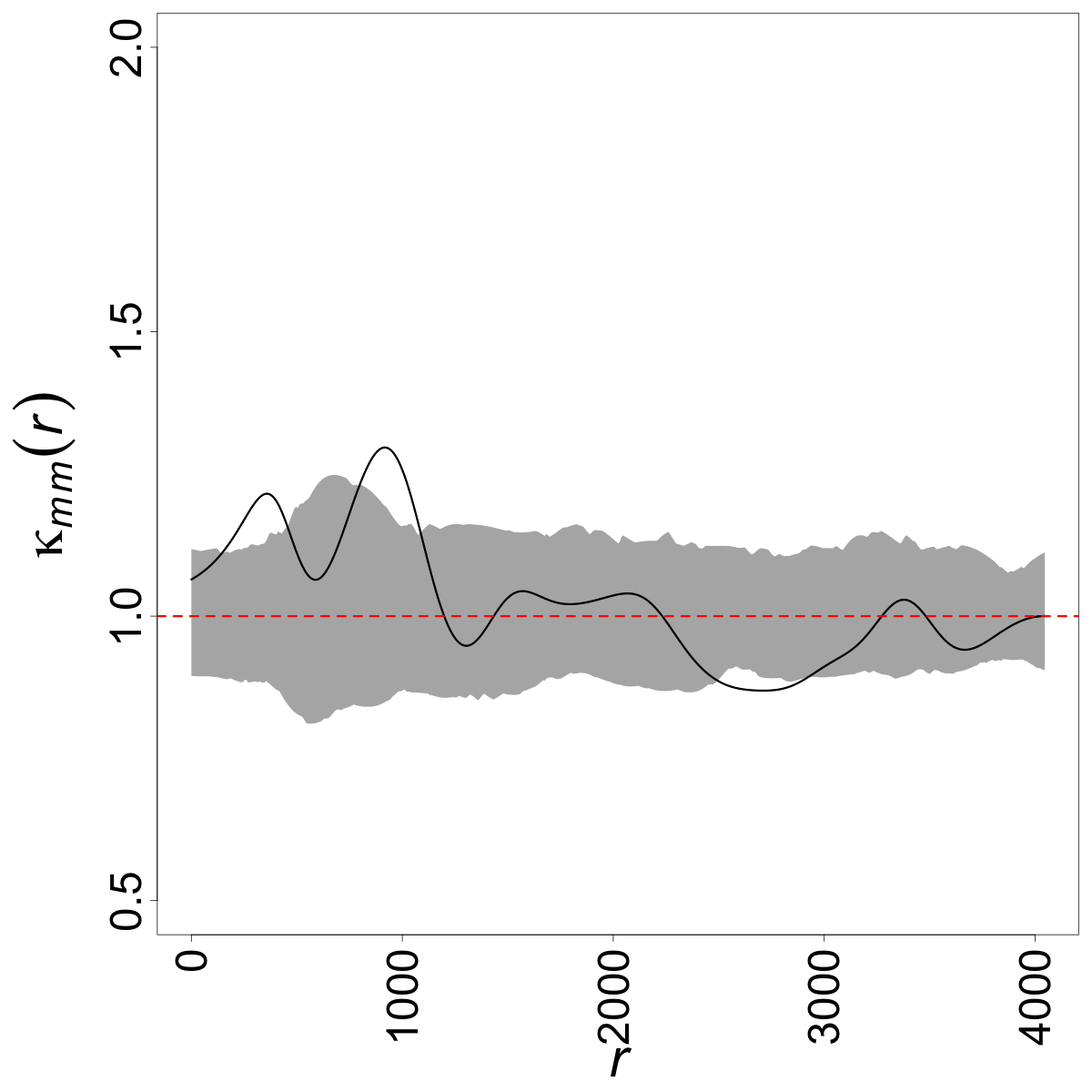}
    \includegraphics[scale=0.07]{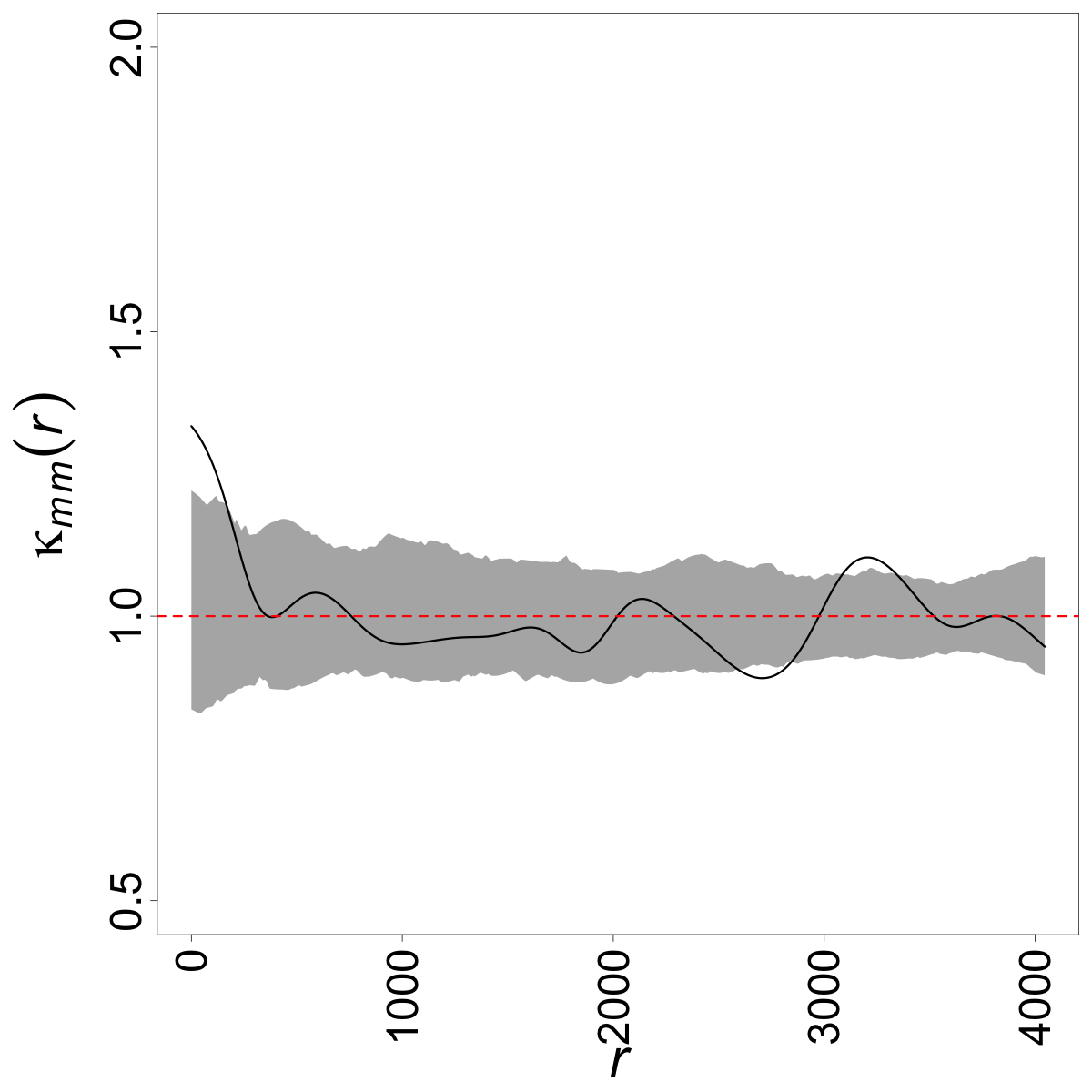}
    \includegraphics[scale=0.07]{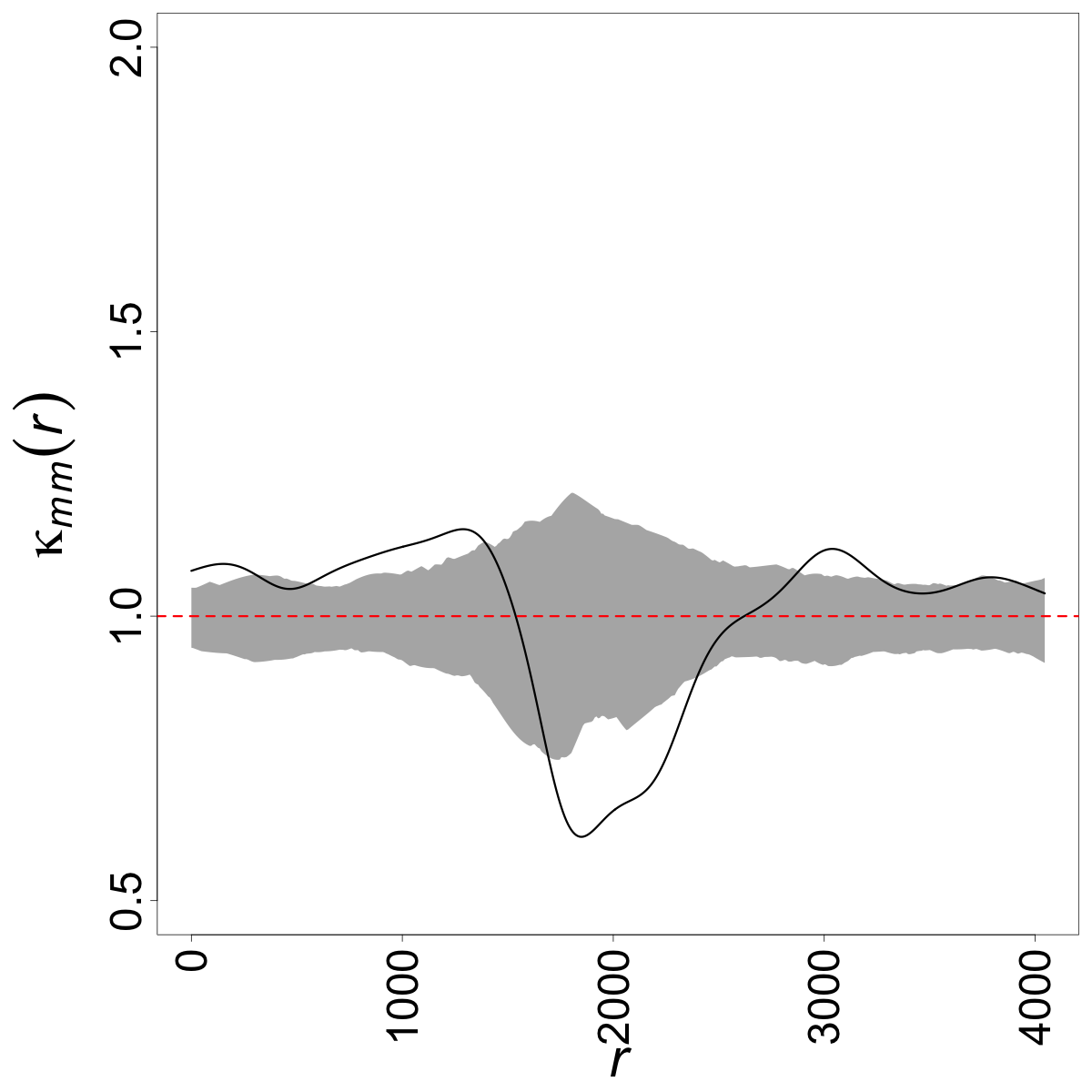}

    \caption{
    Stoyan's mark correlation function for the five species of trees alongside the street network of  Vancouver, Canada; the real-valued mark is dbh. From left to right: Arnold, Populus, Involucrata, Aquifolium, Bignonioides. 
    The top row considers the network; the bottom row ignores it. 
    }
    \label{fig:Vantrees}
\end{figure}

Now we employ Stoyan's mark correlation functions $\kappa_{mm}(r)$ and $\kappa^{\LL}_{mm}(\rL)$ to investigate the distribution of diameter at breast height (dbh) for the five species of public trees planted along the street network of Vancouver, Canada. Our aim is to uncover potential interactions among dbh values for trees of the same species. Similarly, we rely on $95\%$ pointwise envelopes obtained from 199 point patterns where spatial locations are the same as the original data, but dbh values are randomly allocated.
Looking at Figure \ref{fig:Vantrees}, one can evidently see that $\kappa_{mm}(r)$ and $\kappa^{\LL}_{mm}(\rL)$ give rise to different conclusions, highlighting the importance of taking the underlying network into account. In the case of trees of type Arnold, both $\kappa_{mm}(r)$ and $\kappa^{\LL}_{mm}(\rL)$ stay above the envelope but with different degrees and turning points, highlighting that for a pair of nearby trees, at least one of them has a large dbh. Interestingly, for distances around $r=3$km, $\kappa_{mm}(r)$ stays below the corresponding envelope, while this is not the case for $\kappa^{\LL}_{mm}(\rL)$, pointing to different conclusions. Similar conclusions hold for trees of type Populus; on the one hand $\kappa^{\LL}_{mm}(\rL)$ stays above the corresponding envelope for distances around $\rL=2$km while this is not the case for $\kappa_{mm}(r)$, on the other hand, $\kappa_{mm}(r)$ stays below the envelope for distances $r=2.5$km while $\kappa^{\LL}_{mm}(\rL)$ remains inside its envelope. Similar conclusions hold for other species, pointing to the impact of the underlying network. The main difference between the distribution of dbh values for the considered species might be the fact that, for small interpoint distances, mark correlation functions for all species have larger values than the expected value under mark independence, except type Populus, for which mark correlation functions stay below the corresponding envelopes for very small interpoint distances. Note that this means for a pair of trees of the type Populus with very small interpoint distances, the dbh value for at least one of them is pretty small.

\section{Discussion}\label{sec:final}

Methods for analyzing marked spatial point processes, when each point is augmented by some auxiliary quantity, have recently witnessed impressive developments. Apart from some extensions of the current methodological toolboxes to the case of point processes with events on linear networks/graphs \citep{jammalamadaka2013statistical, BaddeleyJammalamadakaNair2014}, parts of the recent literature proposed generalizations to more challenging scenarios where marks are non-scalar. Extensions to the settings where marks are function-valued were considered, followed by some recent contributions covering functional characteristics, where marks are inherently structured objects including compositions, densities, and graphs \citep{EckardtMariCMSPP}. In this paper, we have restricted our focus to general marked summary characteristics when marks are integer/real/object-valued and points live on either planar spaces or linear networks. By combining the general marked characteristic ideas with linear network settings, we have not only reviewed the current methodologies for point processes in $\R^2$ but, in particular, introduced novel characteristics for marked point processes on linear networks.


Despite all methodological progress, most of the proposed methodologies are limited to stationary point processes, imposing two restrictions: (i) 
it is quite rare to have stationary processes in practice, and (ii)
stationarity for point processes on linear networks is quite limited and challenging \citep{baddeley2017stationary,cronie2020inhomogeneous}. In fact, among all developed methods, only the cross/dot-type summary characteristics and mark-weighted summary characteristics are defined for both homogeneous and inhomogeneous point processes. This fact could limit the analysis of marked point processes, especially when marks are real-valued and/or object-valued \citep{Eckardt2023MultiFunctionMarks}. Further, the growing availability of complex point patterns calls for suitable methods to investigate the dependencies between different types of object-valued marks, e.g. compositions and shapes, which is not available under the present methodological toolbox. Besides, there are various introduced local point characteristics in the literature \citep{Anselin1995, LISABundlesCressie}, but their extensions for marks have not been established so far; current mark characteristics do not disclose local behaviors. In addition, although still recognized within a small niche of research, spatial frequency domain methods might become useful tools for modern mark scenarios and help avoid burdensome computations.  

Most of the proposed marked characteristics are not available in space-time settings where marks are simultaneously both space- and time-dependent; an application of such a case is trajectories \citep{moradi2018spatial}. Further, in the case of function-valued marks, marks might undergo abrupt and/or gradual changes, which may not be accounted for in the current settings \citep{moradi2023hierarchical}. Apart from the discussed methods, suitable extensions of point-to-mark tools \citep{Schlather2004} to more challenging state spaces and/or marks have not been established. We further highlight the lack of methods for marked point processes on a sphere. 
As a final limitation, it would also be interesting to develop methodologies and test functions that can incorporate negative marks. 



\section*{Acknowledgement}

The authors gratefully acknowledge financial support through the German Research Association. Matthias Eckardt was funded by the Walter Benjamin grant 467634837 from the German Research Foundation.

\bibliography{Mppp}
\bibliographystyle{chicago}

\end{document}